\documentclass[12pt,a4paper,notitlepage]{article}
\usepackage[utf8]{inputenc}
\usepackage[english]{babel}
\usepackage[T1]{fontenc}
\usepackage{hyperref}
\usepackage{amsmath}
\usepackage{stmaryrd}
\usepackage{amsfonts}
\usepackage{amssymb}
\usepackage{color} 
\usepackage{caption}
\usepackage{amsmath}
\usepackage{relsize,exscale}
\usepackage{array,multirow,makecell}

\usepackage[toc,page]{appendix}

\setcellgapes{1pt}
\makegapedcells
\newcolumntype{R}[1]{>{\raggedleft\arraybackslash }b{#1}}
\newcolumntype{L}[1]{>{\raggedright\arraybackslash }b{#1}}
\newcolumntype{C}[1]{>{\centering\arraybackslash }b{#1}}

\newcommand{\G}{\mathcal{G}}
\newcommand{\Q}{\mathrm{Q}}
\newcommand{\F}{\mathrm{F}}
\newcommand{\B}{\mathfrak{B}}

\newtheorem{definition}{Definition}
\newtheorem{proposition}{Proposition}

\newtheorem{corollary}{Corollary}

\newcommand{\sym}{\mathrm{Sym}}

\newtheorem{lemma}{Lemma}
\usepackage{enumerate}
\usepackage{subfig}
\usepackage{graphicx}

\newcommand{\beq}{\begin{equation}}
\newcommand{\eeq}{\end{equation}}
\newcommand{\bea}{\begin{eqnarray}}
\newcommand{\eea}{\end{eqnarray}}
\definecolor{mygray}{gray}{0.3}

\newcommand{\bes}{\begin{eqnarray}}
\newcommand{\ees}{\end{eqnarray}}

\newcommand\restr[2]{{
  \left.\kern-\nulldelimiterspace 
  #1 
  \vphantom{\big|} 
  \right|_{#2} 
  }}

\usepackage{multicol}
\usepackage{amssymb}

\makeatletter
\begin{document}
\begin{center}
\textbf{\Large{Nonperturbative renormalization group beyond
melonic sector}}\\
\medskip
\textbf{\large{The Effective Vertex Expansion method for group
fields theories}}
\vspace{15pt}

{\large Vincent Lahoche$^a$\footnote{vincent.lahoche@cea.fr}  \,\,and 
Dine Ousmane Samary$^{a,b}$\footnote{dine.ousmanesamary@cipma.uac.bj}
 }
\vspace{15pt}

a)\,  Commissariat à l'\'Energie Atomique (CEA, LIST),
 8 Avenue de la Vauve, 91120 Palaiseau, France

b)\, Facult\'e des Sciences et Techniques (ICMPA-UNESCO Chair),
Universit\'e d'Abomey-
Calavi, 072 BP 50, Benin
\vspace{0.5cm}
\end{center}
\begin{center}
\textbf{Abstract}
\end{center}
Tensor models admit the large $N$ limit dominated by the graphs
called melons. The melons are characterized by the Gurau number
$\varpi=0$ and the amplitude of the Feynman graphs are
proportional to $N^{-\varpi}$. Other leading order contributions
i.e. $\varpi> 0$ called pseudo-melons can be taken into account
in the renormalization program. The following paper deals with
the renormalization group for a $U(1)$-tensorial group field theory
model taking into account these two sectors (melon and
pseudo-melon).  It generalizes a recent work (Lahoche and Ousmane Samary, arXiv:1803.09902), in which only the melonic sector has been studied.
Using the power counting
theorem the divergent graphs of the model are identified. Also, the effective vertex expansion is used to generate in detail the combinatorial analysis of these two leading
order sectors.
We obtained the  structure equations,
which help to improve the truncation in the Wetterich equation.
The set of Ward-Takahashi identities is derived and their
compatibility along the flow provides a nontrivial constraint
in the approximation schemes. In the symmetric phase the
Wetterich flow equation is given and the numerical solution is studied.
  \newpage
\tableofcontents
\bigskip
\section{Introduction}
The consistent formulation of the quantum theory of gravity (QG)
is one of the fundamental and tedious problems of modern physics,
which remains unsolved, and most probably intertwines between
quantum mechanics  and general relativity (GR). Its has
evolved a lot since the past two decades due to the appearance
of new background independent approaches such as loop quantum
gravity, dynamical triangulations, and noncommutative geometry (see
\cite{Rovelli:1997yv}-\cite{Aastrup:2006ib} and references
therein). Recently, tensor models (TMs) and group field
theories (GFTs) are developed as new way to investigate this
conundrum question \cite{Oriti:2006ar}-\cite{Freidel:2009hd}. TMs
generalize matrix models and are considered as a convenient
formalism for studying random geometry in dimensions $D\geq 3$.
Its provides a well-defined framework for addressing QG in higher
dimensions and its cortege of consequences on integrable systems
\cite{Rivasseau:2016zco}. GFTs are quantum field theories over
the group manifolds and are considered as a second quantization
of loop quantum gravity \cite{Oriti:2014yla}. GFT is
characterised by a specific form of nonlocality in their
interactions, with the basis variable being a complex field,
function of d-group elements \cite{Oriti:2006ar},
\cite{Oriti:2018dsg}. Recently TMs and
GFTs are merged to provide the so-called
tensorial group field theory (TGFT)
\cite{Carrozza:2012uv}-\cite{Carrozza:2017vkz}. It can also be
viewed as a new proposal for quantum field theories based on a
Feynman path integral, which generates random graphs describing
simplicial pseudo-manifolds. It aims at providing a content to a
phase transition called geometrogenesis scenario by relating a
discrete quantum pregeometric phase of our spacetime to the
classical continuum limit consistent with Einstein GR
\cite{Oriti:2013jga}-\cite{Wilkinson:2015fja}. In short, within
this approach, our spacetime and its geometry has to be
reconstructed or must emerge from more fundamental and discrete
degrees of freedom. Its very encouraging features such as
renormalization of large class of models and asymptotic freedom
on the one hand \cite{BenGeloun:2012yk}-\cite{Carrozza:2014rba},
and a coexistence with a Wilson-Fisher fixed point, on the other
hand \cite{Geloun:2016qyb}-\cite{Lahoche:2016xiq}, ensure not
only the quantum consistency at macroscopic scales but also, the
possible existence of the condensate phase
\cite{Gielen:2014uga}.

The renormalization of TGFT models started with the work given
in \cite{BenGeloun:2012pu}-\cite{BenGeloun:2011rc}.  The multiscale analysis and power
counting theorem are used to prove that the $U(1)$-tensor models in three and four dimensions
 are just renormalizable. Very quickly several other
interesting models are studied and have proved to be just and
super-renormalizable
\cite{Carrozza:2012uv}-\cite{Geloun:2011cy}. The classification of the renormalizable TGFT models framework is also  investigated \cite{Geloun:2013saa}.  The computation of
the $\beta$-functions to prove the asymptotic freedom and safety
is also given \cite{BenGeloun:2012yk}-\cite{Carrozza:2014rba}.
Very recently the functional renormalization group (FRG) method is
introduced in the context of TGFT and allowed   to solve the flow equations of these various models
\cite{Geloun:2016qyb}-\cite{Lahoche:2016xiq}. The occurrence of
nonperturbative fixed points and their critical behavior in the
UV and IR is studied. The confirmation of the asymptotic
freedom and safety are also given.  The FRG
for TGFT \cite{Benedetti:2014qsa} derived from the method
used in the case of matrix models
\cite{Wilson:1971dh}-\cite{Eichhorn:2013isa} can be simply
applied when the dimension of tensors is not very high $(d=3,
d=4, d=5)$. In the case where $d\geq 6$ due to the combinatoric,
other technical methods require to be proposed. In
\cite{Benedetti:2015yaa} this question is solved and turned on
a new  way for  investigating the FRG to higher dimensions tensors models \cite{Carrozza:2016tih}-\cite{Carrozza:2017vkz}.
FRG can be roughly described as a flow in a certain infinite
dimensional functional space for actions, the theory space.
The scale plays the role of time for this flow. Its allows the
construction of a set of effective action $\Gamma_s,\,\,
-\infty\leq s\leq +\infty$, which interpolates between the
classical action $S$ and the full effective action $\Gamma$ such
that this full effective action is obtained for the value
$s=-\infty$ \cite{Wetterich:1989xg}. $\Gamma$ is also called the
generating functional of one-particle irreducible vertices. At
the same time when $s$ walks $\mathbb{R}$ the flow equations
enable us to interpolate smoothly between the UV laws and the IR
phenomena for our  systems.
The flow equations are described by the Wetterich equation in which
the choice of IR regulator and the full effective action remains
the only condition to provide solutions and probably maybe help to derive the
fixed points.

The Feynman graphs of TM can be organized as a series in $1/N$
and therefore the class of combinatorial objects can be
selected.
In the large $N$ limit the dominant graphs are called melon 
\cite{Gurau:2011xq}-\cite{Gurau:2010ba}. This limit allows us to
understand the statistical physics properties such as continuum
limits, phase transitions and critical exponents. Taking into
account the Gurau $1/N$-expansion the amplitudes of the Feynman
graphs are proportional to $N^{-\varpi}$, ($\varpi$ is the Gurau
degree), and the melonic contributions are characterized by
$\varpi=0$. However, it might be possible to consider the
non-melonic leading order contributions $\varpi\neq 0$, which we
call in this work the pseudo-melonic graphs. The canonical
dimension of the melons and the pseudo-melons is identified for the $n$-point graphs.
The  new leading order contributions (the pseudo-melons) modify drastically the power counting
theorem and the renormalization properties of the class of
models studied in this direction. In the nonperturbative
analysis of the TGFT with melonic and pseudo-melonic graphs the
FRG should be carefully use. There are several reasons to
consider the mixing melon and pseudo-melon see \cite{Bonzom:2015axa}-\cite{Bonzom:2012hw} for more explanation. The combinatorial analysis of these two sectors generated
an intermediate sector between melon and pseudo-melon and should be taken into account in the renormalization program. This point will be discussed in detail throughout this work.

Recently one new breakthrough is done in the context of FRG of TGFT models,  to improve the truncation
and to choose the regulator in the appropriate way \cite{Lahoche:2018vun}. This is
possible by adding in the Wetterich equation, the so-called structure
equations and the Ward-Takahashi (WT) identities. In the case of symmetric phase a
nontrivial UV attractive fixed point is given.
The WT identities are used to define the nontrivial constraint
on the flows and therefore the method proposed in
\cite{Lahoche:2018vun} is totally different from the usual FRG
method. Despite all the results in this recent contribution, a
lot of questions need to be addressed. First the non-melonic
leading order contribution should been taken into account in the
flow equations. The non-symmetric phase needs also to be
scrutinized using probably the intermediate field representation. The
purpose of the following work is to provide the FRG by taking
into account the leading order contribution (the melonic and
pseudo-melonic graphs) in the flow equations. The new power
counting theorem is derived and the classification of the graphs
that contributes to these two sectors is given. The structure
equations and the set of WT identities are used to provide a
nontrivial constraint on the reliability of the approximation
schemes, especially on the truncation and the choice of the
regulator. The comparisons between these new results and what
 we obtain in the case of usual  truncation is also given.

The paper is organized as follows: In section \eqref{sec2} 
we provide the useful definitions and notations, which will be used  throughout
the paper. Particularly we give the definition of our model and its symmetries, and then  introduce the FRG method by giving  the Wetterich equation. In section \eqref{sec3} the effective vertex expansion is studied. We identified the renormalization sector by using the power counting theorem. We also computed the canonical dimension of the $n$-point Feynman graphs that contribute to melons and pseudo-melons independently. In section \eqref{sec4} the same analysis as the previous section is given by mixing these two sectors. By contracting the elementary melon we generate the family of six point vertices's in which we considered only the non-branching sector.  We identified one new leading order contribution that we called intermediate sector. The structure equations for melons pseudo-melons and intermediate graphs  are also given carefully. Section \eqref{sec5} is devoted to the FRG analysis of the model using not only the structure equations but also the set of WT-identities as   nontrivial constraints  in the approximation schemes. The phase diagrams around the fixed points are also built and examined. We then provide discussion between our method and what we obtained in the case of ordinary truncation. In section \eqref{sec6} discussions and conclusion of our work is given. The set of two Appendices is given. In Appendix \eqref{AppB} the computation of the useful formulas concerning convergent sums that we used in the core of this paper is given. In Appendix \eqref{AppA} the usual FRG analysis for a model is given. chosen the truncation in appropriate way and the regulator the flow of the couplings and mass are given.  The proof of the asymptotical safety is also scrutinized.

\section{Flowing on tensorial group field theory space
}\label{sec2}

A tensorial group field theory (TGFT) is a field theory defined
on a direct product of group manifolds. In this paper we focus
on an Abelian group field theory, defined on $d$-copies of the
unitary group $U(1)$ isomorphic to the complex numbers of module
$1$. We consider a pair of fields, say $\phi$ and $\bar\phi$ on
$U(1)^d$
\begin{equation}
\phi,\bar\phi:U(1)^d\to \mathbb{C}\,.
\end{equation}
The dynamics of the TGFT model is governed by the
\textit{classical action} $S(\phi, \bar{\phi})$ chosen to be of
the form
\begin{equation}
S(\phi, \bar{\phi})=\int d\textbf{g}\,
\bar{\phi}(\textbf{g})(-\Delta_{\textbf{g}}+m^2)\phi(\textbf{g})+S_{\text{int}}(\phi,
\bar{\phi})\,,\label{classicalaction}
\end{equation}
where $\Delta_{\textbf{g}}$ denotes the Laplace-Beltrami
operator on the torus $U(1)^d$ and
$\textbf{g}:=(g_1,\cdots,g_d)\in U(1)^d$. For a tensorial
theory, the interaction $S_{int}$ is a sum of \textit{connected
tensorial invariants} $S_{\text{int}}(\phi,
\bar{\phi})=\sum_n\mathcal{V}_n(\phi, \bar{\phi})$ made with an
equal number $n$ of fields $\phi$ and $\bar{\phi}$, whose
arguments are identified and summed only between $\phi$ and
$\bar{\phi}$. A generic interaction term in $\mathcal{V}_n$ is
then of the form:
\begin{equation}
\upsilon_h(\phi, \bar{\phi})=\lambda_h\int \prod_{p=1}^n
d\textbf{g}_p\,d \bar{\textbf{g}}_p\, \prod_{p=1}^n
\phi(\textbf{g}_p)\bar{\phi}(\bar{\textbf{g}}_p)
\prod_{p=1}^n\prod_{i=1}^d\,\delta(g_{ip}-\bar{g}_{ih_i(p)})\,,\label{eqexample}
\end{equation}
where $\delta$ denotes the Dirac delta distribution over $U(1)$
and $\lambda_h$ denotes the \textit{coupling constant}. The
interaction as well as the coupling are indexed with a set of
maps $h:=\{h_i\,,i=1,\cdots,d\}$ such that for any
$p\in\llbracket 1,n\rrbracket$; $h_i(p)\in \llbracket
1,n\rrbracket$. We can then associate black and white nodes
respectively for fields $\phi$ and $\bar{\phi}$, and a link
between black and white nodes for each delta, labeled by a color
index running from $1$ to $d$. As a result, each interaction can
be then labeled from an unique \textit{colored bipartite regular
graph} rather than with the map $h$; and $\mathcal{V}_n$ may be
decomposed as a sum of terms indexed with such a graphs
$\gamma_n$ with $n$ white nodes:
\begin{equation}
\mathcal{V}_n(\phi,
\bar{\phi})=\sum_{\gamma_n}\upsilon_{\gamma_n}(\phi,
\bar{\phi})\,.
\end{equation}
For the rest of this paper, we call \textit{valence} the integer
$n$. Such graphs are called \textit{tensorial invariants
bubbles}, or simply \textit{bubbles}. As an example, for $n=4$
and $d=5$ we get:
\begin{equation}
\mathcal{V}_4(\phi, \bar{\phi})=\sum_{i=1}^d
\,\lambda_{4,i}\,\vcenter{\hbox{\includegraphics[scale=0.8]{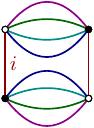}
}}+\sum_{j<i}^d
\,\lambda_{4,ij}\,\vcenter{\hbox{\includegraphics[scale=0.8]{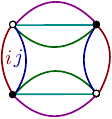}
}}\,,
\end{equation}
where each color on the graphs corresponds to one of the integer
of the set $\llbracket 1,d\,\rrbracket$, and $\lambda_{4,i}$,
$\lambda_{4,ij}$ denote {coupling constants}. Note that in this
representation we identified the diagram with the interaction
itself. For convenience, in the rest of this paper, we do not
work on $U(1)^d$ but on its Fourier dual space $\mathbb{Z}^d$.
Fixing the group representation, any element $g\in U(1)$ can be
uniquely represented as a complex number of module $1$:
$g=e^{i\theta}$ with $\theta\in [0,2\pi[$. The fields can then
be viewed as functions over the complex unit circle, depending
on $d$--angular coordinates $\phi(\theta_1,\cdots,\theta_d)$ and
$\bar{\phi}(\theta_1,\cdots,\theta_d)$, so that their Fourier
decomposition writes as
\begin{equation}
\phi(\theta_1,\cdots,\theta_d)=\sum_{\vec{p}\in\mathbb{Z}^d} \,
T_{\vec{p}}\, e^{i\sum_{j=1}^d \,\theta_jp_j}\,,\quad
\bar{\phi}(\theta_1,\cdots,\theta_d)=\sum_{\vec{p}\in\mathbb{Z}^d}
\, \bar{T}_{\vec{p}}\, e^{-i\sum_{j=1}^d \,\theta_jp_j}\,.
\end{equation}
$T$ and $\bar{T}$ are maps from $\mathbb{Z}^d$ to $\mathbb{C}$.
They are $d$-\textit{tensors} with infinite size. In Fourier
components, the classical action \eqref{classicalaction}
becomes, for $d=5$:
\begin{equation}
S(T,\bar{T})=\sum_{\vec{p}}
\,\bar{T}_{\vec{p}}\,(\vec{p}\,^2+m^2\,)T_{\vec{p}}\,+\lambda_{4,1}\sum_{i=1}^d\vcenter{\hbox{\includegraphics[scale=0.8]{melon1.pdf}
}}+\lambda_{4,2}\sum_{j<i}^d
\vcenter{\hbox{\includegraphics[scale=0.8]{PS1.pdf}
}}\,+\cdots\,,
\end{equation}
where we have chosen the same coupling for the two sets of
interactions pictured above. In Fourier space, the Dirac delta
occurring in the interactions like in \eqref{eqexample} become
discrete Kronecker delta. Because their explicit expressions
shall be useful in the next sections, we give the decomposition
of the following two diagrams:
\begin{equation}
\vcenter{\hbox{\includegraphics[scale=0.7]{melon1.pdf}
}}=:\sum_{\{\vec{p}_i\}}
\mathcal{V}^{(4,1)\,i}_{\vec{p}_1,\vec{p}_2,\vec{p}_3,\vec{p}_4}
T_{\vec{p}_1}\bar{T}_{\vec{p}_2}T_{\vec{p}_3}\bar{T}_{\vec{p}_4}\,,\,\,\vcenter{\hbox{\includegraphics[scale=0.7]{PS1.pdf}
}}=:\sum_{\{\vec{p}_i\}}
\mathcal{V}^{(4,2)\,ij}_{\vec{p}_1,\vec{p}_2,\vec{p}_3,\vec{p}_4}
T_{\vec{p}_1}\bar{T}_{\vec{p}_2}T_{\vec{p}_3}\bar{T}_{\vec{p}_4}\,,
\end{equation}
where:
\begin{equation}
\mathcal{V}^{(4,1)\,i}_{\vec{p}_1,\vec{p}_2,\vec{p}_3,\vec{p}_4}=\delta_{p_{1i}p_{4i}}\delta_{p_{2i}p_{3i}}\prod_{j\neq
i} \delta_{p_{1j}p_{2j}}\delta_{p_{3j}p_{4j}}\,,
\end{equation}
and:
\begin{equation}
\mathcal{V}^{(4,2)\,ij}_{\vec{p}_1,\vec{p}_2,\vec{p}_3,\vec{p}_4}
=\prod_{l=i,j}\delta_{p_{1l}p_{4l}}\delta_{p_{2l}p_{3l}}\prod_{k\neq
i,j} \delta_{p_{1k}p_{2k}}\delta_{p_{3k}p_{4k}}\,.
\end{equation}
In the literature, the first diagrams are known as
\textit{melonics}. For our purpose, we then denote as
\textit{pseudo-melonics} the second ones with two weak edges
rather than one. Because of their importance in the rest of this
paper, we will denote by $\mathfrak{B}_i$ the four-valent melonic
diagrams and $\mathfrak{B}_{ij}$ the four-valent pseudo-melonic
ones. The indices $i$ and $ij$ refer to the \textit{weak
edges} in both cases. \\

\noindent
Among their properties, the tensorial interactions have revealed
a new and nontrivial notion of locality, said
\textit{traciality}, which is the only one appropriate to deal with nonlocal structure of the interactions over the group
manifold. In particular, traciality allowed to renormalize the
quantum field theory version of these classical theory, and it
plays an important role in the building of their
\textit{renormalization group flow} \cite{Carrozza:2013mna}.
Without additional gauge invariance like \textit{closure
constraint}, traciality reduces to tensoriality
\cite{Lahoche:2015ola}. Then we retain the following definition:
\begin{definition}
A connected tensorial invariant bubble interaction is said to be
local. In the same footing, any interacting action expanded as a
sum of such diagrams is said to be local. \label{deflocal}
\end{definition}

\noindent
The quantum theory is then defined from the partition
function\footnote{Strictly speaking the term "quantum" is
abusive, we should talk about statistical model, or quantum
field theory in the Euclidean time.}:
\begin{equation}
Z(J,\bar{J})=\int
dTd\bar{T}\,e^{-S(T,\bar{T})+J\bar{T}+\bar{J}T}\,,
\end{equation}
where $dT$ (respectively $d\bar{T}$) is the standard Lebesgue measure
for path integration and
$J\bar{T}:=\sum_{\vec{p}}\,J_{\vec{p}}\bar{T}_{\vec{p}}$.
Because of the ultraviolet (UV) divergences, we introduce a
regularization which suppresses the high momenta contributions.
There are different choices of regularization functions. The
most common for renormalization are Schwinger and sharp
regularizations. For our purpose, it is suitable to consider the
sharp regularization, such that the UV regularized free
two-point functions is:
\begin{equation}
C_\Lambda(\vec{p},\vec{p}\,^\prime):=\frac{\Theta(\Lambda^2-\vec{p}\,^2)}{\vec{p}\,^2+m^2}\,\delta_{\vec{p}\vec{p}\,^\prime}\,.
\end{equation}
where $\Theta$ is the Heaviside step function. The presence of
the Laplacian propagator in the classical action
\eqref{classicalaction} generates a canonical notion of
\textit{scale} over the Feynman graphs. For TGFTs, these graphs
are dual to the simplicial topological manifold, and then generate
discretization of topological spaces from the perturbative
expansion itself. An example of such a Feynman graph is given in
Figure \eqref{fig1}, the dotted edges being Wick contractions
between $T$ and $\bar{T}$ fields. We conventionally attribute
the color $0$ to these edges.\\
\begin{center}
\includegraphics[scale=1]{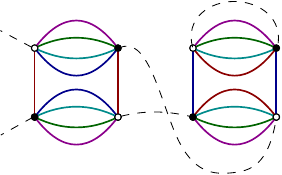} 
\captionof{figure}{A typical Feynman graph contributing to the
perturbative expansion of the connected two-point functions. The
dotted edges correspond to free propagator contractions between
pair of fields.} \label{fig1}
\end{center}
An important notion for tensorial Feynman diagrams is the notion
of \textit{faces}, which we recall in the following definition:
\begin{definition}
A face is defined as a maximal and bicolored connected subset of
lines, necessarily
including the color 0. We distinguish two cases:\\

\noindent
$\bullet$ The closed or internal faces, when the bicolored
connected set correspond to a cycle.\\

\noindent
$\bullet$ The open or external faces when the bicolored
connected set does not close as a cycle.\\

\noindent
The boundary of a given face is then the subset of its dotted
edges, and its length is defined as the number of internal
dotted edges on its boundary.
\end{definition}

\noindent
To complete this definition, we provide what we call
internal/external edges and interior/boundary vertices:
\begin{definition}\label{definitionTomTom}
On a given Feynman graph, the set of edges split into internal
and external edges. External edges come from the Wick
contraction with external fields and internal edges come from
the Wick contractions between vertex fields. Moreover, a vertex
is said to be a boundary vertex if at least one of the external
edges is hooked to him. It is an interior vertex otherwise.
Finally, we define the interior of a Feynman diagram as the set
of internal vertices with dotted edges.
\end{definition}

\noindent
The quantum fluctuation can then be integrated out from higher
to lower scales, generating a sequence of effective theories
which describes a curve into the theory space. Along the
trajectories, all the coupling constants move from their initial
definition, their running describing the \textit{renormalization
group flow}. The FRG is a specific method to build such a
running, well adapted to the TGFT context. To parametrize the
flow, we then introduce a real parameter
$s\in\,]-\infty,+\infty[$ and a one parameter family of models
$Z_s$ such that:
\begin{equation}
Z_s[J,\bar{J}]:=\int
dTd\bar{T}\,e^{-S(T,\bar{T}\,)-R_s(T,\bar{T}\,)+\bar{J}T+\bar{T}J}\,.
\end{equation}
The regulator term $R_s(T,\bar{T}\,)$ depends on the running
scale $k=e^s$ such that $\Lambda\geq k\geq 0$, and introduces a
dynamical splitting into high and low scale. It acts as a mass
term which decouples the low scale contributions from the long
distance physics, allowing to build an effective action
$\Gamma_s$ for the long distance observable, such that higher scale
fluctuations have been integrated out and the coupling constants
moved from their initial values to their effective values at the
given scale $k$. \\

\noindent
To make this more concrete, the regulator is chosen of the form
\cite{Wetterich:1989xg}-\cite{Schnoerr:2013bk}.
\begin{equation}
R_s(T,\bar{T}\,)=\sum_{\vec{p}\in\mathbb{Z}^d}
\,r_s(\vec{p}\,^2) \bar{T}_{\vec{p}}\,T_{\vec{p}}\,,
\end{equation}
so that the \textit{regulator function} $r_s(\vec{p}\,)$ may be
suitably introduced in the kinetic part of the action, providing
us the effective Gaussian propagator :
\begin{equation}
C_s(\vec{p}\,^2)=\frac{1}{\vec{p}\,^2+m^2+r_s(\vec{p}\,^2)}\,,
\end{equation}
where we assumed $\Lambda\gg k$ so that the UV regulator
disappears. The regulator function satisfies some properties
ensuring that fluctuations are well integrated, among them (see
\cite{Litim:2000ci} for more details):
\begin{itemize}
\item $r_{s}\geq 0$,
\item $\lim_{s\rightarrow -\infty}r_{s}=0$,
\item $\lim_{s\rightarrow \infty}r_{s}=\infty\,.$
\end{itemize}
The last condition ensures that the initial conditions are those
imposed by the classical action itself. Moreover, the second
condition ensures that all the fluctuations are integrated out
in the deep infrared (IR). The first condition finally protects
the flow from singularities. Note that UV and IR correspond
respectively to the ending points of the domain of $s$:
$k=\infty$ in the UV sector and $k=0$ in the IR. The central
object in the FRG approach is the \textit{averaged action}
$\Gamma_s$, defined as the (slightly modified) Legendre
transform of the free energy $W_s:=\ln(Z_s)$:
\begin{equation}\label{Legendre}
\Gamma_s[M,\bar{M}]+R_s[M,\bar{M}]=\bar{J}M+\bar{M}J-W_s[J,\bar{J}]\,,
\end{equation}
where $M$ (respectively  $\bar{M}$) is the means field:
\begin{equation}
M:=\frac{\partial W_s}{\partial \bar{J}}\,\qquad
\left(\text{respectively}\,\bar{M}=\frac{\partial
W_s}{\partial{J}}\right)\,.\label{means}
\end{equation}
The presence of the regulator on the left-hand side ensures the
following initial conditions hold:
\begin{equation}
\Gamma_{s=-\infty}=\Gamma\,,\qquad \Gamma_{s=+\infty} =S\,,
\end{equation}
where $S$ is the classical action and $\Gamma$ the full
effective action for $r_s=0$. The renormalization group flow is
then described from an order one nonlinear differential
equation for $\Gamma_s$ \cite{Berges:2000ew}:
\begin{equation}
\dot{\Gamma}_s=\sum_{\vec{p}}\dot r_s(\vec{p}\,^2)
G_s(\vec{p},\vec{p})\,,\label{Wett}
\end{equation}
where the dot designates the derivative with respect to $s$ i.e.
$\dot{\Gamma}_s:=\frac{\partial \Gamma_s}{\partial s}$ and
\begin{equation}
G_s^{-1}:=\frac{\partial^2\Gamma_s}{\partial M\partial\bar
M}+r_s=:\Gamma^{(2)}_s+r_s\,,\label{propar}
\end{equation}
is the effective two-point function. Moreover, the regulator is
chosen such that only a finite window of momenta contributes to
the sum, ensuring that it is finite both in UV and IR. This
equation is both simple and complicated. Simple, because it is
of order one with respect to the flow parameter $s$. Moreover,
it only involves a single effective loop rather than a
multiloop expansion as in the standard Wilson-Polchinski approach,
and is then well adapted to nonperturbative considerations.
However, this equation is also highly nonlinear, and except for
very special cases, it is impossible to solve it exactly.
Extracting from it a nonperturbative information on the
renormalization group flow then requires approximations. The most
popular for TGFT are truncations
\cite{Geloun:2016qyb}-\cite{Lahoche:2016xiq}. With this method,
the flow in the full theory space is projected into a reduced
dimensional subspace. Another method, already considered in
\cite{Lahoche:2018vun} and that we will use in this paper may be
called \textit{effective vertex method}. Taking successive
derivatives of $\Gamma_s$ in flow equation \eqref{Wett}, we then
get an infinite hierarchical system, expressing
$\dot{\Gamma}^{(n)}_s$ in terms of $\Gamma^{(n+2)}_s$ and
$\Gamma^{(n+1)}_s$. The effective vertex method stops the
infinite hierarchical tower of equation expressing
$\Gamma^{(n+2)}_s$ and $\Gamma^{(n+1)}_s$ in terms of
derivatives up to a certain $n$, so that the system of equation
becomes closed. The importance of this new method comes from the
fact that we keep the full momentum dependence on the effective
vertex. We will compare the two methods in the last section, and
point out the new behavior coming from the full momentum
dependence. \\

\noindent
Now let us give an important aspect about the construction of
the flow and the difference between symmetric and non-symmetric
phases. Usually, these terms refer to the value of the mean
field $M$. For $M=0$, the model is said to be in the symmetric
phase. Otherwise the model is said to be in the non-symmetric
phase. For tensorial group field theories, we adopt another
definition, already considered in \cite{Lahoche:2018vun}:
\begin{definition} \textbf{Symmetric and non-symmetric phases}
As long as the effective two-points function $G_s$ remains
diagonal: $G_s(\vec{p},\vec{q}\,)\propto
\delta_{\vec{p}\,\vec{q}}$, the theory is
said to be in the symmetric or perturbative phase. If this is
not the case the theory is said to
be in the non-symmetric or nonperturbative regime.
\end{definition}
This definition comes from the fact that in a perturbative
regime, all the 1PI correlations functions of the form
$\Gamma_s^{(2n+1)}$ vanish. Moreover, the conservation of the
external momenta running along the boundaries of the external
faces ensure the presence of a global conservation Kronecker
delta $\delta_{\vec{p}\,\vec{q}}$. For instance, let us consider
the violet external face on the Feynman graph given on Figure
\eqref{fig1}:
\begin{equation}\label{VincentL}
\includegraphics[scale=1]{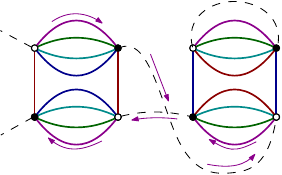} \,.
\end{equation}
The path of the boundary of the violet face in equation
\eqref{VincentL} is indicated with arrows. Each propagator shares
a Kronecker delta, as well as each colored edge. Then, the
momentum is already conserved along the path. The same property
is true for all colors. Now, let us consider
$\Gamma_{s;\vec{p}\,\vec{q}}^{(2)}\,(M,\bar{M})$ for a non
vanishing mean field. Expanding it in powers of these fields,
we get:
\begin{equation}
\Gamma_{s;\vec{p}\,\vec{q}}^{(2)}(M,\bar{M})=\Gamma_{s;\vec{p}\,\vec{q}}^{(2)}\,(0,0)+\sum_{\vec{p}_1,\vec{p}_2}\Gamma_{s;\vec{p}\,\vec{q};\vec{p}_1,\vec{p}_2}^{(4)}(0,0)\,M_{\vec{p}_1}\bar{M}_{\vec{p}_2}+\cdots\,.\label{meanfieldexp}
\end{equation}
Perturbatively, the first term of the expansion
$\Gamma_{s;\vec{p}\,\vec{q}}^{(2)}(0,0)$ is proportional to
$\delta_{\vec{p}\,\vec{q}}$. However, it is not the case of the
next term. Indeed, a leading order contribution to the second
term could be, at the first order in perturbative expansion for
$\Gamma_{s;\vec{p}\,\vec{q};\vec{p}_1,\vec{p}_2}^{(4)}(0,0)$:
\begin{equation}
\vcenter{\hbox{\includegraphics[scale=1]{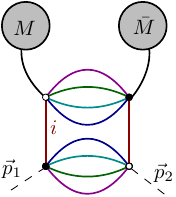}}} \propto
\sum_{\vec{k}\in\mathbb{Z}^{d-1}} M_{\vec{k},p_{1i}}
\bar{M}_{\vec{k},p_{2i}}\,\prod_{j\neq i}
\,\delta_{p_{1j}p_{2j}}\,,
\end{equation}
where the fat black edges correspond to Kronecker delta
contractions. As a result, except for very special mean field
configurations ensuring that $$
\sum_{\vec{k}\in\mathbb{Z}^{d-1}} M_{\vec{k},p_{1i}}
\bar{M}_{\vec{k},p_{2i}}\propto \delta_{p_{1i}p_{2i}}\,,$$
$\Gamma_{s;\vec{p}\,\vec{q}}^{(2)}$ cannot be diagonal for non
vanishing means fields\footnote{Note that the last condition
holds for theory with \textit{closure constraint}, a gauge
invariance ensuring that, ‘‘on shell", $\sum_{i=1}^d p_i=0$. If
$M$ and $\bar{M}$ are on shell, the product $M_{\vec{k},p_{1i}}
\bar{M}_{\vec{k},p_{2i}}$ is therefore proportional to $
\delta_{p_{1i}p_{2i}}$.}. In the rest of this paper, we consider
only the symmetric phase and the effective vertices then can be
considered as the first terms correction in means field
expansion like in \eqref{meanfieldexp}. \\

\noindent
Finally let us end this section by introducing the notion of
\textit{boundary graph}:
\begin{definition}
Consider $\mathcal{G}$  as a connected Feynman graph with $2N$ external
edges. The boundary graph $\partial\mathcal{G}$ is obtained from
$\mathcal{G}$ keeping only the external blacks and whites nodes
hooked to the external edges, connected together with colored
edges following the path drawn from the boundaries of the
external faces in the interior of the graph $\mathcal{G}$.
$\partial\mathcal{G}$ is then a tensorial invariant itself with
$N$ black (respectively  whites) nodes. An illustration is given on
Figure \eqref{fig2}.
\end{definition}

\begin{center}
$\vcenter{\hbox{\includegraphics[scale=1]{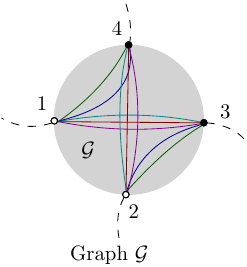} }}\quad
\to \quad\vcenter{\hbox{\includegraphics[scale=1]{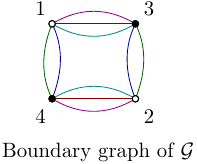}
}} $
\captionof{figure}{An opening Feynman graph with four external
edge and its boundary graph. The strand in the interior of
$\mathcal{G}$ represents the path following by the external
faces.}\label{fig2}
\end{center}

\section{Melonic and pseudo-melonic sectors}\label{sec3}

This section addresses the renormalization of a TGFT model, mixing standard melons and a new interacting sector called pseudo-melonic.  First , we recall some properties of renormalizable theories and consider the purely melonic sector, which has been shown to be just-renormalizable for $d=5$. Second, we define the new family that we call pseudo-melons, and show that it is power-counting renormalizable for $\phi^6$ interactions. 

\subsection{Renormalizable sectors}

Basically, a renormalizable sector is a proliferating family of
divergent graphs having the same combinatorial structure and the
same power counting, such that their divergences can be canceled
from a finite set of counter-terms. For our model with a kinetic
Laplacian term as a boundary condition in the UV, the degree of
divergences $\omega$ of a Feynman graph with $L$ internal
propagator edges and $F$ internal faces is:
\begin{equation}
\omega=-2L+F\,.
\end{equation}
We recall the well-known classification criterion. Let us
consider a given sector. \\

\noindent
$\bullet$ If the degree of divergence depends only on the number
of external edges, and decrease with him, the theory is said to
be \textit{ superficially just-renormalizable}. \\

\noindent
$\bullet$ If the degree of divergence depends on the number of
external edges, and decreases both with the number of vertices
and external edges, the theory is said to be
\textit{superficially super renormalizable}.\\

\noindent
$\bullet$ If the degree of divergence depends on the number of
external edges, and increases with the number of vertices and/or
the number of external edges, the theory is said to be
\textit{superficially non-renormalizable}. \\

\noindent
The adjective ‘‘superficially" refers to the fact that such a
classification remains heuristic without a rigorous proof for
finiteness of renormalized amplitudes. The renormalization of
TGFT models by considering only the melonic sector is given in
\cite{Carrozza:2012uv}-\cite{Lahoche:2015ola} and references
therein. We expect that the new sectors that we will consider in
this paper require minimal modifications on the proofs given in
these references, and we will only proof the key properties
allowing to extend them trivially. Moreover, we left the
adjective ‘‘superficially" for the rest of this paper.\\

\noindent
We now give a precise definition of a sector. 
\begin{definition} \textbf{Families and sectors.}\\

\noindent
$\bullet$ A family $\mathcal{F}$ is a set of proliferating
connected non-vacuum graphs which have the same degree of
divergence and the same boundary graph. A family is then labeled
with a couple $\mathcal{F}=(\omega, \gamma_n)$ with $n>1$,
$\gamma_n$ being a connected invariant bubble of valence $n$.
The leading family $\mathcal{F}_{\gamma_n,\omega_n}$ has the
maximal divergence degree $\omega_n$ i.e.
$\omega_n=\max_{\gamma_n} \omega$. \\

\noindent
$\bullet$ A leading sector $\mathcal{S}$ is a set of leading
families $\mathcal{S}=\{\mathcal{F}_{\gamma_n,\omega_n}\}$ such
that each boundary graphs $\gamma_n$ are sums of the boundary
graphs having smallest valence -- up to an eventual prescription
for the sum. The set of family whose boundaries graphs have the
smallest valence is called root set, and their elements root
families. \\

\noindent
$\bullet$ A sub-leading sector is a sector whose families have
the same boundary graphs as the families of the leading sector,
but smallest degree of divergence.
\end{definition}
Then all the diagrams in a given family behave like
$\Lambda^\omega$ with the UV cutoff $\Lambda$. Moreover, we
recall the definition of the sum of connected invariants:
\begin{definition}
Let $\gamma_n$ and $\gamma_m$ be two bubbles with valence $n$
and $m$ respectively. Let $n_1\in \gamma_n$ and $n_2\in
\gamma_m$  be two black and white nodes. The sum $\gamma_n
\ast_{n_1n_2} \gamma_m$ is the connected bubble of valence
$n+m-2$ obtained from $\gamma_n$ and $\gamma_m$ as:\\

\noindent 
$\bullet$ Drawing an edge between $n_1$ and $n_2$\\

\noindent
$\bullet$ Contracting it, deleting the end nodes $n_1$ and $n_2$
and connecting together the colored edges hooked to them
following their respective colors. \\

\noindent
Figure \eqref{fig3} provides an example.
\end{definition}

\begin{center}
$\vcenter{\hbox{\includegraphics[scale=0.8]{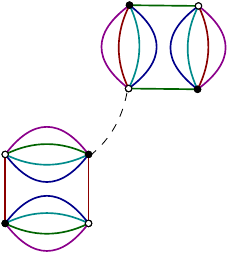}
}}\,\to\,\vcenter{\hbox{\includegraphics[scale=0.8]{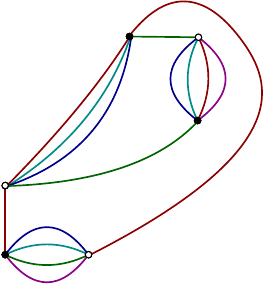}
}}\,\to\,\vcenter{\hbox{\includegraphics[scale=0.8]{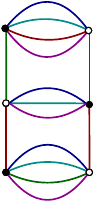}
}}$
\captionof{figure}{The sum of two connected melonic bubbles.}
\label{fig3}
\end{center}
Among the motivations for these definitions, we recall that the
sum of connected tensorial invariant\footnote{which is nothing
that contraction of a $0$-dipole, see definition \eqref{kdipole}.}
does not change the \textit{Gurau degree} characterizing the
tensorial invariant of colored random tensor models
\cite{Gurau:2011xq}-\cite{Gurau:2010ba}. The sectors could be
then labeled with their Gurau degree. \\

\noindent
From the definition of families and sectors, we define the
notion of divergent families and sectors as follows:
\begin{definition}
A family is said to be divergent if $\omega \geq 0$. Let
$\mathcal{S}$ be a sector. The divergent sector
$\mathcal{S}_D\subset \mathcal{S}$ is the subset
$\mathcal{S}_D=\{(\omega,\gamma_n)\vert\, \omega\geq 0\}$ of
divergent families included into $\mathcal{S}$. Note that
$\mathcal{S}_D$ can be an empty set. If
$\mathcal{S}_D=\emptyset$, the sector $\mathcal{S}$ is said to
be safe.
\end{definition} 
Among the families and sectors, we must make the difference
between the case for which the divergent degree increases,
decreases or is constant with respect to the number of vertices.
Note that when the divergent degree depends on the number of
vertices, the families have a short length. In contrast, when a
family is made of an infinite number of graphs, the divergent
degrees, for fixed $n$ does not depend on the number of
vertices. This consideration leads to the notion of
\textit{superficially renormalizable sector}:

\begin{definition}
Let $S_D\subset \mathcal{S}$ the divergent sector on
$\mathcal{S}$ and $\Vert\mathcal{S}_D\Vert$ the number of
elements in $\mathcal{S}_D$. If $\Vert\mathcal{S}_D\Vert <
\infty$ and $\forall \mathcal{F}\in \mathcal{S}_D$, $\Vert
\mathcal{F}\Vert =\infty$, the sector is said to be
superficially renormalizable. If $\omega=0$, the corresponding
family is said to be superficially just renormalizable.
\label{defrensector}
\end{definition}
\noindent
Note that the sub-leading order sectors must have a divergent
sector, requiring to be separately renormalized. The
power-counting renormalizability is a first requirement to prove
that a given theory is renormalizable. In particular, a
renormalizable theory requires the definition of a finite set of
\textit{local counter-terms} canceling the infinities coming
from the divergent families. Then, renormalization requires a
\textit{locality principle}, allowing to localize and subtract
the divergences occurring in a Feynman diagram (including
eventually the full diagram itself). For tensorial theories, the
relevant locality principle has been given in the definition
\eqref{deflocal}. Now we have to show that the
\textit{renormalized amplitudes} whose divergences have been
subtracted with appropriate counter-terms, are finite at all
orders. Technically such a realization requires an appropriate
slicing on the graphs, allowing to subtract only the dangerous
parts of the divergent graphs. All these technical subtleties
have been already considered in the literature for tensor field
theories \cite{Carrozza:2013mna,Lahoche:2015ola}, especially for the
melonic sector. Beyond the melonic sector, other leading order
sectors have been considered in \cite{Carrozza:2017vkz} called
necklaces. As mentioned above, the sector that we consider in
this paper in addition to the melons, the \textit{pseudo-melonic
sector} has many similarities with the melons, and a rigorous
proof of its renormalizability is a small modification of the
proofs given in \cite{Carrozza:2014rba}-\cite{Carrozza:2017vkz}
for melonic sector. We will briefly recall some of these
requirements, and then we will show in a second time that (the
pseudo-melonic sector) satisfies many of them. Let us mention that
our presentation is highly sketched with respect to a rigorous
treatment, and have to be completed with the bibliographic
details for unfamiliar readers.\\

\noindent
Then let us recall some useful definitions that we will use in
the rest of this work. We start by giving the definition of
$k$-dipoles and their contractions:
\begin{definition}
Let $\mathcal{G}$ a connected Feynman graph. A $k$-dipole is
made with two black and white nodes $n$ and $\bar{n}$ joining
together with a dotted edge and $k$ colored edges. Contracting a
$k$-dipole corresponds to delete the $k+1$ edges between the
nodes as the nodes themselves, and connected together the
remaining colored edges following their respective colors. An
example is pictured on Figure \eqref{fig4}.
\end{definition}\label{kdipole}

\begin{center}
$\vcenter{\hbox{\includegraphics[scale=1.2]{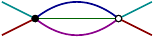}
}}\,\to\,\vcenter{\hbox{\includegraphics[scale=1.2]{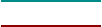}
}}$
\captionof{figure}{A $3$-dipole (on the left) and its
contraction (on the right).}\label{fig4}
\end{center}

\noindent
Another important aspect, especially concerning the power
counting theorem is the notion of \textit{contractibility},
playing an important role in the localization and subtraction of
divergences. A first important remark is that the power counting
for a given graph $\mathcal{G}$ increase exactly of $2$ under
the contraction of spanning tree edge. Indeed, if an edge
$l\in\mathcal{T}$ for $\mathcal{T}\subset \mathcal{G}$, be a
spanning tree of $\mathcal{G}$ , then, the graph $\mathcal{G}/l$
obtained from $\mathcal{G}$ by contracting the edge $l$ does not
change the number of internal faces. However, the number $L$ of
edges decreases to $L-1$. Contracting all the edges of the
spanning tree, we then get a connected graph, said
\textit{rosette} and denoted by $\bar{\mathcal{G}}$. The
divergence degrees of $\mathcal{G}$ and of one of its rosettes
$\bar{\mathcal{G}}$\footnote{In general, there are more than one
spanning tree in a given graph.} are then related by:
\begin{equation}
\omega(\mathcal{G})=\omega(\bar{\mathcal{G}})-2(V(\mathcal{G})-1)\,,
\end{equation}
$V(\mathcal{G})$ being the number of bubble vertices in the
initial graph $\mathcal{G}$, and $V(\mathcal{G})-1$ is the
number of edges in a spanning tree of $\mathcal{G}$. A rosette
is then said to be \textit{contractible} as soon as the
following definition holds:

\begin{definition}
Consider the family $\mathcal{F}$, $\mathcal{G}\in\mathcal{F}$
and $\bar{\mathcal{G}}$ the corresponding rosette.
$\bar{\mathcal{G}}$ is said to be contractible if there exist
$k>0$ such that all the dotted edges in $\bar{\mathcal{G}}$ can
be successively contracted by $k$-dipole contraction.
\end{definition}

\noindent
This definition makes sense due to the fact that for a family
whose rosettes are $k$-dipole contractible, the degree of
divergence for rosettes can be easily computed leading to the
following divergence degree of $\mathcal{G}$:
\begin{equation}
\omega(\mathcal{G})=-2L(\mathcal{G})+k(L(\mathcal{G})-V(\mathcal{G})+1).
\label{powercountingkdip}
\end{equation}
Indeed, contracting one $k$ dipole is equivalent to removing one
dotted edge, such that the divergent degree is increased by $2$.
In the same time, we remove $k$ internal faces, and therefore
$\omega(\bar{\mathcal{G}})=(k-2)L(\bar{\mathcal{G}})$. Finally a
relation between the number of internal edges and vertices can
be easily found, expressing $L$ in terms of a sum involving the
number of vertices with a given valence \cite{Carrozza:2013wda}.
Then, the renormalizability criteria may be directly
investigates. \\

\noindent
First let us consider an example of the melonic sector and let us
adopt the following definition useful particularly in the next
section and also compatible with \cite{Gurau:2011xq}-\cite{Gurau:2010ba}:
\begin{definition}
The melonic sector $\mathcal{S}_M$ is the sector in which, the
root families have the set $\{\B_i\}$ as boundary graphs. Its
elements are called melonic families. Moreover, among all the
families having boundary graphs of valence $n$, the melonic
families optimized the power counting. We call melonic bubbles
and denote as $\mathbb{M}$ the set of all the tensorial
invariants obtained as sums of elementary melons in the set
$\{\B_i\}$.
\end{definition}
The melonic sector has then the set $\{\B_i\}$ as smallest
boundary graphs. The melonic graphs satisfy a recursive
definition, and the non-vacuum diagrams can be obtained from the
vacuum ones. Note that the vacuum diagrams as well as two-point
diagrams are not in the sector $\mathcal{S}_M$.

\begin{definition} \textbf{Vacuum melonic diagrams}\\

\noindent
$\bullet$ For a purely quartic model containing only the set
$\{\B_i\}$ as interactions, the vacuum melonic diagrams are
recursively obtained from the elementary diagrams:
\begin{equation}
\vcenter{\hbox{\includegraphics[scale=0.8]{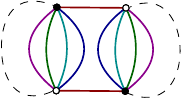}
}}
\end{equation}
by replacing any of the dotted edge as:
\begin{equation}
\vcenter{\hbox{\includegraphics[scale=1]{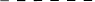} }}\,\to\,
\vcenter{\hbox{\includegraphics[scale=0.8]{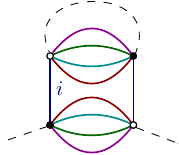}
}}
\end{equation}
for arbitrary $i$. Figure \eqref{fig5} provides us an example of
such quartic vacuum melons. \\

\noindent
$\bullet$ For models involving higher melonic bubbles, obtaining
as sums of elementary quartic melon, the vacuum diagrams are
obtained from quartic vacuum diagrams by contraction of some
$0$-dipole. As illustration the figure \eqref{fig5} provides an example of such a contraction.
\end{definition}

\begin{center}
$\vcenter{\hbox{\includegraphics[scale=0.7]{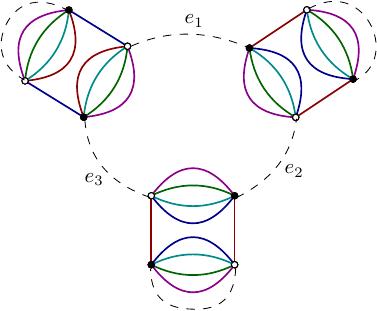}
}}\,\to\,\vcenter{\hbox{\includegraphics[scale=0.7]{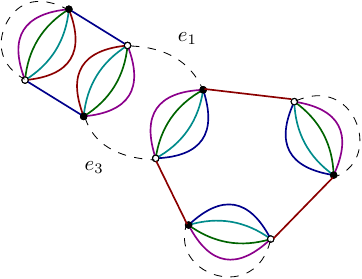}
}}$
\captionof{figure}{A quartic vacuum melonic diagrams with three
vertices (on left), and a vacuum melonic diagram with one
$3$-valent melonic interaction obtained from the first diagram
by the contraction of the edge $e_2$ (on right).}\label{fig5}
\end{center}

\noindent
Note that the elementary procedure allows to  replace one edge with a
two-point tadpole does not change the degree of divergence: To be more precise we
add two dotted edges and therefore decrease the degree by 4, but this
variation is exactly compensated from the creation of four
internal faces. It is easy to check that $\omega=5$ for the
quartic vacuum diagrams, and this quantity increases  exactly by
$2(n-2)$ when we add an $n$ valent melonic interaction with
$n>2$. We denote by $\mathcal{S}_{M_V}$ the set of vacuum
quartic melonic diagrams. The 1PI two-points melonic diagrams
may be obtained from vacuum melons from the deletion of one
dotted edge. Obviously, deleting a dotted edge along a face of
length higher to $1$, we obtain a 1PI graph. Then, to build a
two-point 1PI graph, we have only the choice to delete a tadpole
edge. The deletion suppresses five internal faces, and one internal
edge. The degree of divergence for melonic two-point graphs is
then equal to $2$. We then defined $\mathcal{S}_{M_{2}}$ as the
set of quartic melonic $2$-point graphs. The 1PI $4$-points
graphs will be then obtained from the deletion of another
tadpole dotted edge, and it is easy to check that the optimal
way is to delete an edge in the boundary of one external faces.
A four-point graph has then two external vertices, $2(d-1)$
external faces of length $2$, whose boundaries are pairs of
external edges hooked to the same external vertex, and two
external faces running through the interior of the diagram.
Figure \eqref{fig6} provides an example with two quartic vertices.

\begin{center}
\includegraphics[scale=0.8]{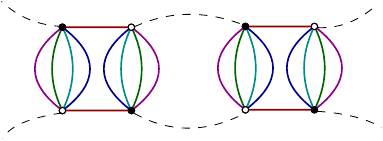} 
\captionof{figure}{A $4$-point melonic diagram with two
vertices. We have $2\times 4$ external faces per external
vertex, and two external red faces running through the internal
dotted edges.}\label{fig6}
\end{center}

\noindent
For 1PI melonic diagrams with more external edges, we get
\cite{Carrozza:2013wda}-\cite{Samary:2012bw}:
\begin{proposition}
Let $\mathcal{G}$ be a 1PI $2N$-points diagram with more than
one vertex. It has $N$ external vertices, $d-1$ external faces
per external vertex, and $N$ external faces of the same color
running through the interior of the diagram.
\label{propmelonfaces}
\end{proposition}

\noindent
Among the interesting properties of melonic diagrams, their
rosettes are $d-1$-dipole contractible, so that formula
\eqref{powercountingkdip} becomes:
\begin{equation}
\omega(\mathcal{G})=-2L(\mathcal{G})+(d-1)(L(\mathcal{G})-V(\mathcal{G})+1)\,.
\label{powercountingmelns}
\end{equation}

\noindent
Counting the number of edges in a Feynman graph with $N$
external edges, and $V_k$ vertices of valence $2k$, we get
$2L=\sum_k 2k V_k-N$ and setting $d=5$:
\begin{equation}
\sum_k2(k-2)V_k(\mathcal{G})
+4-N(\mathcal{G})\,,\label{countingmelons}
\end{equation}
so that the melonic interactions are just-renormalizable for the
quartic set $\{\B_i\}$. Then, the dangerous sector
$\mathcal{S}_{D,M}$ for melonic graph only contains the leading
families with boundary graphs $\gamma_2\in\{\B_i\}$. This
family is built of graphs with arbitrary size, meaning that each
leading family is an infinite set. $\mathcal{S}_{D,M}$ is then a
just-renormalizable divergent sector from definition
\eqref{defrensector}. To this set, we have to add the infinite set
$\mathcal{S}_{M_{2}}$ of two-points melonic graphs, which have
power counting $\omega=2$. To illustrate how the renormalization
works, let us consider the four point graph pictured on Figure
\eqref{fig6}. Let
$\mathcal{A}_{\vec{p}_1,\vec{p}_2,\vec{p}_3,\vec{p}_4}$ be the
amplitude of the corresponding diagram. Assuming that the red
edges correspond to  component $1$ in the momentum space, we
get:
\begin{equation}
\mathcal{A}(\vec{p}_1,\vec{p}_2,\vec{p}_3,\vec{p}_4)=-4\lambda^2\,A(p_{11}^2,p_{21}^2)\,\sym\mathcal{V}^{(4,1)\,1}_{\vec{p}_1,\vec{p}_2,\vec{p}_3,\vec{p}_4}\,,
\end{equation}
where:
\begin{equation}
A(p_{11}^2,p_{21}^2):=\sum_{\vec{q}\in\mathbb{Z}^{d-1}}
\frac{1}{\vec{q}\,^2+p_{11}^2+m^2}\,
\frac{1}{\vec{q}\,^2+p_{21}^2+m^2}\,,
\end{equation}
and 
\begin{equation}
\sym\mathcal{V}^{(4,1)\,1}_{\vec{p}_1,\vec{p}_2,\vec{p}_3,\vec{p}_4}=\mathcal{V}^{(4,1)\,1}_{\vec{p}_1,\vec{p}_2,\vec{p}_3,\vec{p}_4}+\mathcal{V}^{(4,1)\,1}_{\vec{p}_3,\vec{p}_2,\vec{p}_1,\vec{p}_4}\,.
\end{equation}
From power counting, $A(p_{11}^2,p_{21}^2)$ diverges
logarithmically with the UV cutoff. However, any derivative
with respect to the external momenta $p_{11}$ and $p_{21}$ is
convergent. Then, only the first term in the Taylor expansion of
$A(p_{11}^2,p_{21}^2)$ is divergent and requires to be
subtracted:
\begin{equation}
\mathcal{A}(\vec{p}_1,\vec{p}_2,\vec{p}_3,\vec{p}_4)=-4\lambda^2\,A(0,0)\,\sym\mathcal{V}^{(4,1)\,1}_{\vec{p}_1,\vec{p}_2,\vec{p}_3,\vec{p}_4}+\text{finite}\,.
\label{pertexpmelo}
\end{equation}
Note that
$2\sym\mathcal{V}^{(4,1)\,1}_{\vec{p}_1,\vec{p}_2,\vec{p}_3,\vec{p}_4}$
is nothing but the combinatorial factor coming from the
contraction of four external fields with an elementary
two-valent vertex $\B_1$. The divergent term can be then exactly
canceled adding in the original action the following one-loop
counter-term:
\begin{equation}
\delta_1S=2\lambda^2\,A(0,0)\sum_{\{\vec{p}_i\}}
\mathcal{V}^{(4,1)\,1}_{\vec{p}_1,\vec{p}_2,\vec{p}_3,\vec{p}_4}
T_{\vec{p}_1}\bar{T}_{\vec{p}_2}T_{\vec{p}_3}\bar{T}_{\vec{p}_4}\,.
\end{equation}
The counter-term is then factorized as an elementary bubble. For
two-points graphs, the the first derivative of the divergent
function has to be kept, leading to a wave function
counter-term. This is in this sense that the connected tensorial
invariant can be viewed as locals. For the melonic sector, this
property may be generalized for all orders: Any divergent
diagram may be factorized as an elementary melonic bubble times
a divergent contribution (see \cite{Carrozza:2013wda}-\cite{Samary:2012bw}).\\

\noindent
In standard field theory, interactions can be classified
following their dimensions. However, in TGFT, there is no
background space-time, and roughly speaking the action
\eqref{classicalaction} is dimensionless. An appropriate notion
of \textit{canonical dimension} appears from renormalization
group flow considerations. Renormalizable interaction in
particular has zero dimension, a property corresponding to the
marginal behavior of its renormalization group flow in the
vicinity of the Gaussian fixed point. Such a property is
recovered for tensor field theories, where just-renormalizable
interactions have a marginal behavior. We can then define the
\textit{flow dimension} as follows:

\begin{definition}
Let $\mathcal{S}_D$ a renormalizable leading sector and
$\mathcal{F}\in \mathcal{S}_D$ a divergent family with
$\omega=0$. The flow dimension of the corresponding boundary
graph $\gamma_n$ is then fixed to zero. \label{defflowdim}
\end{definition}

\noindent
In words, the renormalizable interaction scales logarithmically
with the UV cutoff, which is another way to say that their
weight on the power counting is zero -- as in equation
\eqref{countingmelons} for quartic melonic diagrams. The flow
dimension of renormalizable couplings being fixed for all
considered renormalizable sectors, we can associate a dimension
for all couplings as their optimal scaling. Moreover, this
scaling may be directly read on the power counting as well, if
it exists. This is explicitly the case on equation
\eqref{countingmelons}. Without explicit power counting, the
flow dimension is obviously closely related to the degree of
divergence. Moreover, $\omega$ is a sum and difference of
integers, the number of dotted edges and faces, respectively
weighted with $2$ and $1$. Added an edge decrease of $2$ the
degree of divergence, while added a face increase it by one. It
is then coherent to associate a \textit{canonical dimension} $2$
for each edge of a theory with Laplacian type propagator, and a
dimension $1$ for each face. From these considerations, we
associate a canonical dimension $2$ for mass and Laplacian. One
more time, the dimension of these two operators being fixed, the
canonical dimensions for other tensorial operators may be
deduced. Moreover, canonical and flow dimension seem to be
closely related, and to make contact between them, we have to
show that renormalizable interactions have zero canonical
dimension. In the literature \cite{Lahoche:2018vun}, the \textit{dimension
estimate} $\tilde{d}_b$ of a bubble $b$ provides an upper bound
for canonical dimension $d_b$:

\begin{definition}\label{canonicaldim}
Let $b$ be a connected tensorial bubble, and $\mathbb{G}_2(b)$
the set of one-vertex two-point graphs obtained from $b$. The
estimate canonical dimension $\tilde{d}_b$ of the bubble $b$ is
then defined as:
\begin{equation}
\tilde{d}_b=2-\underset{\mathcal{G} \in \mathbb{G}_2(b)} \max
\omega(\mathcal{G})\,,\qquad \tilde{d}_b\geq d_b\,.
\end{equation} 
\end{definition}

\noindent
Note that $\tilde{d}_b=2$ if $ \mathbb{G}_2(b)=\emptyset$,
meaning that estimate canonical dimension for two-point bubbles
is then fixed to $2$, with respect to  our preliminary
discussion. This statement may be checked as follows. Let us
consider a 1PI $2n$-point leading graph $\mathcal{G}_n$ in a
renormalizable sector, built of bubbles $\gamma_n$ with zero
flow dimension. A leading order $2n$-point graph $\mathcal{G}_n$
may be obtained from a leading two-points graph $\mathcal{G}_2$
from deleting $n-1$ internal edges. Let $\omega(\mathcal{G}_2)$
be the divergent degree of the two-point graph. Consistency with
renormalizability then requires that the leading quantum
corrections for considered renormalizable sectors behave like
$\Lambda^{2}$, then we come to \\
\begin{equation}
\omega(\mathcal{G}_2)=2\,.
\end{equation}

\noindent
Deleting a $k$-dipole leads to  removing one edge and at most $d$ internal
faces. However, the number of deleted faces can be reduced if we
choose to delete an edge on the boundary of a maximal number of
external faces. We restrict our attention on the cases of
$k$-dipole contractible families. Then, deleting $n-1$ edges correspond to 
deletes $k(n-1)$ faces, and :
\begin{equation}
\omega(\mathcal{G}_2)-(k-2)(n-1)=2-(k-2)(n-1)=0\, \to\,
k=2+\frac{2}{n-1} \,.
\end{equation}
Note that this equation make sense only for $n=2$ and $n=3$. For
$n>3$, the equation $2-(k-2)(n-1)=0$ has no integer solutions.
Then, we expect that the only $k$-dipole just-renormalizable
models are given when  $n=2,k=4$ or $n=3,k=3$. The first condition 
corresponds to the quartic melonic sector. The second condition will  be described later  in the next subsection. \\

\noindent
The canonical dimension of the different boundaries  bubbles for a given sectors   have to be closed. From the bubbles with valence $n$
in a $k$-dipole contractible family, $n-1$ valent bubbles may be
obtained from optimal contraction of a $k$-dipole on a just
renormalizable interaction $\gamma_n$. Creating such a dipole
increases the number of $0$-edges of $1$, and  then creates $k$
internal faces. The scaling is then  given by:
\begin{equation}
\omega(\gamma_{n-1})=\omega(\gamma_n)+(k-2)=\frac{2}{n-1}\,,
\end{equation}
implying that $\omega(\gamma_{1})=2$ from construction.
$\omega(\gamma_{n-1})$ is nothing but the optimal scaling, that
is, what we call flow dimension. Moreover, it is easy to check
that $\omega(\gamma_2)=2-2/(n-1)$, then:
\begin{equation}
\omega(\gamma_{n-1})=2-\omega(\gamma_2)\,,
\end{equation}
meaning that the flow dimension for couplings in a renormalizable
sector coincides with the estimated canonical dimension. The
canonical dimension will be investigated in full detail for the
two considered renormalizable sectors of this paper at the end
of the next section.

\subsection{Pseudo-melonic sector}\label{Pseudomelonic}

In this subsection we consider a new leading order sector that we call
\textit{pseudo melonic}. The power counting theorem and the classification of the graphs that contribute to this sector are given. Let us start by the following  definition:

\begin{definition}
The pseudo-melonic sector $\mathcal{S}_{PM}$ is the family set of graphs
whose root familiy boundaries are in the set $\{\B_{ij}\}$. We
call pseudo-melonic bubbles and denote as $\mathbb{PM}$ the set
of all pseudo-melonic tensorial invariants obtained as sum of
the elementary pseudo-melons in $\{\B_{ij}\}$.
\end{definition}

An elementary investigation suggests that $d-2$ dipoles will play
the same role for pseudo-melonic graphs as $d-1$ dipoles for
the melonic ones. We then give the following lemma which provides 
 the effect of a $d-2$ dipole contraction:

\begin{lemma}
The set $\mathbb{PM}\cup \{\gamma_1\}$ is stable under $d-2$
dipole contraction.
\end{lemma}
\textit{Proof.} We prove this lemma recursively. Let
$\gamma_n\in \mathbb{PM}$ for $n\geq 2$. First of all,  for $n=2$, $\gamma_2\in
\{\B_{ij}\}$ and the contraction of any $d-2$ dipole leads to
the elementary graph $\gamma_1$ with valence one. We then consider
the case $n>2$. From its definition, $\gamma_n$ is a connected sum
of two-valent bubbles. Then, $\gamma_n$ can be considered as  a tree made with $n-1$
elementary pseudo-melons. Assuming that the property holds for
$n=n_0$, any $n_0+1$ valent bubbles can be obtained from 
$\gamma_{n_0-1}\in \mathbb{PM}$ as sum of  an elementary pseudo
melon $\gamma_2$. Let $\gamma_{n_0}=\gamma_{n_0-1}\ast\gamma_2$.
We have to distinguish three cases. First of all, the $(d-2)$-dipole is on the added $\gamma_2$. Contracting this dipole leads
to a $\gamma_1$. But $\gamma_{1}$ is the identity element of the
$\ast$--algebra on $\mathbb{PM}$, then:
$\gamma_{n_0-1}\ast\gamma_{1}=\gamma_{n_0-1}$. The second case
is when a $(d-2)$-dipole is contracted on the component
$\gamma_{n_0-1}$ itself.Moreover from induction hypothesis, the
contraction leads to a connected $\gamma_{n_0-2}\in\mathbb{PM}$
of valence $n_0-2$, summed with $\gamma_2$, and
$\gamma_{n_0-2}\ast\gamma_2\in\mathbb{PM}$. The third and last
possibility is to contract a $(d-2)$-dipole corresponding to a
dotted edge between $\gamma_{n_0-1}$ and $\gamma_2$. The only
possibility is that $\gamma_{n_0-1}$ and $\gamma_2$ share $(d-2)$
internal faces of length $2$ as on figure \eqref{fig7} below. But
it is easy to check that the contraction of the two dotted edges
bounded the faces yield again the pseudo melonic bubble
$\gamma_{n_0-1}\in\mathbb{PM}$.

\begin{center}
\includegraphics[scale=0.7]{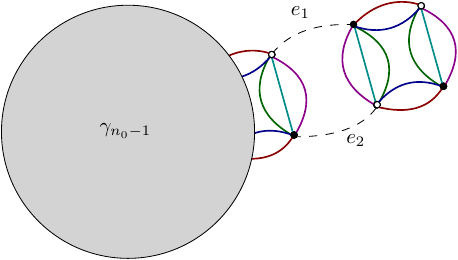} 
\captionof{figure}{Illustration of the third configuration: The
two pseudo-melonic connected tensorial invariants
$\gamma_{n_0-1}$ and $\gamma_{2}$ share three internal faces of
length $2$. Contracting $e_1$, we get a $3$-dipole, whose
contraction leads to $\gamma_{n_0-1}$ itself.}\label{fig7}
\end{center}
\begin{flushright}
$\square$
\end{flushright}

\begin{definition}
A connected Feynman graph $\mathcal{G}$ is said pseudo-melonic
if it contains only pseudo melonic tensorial bubbles. Now contracting
the $V(\mathcal{G})-1$ dotted edges of a spanning tree leads
to a pseudo-melonic rosette with
$L(\mathcal{G})-V(\mathcal{G})+1$ dotted edges.
\end{definition}

\begin{definition}
For any pseudo melonic graph $\mathcal{G}$, we define
$\rho(\mathcal{G})$ as:
\begin{equation}
\rho(\mathcal{G})=(d-2)(L(\mathcal{G})-V(\mathcal{G})+1)-F(\mathcal{G})\,.
\end{equation}
\end{definition}

\begin{lemma}
For any pseudo-melonic graph, $\rho(\mathcal{G})$ is invariant
under $(d-2)$-dipole contraction and also as well under a tree edge
contraction.\label{lemma2}
\end{lemma}
\textit{Proof.}
Let us consider an edge $e$ in a spanning tree. Contracting it
does not change the number of face as well as the combination   $L-V$, due to the fact that contracting a tree edge decreases both
$L$ and $V$ of one unit. Then, $\rho$ is invariant under a tree
edge contraction.\\

\noindent
Let us consider a ($d-2)$-dipole. Contracting it deletes exactly
$d-2$ internal faces. Moreover $V$ is invariant and $L$
decreases of one unit. Finally, the variation of the first
term exactly compensates the variation of $F$.

\begin{flushright}
$\square$
\end{flushright}

\begin{proposition}
Let $\mathcal{G}$ be a pseudo-melonic non-vacuum graph containing
only two valent vertices. Then:
\begin{equation}
\rho(\mathcal{G}) \in \mathbb{N}\,,
\end{equation}
and 
\begin{equation}
\rho(\mathcal{G})=0\,,
\end{equation}
if and only if $\mathcal{G}/\mathcal{T}$ is $(d-2)$-dipole
contractible for any spanning tree
$\mathcal{T}\subset\mathcal{G}$. \label{proprho}
\end{proposition}
\textit{Proof.} 
We proceed by induction on the number of elementary
pseudo-melonic bubbles. For $V=1$, the proposition can be easily
checked. For $V>1$, we assume the proposition true for $V=V_0$
and consider an arbitrary pseudo-melonic graph $\mathcal{G}$
having $V=V_0+1$ vertices. Choosing an external vertex
$b\in\mathcal{G}$, we have to distinguish three cases (see
Figure \eqref{fig8}a):\\

\begin{itemize}
\item Only one external edge is hooked to the vertex $b$. It is
then connected with one, two or three connected components.

\item Two external edges are hooked to the vertex $b$. Then one
or two connected components can be hooked to him.

\item Three external edges are hooked to $b$, and only one
connected component is hooked to him.
\end{itemize}
\noindent
Note that the last case with four external edges hooked to $b$ is then
excluded from the recursive hypothesis $V>1$. Let us consider
the generic case pictured having $\mathfrak{c}$ connected
components $\mathcal{G}_i$. From recursive hypothesis, these two
connected components satisfy
$\rho(\mathcal{G}_i)\in\mathbb{N}\,\,,i=1,\cdots,c$. Let
$\mathcal{G}^\prime:=\cup_{i=1}^c\mathcal{G}_i$ the graph with
$c$ connected components obtained from $\mathcal{G}$ deleting
the external vertex $b$. We have $F(\mathcal{G}_i)\leq
(d-2)(L(\mathcal{G}_i)-V(\mathcal{G}_i)+1)$,  $\forall i$,
and:
\begin{equation}
F(\mathcal{G}^\prime)\leq
(d-2)(L(\mathcal{G}^\prime)-V(\mathcal{G}^\prime)+1)+(d-2)(\mathfrak{c}-1).
\end{equation}
Moreover, $V(\mathcal{G})=V(\mathcal{G}^\prime)+1$ and
$L(\mathcal{G})=L(\mathcal{G}^\prime)+\ell$, where $\ell$ is the
number of dotted edges hooked the vertex $b$ to
$\mathcal{G}^\prime$. Obviously $\ell\geq \mathfrak{c}$, and:
\begin{equation}
F(\mathcal{G}^\prime)\leq
(d-2)(L(\mathcal{G}^\prime)-V(\mathcal{G}^\prime)+1)+(d-2)(\mathfrak{c}-\ell)\,.
\end{equation}
Moreover, it is easy to check that, from all the possibilities  listed above we have: $F(\mathcal{G})\leq
F(\mathcal{G}^\prime)+(d-2)(\ell-\mathfrak{c})$ and then 
$\rho(\mathcal{G})\geq 0$. \\

\noindent
From lemma \eqref{lemma2}, if $\mathcal{G}/\mathcal{T}$ is a
$(d-2)$-dipole contractible rosette,
$\rho(\mathcal{G}/\mathcal{T})=\rho(\mathcal{G})=0$.
Reciprocally, assuming  that $\rho(\mathcal{G})=0$, then
$\rho(\mathcal{G}/\mathcal{T})=0$ for any spanning tree 
$\mathcal{T}\subset \mathcal{G}$. Moreover, from recursion
hypothesis, one has:
\begin{equation}
\rho(\mathcal{G}_i)=0\,\quad \forall i\,\,\Rightarrow \,\,
F(\mathcal{G}_i)=(d-2)(L(\mathcal{G}_i)-V(\mathcal{G}_i)+1)\,\quad\forall
i\,,
\end{equation}
and for each $i$, and any spanning tree $\mathcal{T}_i\subset
\mathcal{G}_i$, $\mathcal{G}_i/\mathcal{T}_i$ is $(d-2)$-dipole
contractible. Any spanning tree $\mathcal{T}\subset \mathcal{G}$
can be decomposed as follows:
$\mathcal{T}=\cup_{i=1}^{\mathfrak{c}}
\mathcal{T}_i\cup\{l_1,\cdots,l_{\mathfrak{c}}\}$, where
$\{l_1,\cdots,l_{\mathfrak{c}}\}\subset \{l_1,\cdots,l_{\ell}\}$
is a subset of k${\mathfrak{c}}$ edges hooked to $b$.

\begin{center}
\includegraphics[scale=0.8]{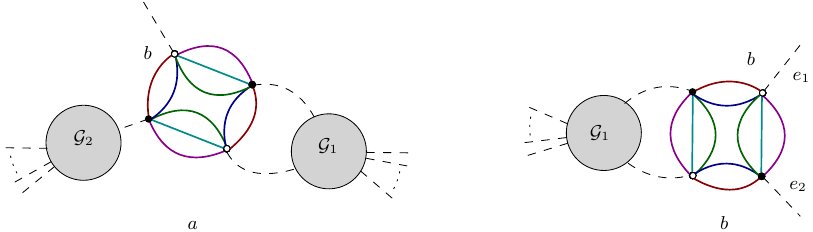} 
\captionof{figure}{a) An illustration for the first item: One
external edge is hooked to $b$, and two connected components are
hooked to him, namely $\mathcal{G}_1$ and $\mathcal{G}_2$,
pictured as gray disks. b) The only configuration creating $d-2$
internal faces. Note that at least one of the two edges $e_1$
and $e_2$ have to be external.}\label{fig8}
\end{center}
\begin{flushright}
$\square$
\end{flushright}

\noindent
Contracting the edges of the trees $\mathcal{T}_i$, we get
contractible rosettes which can be contracted themselves. It
remains at most $\ell-{\mathfrak{c}}$ loops, carrying at most
$d-2$ faces. Optimally, we then have to
$F(\mathcal{G})=F(\mathcal{G}^\prime)+(d-2)(\ell-{\mathfrak{c}})$,
but it is easy to check that there is only one such
configuration, for $\ell-{\mathfrak{c}}=1$, pictured on Figure
\eqref{fig8}b. After contraction of the remaining tree edge, we
get a $(d-2)$-dipole contractible graph, completing the proof.

\begin{flushright}
$\square$
\end{flushright}

\begin{corollary}
For any non-vacuum pseudo-melonic diagram $\mathcal{G}$,
$\rho(\mathcal{G})\in\mathbb{N}$. Moreover,
$\rho(\mathcal{G})=0$ if and only if $\mathcal{G}/\mathcal{T}$
is $(d-2)$-dipole contractible, $\mathcal{T}$ being a spanning
tree of $\mathcal{G}$ containing all the $0$-dipole edges of the
decomposition of all the pseudo-melonic bubbles into sums of
elementary quartic pseudo-melonic bubbles. \label{corolrho}
\end{corollary}
\noindent
\textit{Proof.} Any pseudo-melonic diagram can be obtained from
a purely quartic pseudo-melonic diagram from the contraction of
$0$-dipole edges. However, $0$-dipole contractions do not change
the $\rho(\mathcal{G})$.
\begin{flushright}
$\square$
\end{flushright}

\noindent
As a direct consequence, we deduce that a necklace bubble cannot have more than three colored edges between a pair of black
and white nodes. Otherwise, it will be possible to obtain a
four-dipole, which violates the bound $\rho\geq 0$. Indeed, \\

\begin{corollary}
Let us consider  a non-vacuum pseudo-melonic graph $\mathcal{G}$ having
$V_k(\mathcal{G})$ pseudo melonic bubbles of valence $2k$,
$L(\mathcal{G})$ internal edges and $N(\mathcal{G})$ external
edges. From the relation $2L(\mathcal{G})=\sum_k
2k\,V_k(\mathcal{G})-N(\mathcal{G})$, we then have:\\
\begin{equation}
\omega(\mathcal{G})=\frac{1}{2}\sum_k (2k-6)
V_k(\mathcal{G})+\left(3-\frac{1}{2}N(\mathcal{G})\right)-\rho(\mathcal{G})
\,.\label{corolpower}
\end{equation}
\end{corollary}
\noindent
Then, proposition \eqref{proprho} and corollary \eqref{corolrho}
motivate the following definition:
\begin{definition}
For a fixed configuration $(\{V,k\},N)$, the leading order
graph satisfies $\rho=0$. We call pseudo-melons these leading
families which bound the power counting and build the
pseudo-melonic sector $\mathcal{S}_{PM}$. \\
\end{definition}
From \eqref{corolpower}, it is clear that the corresponding
divergent sector will be of finite size if and only if
$V_k(\mathcal{G})=0$ for $k>3$. In other words:
\begin{proposition}
The pseudo-melonic sector is just-renormalizable up to
$3$-valent pseudo-melonic interactions. Moreover, power counting
graphs with $N>6$ admit the following bound:
\begin{equation}
\omega\leq -\frac{N}{8} \,,\qquad N>6\,.
\end{equation}
\end{proposition}
For a purely $3$-valent model, the power counting reduces to
$\omega=3-N/2$, and vanishes for the $3$-valent leading family. We
then fix to zero the canonical dimension of $3$-valent
pseudo-melonic bubbles. The complete set of renormalizable
$3$-valent interactions is given on Figure \eqref{fig10}. \\

\begin{center}
\includegraphics[scale=0.9]{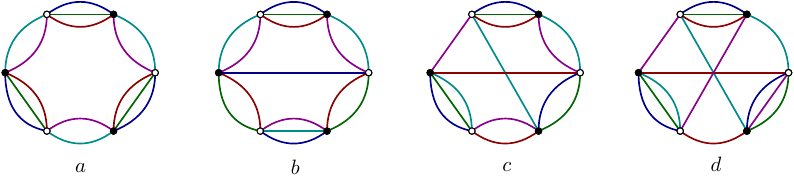} 
\captionof{figure}{The complete set of just-renormalizable
interactions, having zero canonical dimension.}\label{fig10}
\end{center}

\noindent
A direct inspection shows that all the divergences have not taken
into account in the leading divergent set of pseudo-melons.
Indeed, for $\rho=0$, two, four and six-points functions have
respectively $\omega=2$, $\omega=1$ and $\omega=0$. The
remaining functions are power-counting convergent. However,
for $\rho=1$, the two and four point functions remain divergent,
respectively as $\omega=1$ and $\omega=0$; and for $\rho=2$, the
two-point function remains logarithmically divergent $\omega=0$.
We have then two sub-leading order sectors, characterized by
$\rho=1$ and $\rho=2$. Figure \eqref{Fig11} provides some
examples. Moreover, for $\rho>2$, the divergent degree admits
the bounds:
\begin{equation}
\omega\leq -\frac{N}{2}\,,
\end{equation}
ensuring power-counting convergence. The complete
renormalization of the theory then requires to renormalize
sector by sector with appropriate counter-terms. In the rest of
this paper, we will only focus on the leading order sectors,
mixing melons and pseudo melons, which dominate the
renormalization group flow in the deep UV limit.
\begin{center}
\includegraphics[scale=0.9]{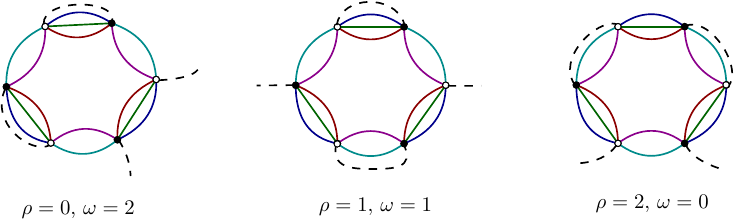} 
\captionof{figure}{One-vertex $2$-point functions examples for
leading and sub-leading orders.} \label{Fig11}
\end{center}
\noindent
From Definition \eqref{canonicaldim}, we then deduce the estimate
canonical dimension for each of the pseudo-renormalizable
interactions, that is, with valence smallest or equal to $3$. We
then get $\tilde{d}_b=0$ for each of the bubbles pictured on
Figure \eqref{fig10}, in agreement with their expected marginal
behavior. Moreover, the dimension of the quartic bubbles in the
set $\{\B_{ij}\}$ is equal to $1$, corresponding to an essential
coupling. To close this section, we will discuss this point with
some detail. Let us consider the quartic melonic sector. From
definition, the coupling $\lambda_{4,1}$ must have a zero flow
dimension, because all leading $4$-point quantum corrections
have zero divergent degree and then behave
logarithmically\footnote{In other words, adding a new vertex has
no additional cost, which is explicit in equation
\eqref{countingmelons}.}. Fixing $d_{4,1}=0$, it is easy to see
that the optimal quantum corrections for two-points functions
and then for mass scales as $\Lambda^2$. The flow dimension for
mass and Laplacian then is fixed to $2$, and then corresponds to
the canonical dimension coming from power-counting itself.
Moving on to the pseudo-melonic sector, the flow dimension for
all the couplings corresponding to the bubbles pictured on
Figure \eqref{fig10} have zero flow dimension. Moreover, optimal
quantum corrections for two-points functions made only with these
bubbles scales as $\Lambda^2$ as well from power counting
\eqref{corolpower}. Finally, optimal quantum corrections for
four-points functions scale as $\Lambda$, in accordance with a
canonical dimension $1$. In general, from power counting and for
a general tensorial invariant of valence $n$:
\begin{equation}
d_{\gamma_n} = \left\{
    \begin{array}{ll}
      4-2n & \mbox{In melonic sector} \\
       3-n & \mbox{In pseudo-melonic sector} 
    \end{array}
\right.\label{dimensiongeneral}
\end{equation}

\section{EVE for a mixing tensorial theory space}\label{sec4}
This section aims at building the effective vertices for a renormalizable sector beyond the
standard melonic Feynman graphs. We deduce the effective vertices for a model including melonic and pseudo-melonic
renormalizable interactions as initial conditions in the $UV$. For the rest of this paper,
we consider $5$-dimensions TGFT with these two leading contributions, and we restrict our
attention on the $UV$ sector $\Lambda\gg k\gg 1$. 

\subsection{Mixing melons and pseudo-melons}\label{sousec4}

In this section we investigate the leading sector for a theory mixing melonic and pseudo-melonic renormalizable sectors. The theory from which we start in the deep UV is the following:
\begin{align}
S_{\text{int}}(T, \bar{T})=\lambda_{4,1}\sum_{i=1}^d\vcenter{\hbox{\includegraphics[scale=0.8]{melon1.pdf} }}+\sum_{j<i}\left(\lambda_{4,2}\,\vcenter{\hbox{\includegraphics[scale=0.8]{PS1.pdf} }}+\lambda_{6,1}\vcenter{\hbox{\includegraphics[scale=0.7]{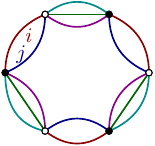} }}\right)\label{initialaction}
\end{align}
Note that we have chosen one among the three type of interactions pictured on Figure \eqref{fig10}. We call this sector \textit{non-branching pseudo-melonic}. Non-branching graphs have been studied in \cite{Carrozza:2017vkz}. The boundary graphs for these restricted sectors $\gamma_n$ are such that:
\begin{equation}
\gamma_n=\underbrace{\B_{ij}\ast\B_{ij}\ast\cdots\ast\B_{ij}}_{n-1\,\text{times}}=\ast_{k=1}^{n-1}\B_{ij}=:\B_{ij}^{(n)}\,.
\end{equation}
Figure \eqref{fig12} provides the structure of general pseudo-melonic interactions. The non-branching pseudo-melonic sector then splits into $d(d-1)$ sectors labeled with a couple $(i,j)$, $i>j$, such that each family of valence $n$ has boundary $\B_{ij}^{(n)}$. The interest of this restriction is that the pseudo-melonic sector is stable in the deep-UV: There are no leading order quantum corrections which generate an effective vertex outside of the non-branching sector from the initial conditions \eqref{initialaction}. In other words, the renormalization group flow is stable on the non-branching sector. 

\begin{center}
\includegraphics[scale=1]{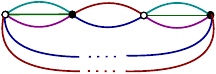} 
\captionof{figure}{Structure of general pseudo-melonic interactions.}\label{fig12}
\end{center}

\noindent
As a first step we will investigate the general structure for the leading order (LO) graphs. Let $\G$ be a Feynman graph obtained from the classical action \eqref{initialaction}. Moreover, let $\mathrm{Q}$ be the map from $\G$ to its \textit{quartic representation} $\mathrm{Q}(\G)$ i.e.  the graph obtained from $\G$ by replacing all the $3$-valent vertices with their connected sum of quartic interactions. 
\begin{definition}
We denote by $\mathbb{G}$ the complete set of connected Feynman graphs obtained from action \eqref{initialaction}.  We call \textit{quartic sector} the connected Feynman graphs with only two-valent vertices. We denote by $\Q(\mathbb{G})$ the corresponding subset. 
\end{definition}

Obviously, for any LO graph $\mathcal{G}$, $\Q(\mathbb{G})$ is a LO graph in the quartic sector. Reciprocally, LO graphs in the full sector can be obtained from contraction of some $0$-dipoles on a given LO quartic graph. We then investigate the quartic sector in the first step. 

\subsubsection{The quartic sector}

The quartic sector is most conveniently studied in the Hubbard–Stratonovich (HS) representation. For tensorial theories, HS representation has been largely discussed in the literature \cite{Carrozza:2012uv}-\cite{Geloun:2011cy}. The procedure can be summarized as follows. In the HS representation, each vertex corresponds to a cycle of edges of color $0$ of the original representation. The \textit{arcs}\footnote{or \textit{corners} in mathematical literatures.} are  the edges of color $0$ in the original representation. For our model, there are two types of edges in the HS representation. Monocolored edges labeled with $i$ ($i=1,\cdots,d$) corresponding to melonic vertices, and bicolored edges of colors $ij$, ($i>j$), corresponding to pseudo-melonic interactions, opening simultaneously both strands of color $i$ and $j$ between two HS vertices. Finally, external edges in the original representation are matched as \textit{cilia} on the vertices in the HS representation, in such a way that  such  a vertex has at least one bear at least one cilium. The exact way to construct the correspondence can be found in \cite{Gurau:2013pca},\cite{Bonzom:2012wa}. Then, a graph in the HS representation is made with three types of bicolored lines joining some vertices with at least one cilium, in such a way that the number of cilium in the HS representation is  half of the number of external edges in the original representation. We call \textit{maps} the graphs in the HS representation and denote their set as $\mathbb{M}$. Note that the correspondence between the two representations is exact, in the sense that we can construct a bijective mapping $\F:\mathbb{G}\to\mathbb{M}$ such that for any quartic Feynman graph $\mathcal{G}$ in the original representation we may associate  a unique graph $\F(\mathcal{G})$ in the HS representation and reciprocally.  The construction of the mapping $\F$ is pictured in Figure \eqref{fig13} for some example, and detailed in \cite{Bonzom:2015axa}. 

\begin{center}
\includegraphics[scale=0.8]{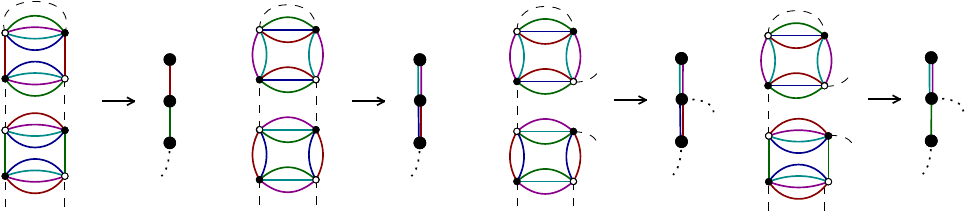} 
\captionof{figure}{HS correspondence between some quartic Feynman graphs.}\label{fig13}
\end{center}

We will investigate the structure of the LO quartic graphs in the HS representation. We have the first result:
\begin{proposition}
Leading order quartic vacuum Feynman graphs are trees in the HS representation, with power counting $\omega=5$. \label{LOvaccum1}
\end{proposition}
\textit{Proof.}
We proceed recursively on the number of colored and bicolored edges. Let $n$ be  the number of colored edges, $m$ the number of bicolored edges and $\mathcal{N}=n+m$ the total number of edges. For $\mathcal{N}=1$, there are two leading configurations, corresponding to $(n=1,m=0)$ and $(n=0,m=1)$:
\begin{equation}
\includegraphics[scale=1]{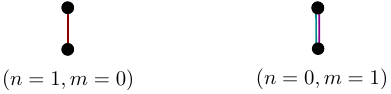} \,.
\end{equation}
For the first one, we have two dotted edges and nine internal faces, then $\omega=-2\times 2+9=5$. For the second case, we have two dotted edges and eight internal faces. Moreover, a quartic pseudo-melonic interaction has canonical dimension $1$, the complete scaling is then the same. Let $\mathcal{M}_{\mathcal{N}}$ be a map of order $\mathcal{N}>1$. Assuming that $\mathcal{M}_{\mathcal{N}}$ is a tree, 
\begin{equation}
\mathcal{M}_{\mathcal{N}}=\vcenter{\hbox{\includegraphics[scale=1]{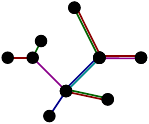} }}\,,
\end{equation}
there are only four moves, to pass from $\mathcal{M}_{\mathcal{N}}$ to $\mathcal{M}_{\mathcal{N}+1}$, listing both for colored and bicolored edges on Figure \eqref{fig14}.

\begin{center}
\includegraphics[scale=0.8]{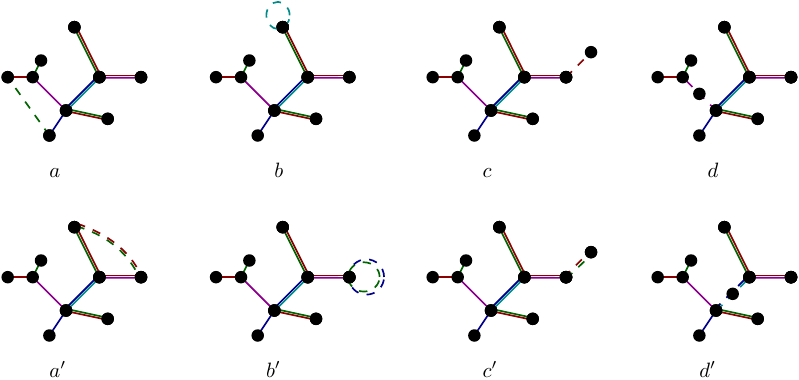} 
\captionof{figure}{Elementary moves on the tree $\mathcal{M}_{\mathcal{N}}$. For $a$, $b$, $c$ and $d$, we add a single--colored edge, while for $a^\prime$, $b^\prime$, $c^\prime$, $d^\prime$ we add a bicolored edge. }\label{fig14}
\end{center}

Investigating separately each case, we get:
\begin{enumerate}
\item We add a colored edge between two vertices, creating a loop a) or we add a tadpole colored edge on a single vertex b). From these moves, we create two internal propagator edges of color $0$, and at most one internal face. The variation of the power counting is then $\delta \omega =-2\delta L+\delta F\leq -3$.

\item We add a bicolored edge between two vertices, creating a loop a') or we add a tadpole bicolored edge on a single vertex b'). From these moves, we create two internal propagator edges of color $0$, and at most two internal faces. Taking into account the canonical dimension of the quartic pseudo-melonic vertices, the variation of power counting is then: $\delta \omega =-2\delta L+\delta F+1\leq -1$.

\item We add a new monocolored leaf c) or a monocolored bridge d). From these moves, we create two internal propagator edges of color $0$ and $4$ internal faces. The variation of power counting is then : $\delta \omega=-2\delta L+\delta F=0$.

\item We add a new bicolored leaf c') or a biocolored bridge d'). From these moves, we create two internal propagator edges of color $0$ and $3$ internal faces. The variation of power counting is then : $\delta \omega=-2\delta L+\delta F+1=0$.
\end{enumerate}

\noindent
As a result, only the moves $c$, $d$, $c^\prime$ and $d^\prime$ do not decrease the power counting. Moreover, all these configurations preserve the tree structure, ensuring that $\mathcal{M}_{\mathcal{N}+1}$ is LO only if it is a tree itself.

\begin{flushright}
$\square$ 
\end{flushright}

\noindent
Non-vacuum 1PI two-point graphs are then obtained from vacuum graphs from cutting an internal dotted edge. Obviously, cutting a dotted edge on the boundary of an internal face with length upper than $1$ creates a 1PR graph. Then we have to cut a dotted edge with ends points hooked on the same vertex, corresponding to the leafs on the HS representation. Opening an internal dotted edge deletes five internal faces, the variation of power counting is then $\delta \omega = -2\delta L+\delta F=2-5=-3$, meaning that leading two point functions scales with $\omega=2$. \\

\noindent
1PI four-point graphs are then obtained from deleting another tadpole. However, we have to distinguish between four-point diagrams with melonic and pseudo-melonic boundaries. \\

\begin{itemize}
\item For melonic boundary, we have to distinguish two cases. The first one when at least one of the two boundary vertices is a melonic vertex, the second one when the two boundary vertex are pseudo-melonic. Let us start with the first case, and assume that the first deleted dotted edge is a tadpole over a melonic vertex. The second move  can delete another tadpole, and will be optimal if the deleted dotted edge is on the boundary of one of the five opened faces from the first move. The second move then discards only four faces, implying $\delta\omega=2-4=-2$, and the LO 1PI four-point graphs are such that  $\omega=0$. When the second deleted tadpole is pseudo-melonic we have four external faces of length zero on the melonic vertex and three on the pseudo-melonic one. We have then two external faces of the same color with length $\geq 1$ passing through the two boundary vertices, and finally an external face of length $\geq 0$ starting and ending on the pseudo-melonic boundary vertex. These two cases are pictured on Figure \eqref{fig15} a and b. The last case is when the two deleted dotted edges are pseudo-melonic tadpoles. In this case, the second move is optimal if the second deleted dotted edge is on the boundary of one of the opened external faces from the first move. There are then three faces of length zero per external pseudo-melonic vertices, two external faces (not necessarily of the same colors) with length $\geq 0$ starting and ending on the same vertex and two external faces of the same color connecting together the boundary vertices. This configuration is pictured on Figure \eqref{fig15}c.

\item For pseudo-melonic boundaries, the two opening tadpoles have to be pseudo-melonic, meaning that the four external dotted edges are hooked on two pseudo-melonic vertices. The optimal cutting deletes a single dotted tadpole edge in the boundary of two external faces. Then, this move delete only $3$ internal faces, such that the total variation for power counting is: $\delta\omega=2-3=-1$, meaning that the LO 1PI four-point graphs with pseudo-melonic boundaries with a power counting  $\omega=1$, in accordance to their proper canonical dimension. This configuration is pictured in Figure\eqref{fig15}d. 
\end{itemize}

\begin{center}
\includegraphics[scale=1]{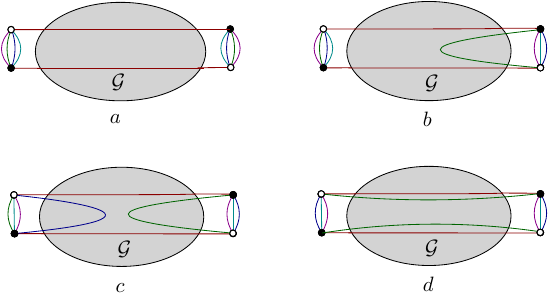} 
\captionof{figure}{The four possible boundaries. The three first ones ($a$) and ($b$) and ($c$) have a quartic melonic boundary while the last one ($d$) has a quartic pseudo-melonic boundary. The target of the external faces running through the interiors of the diagrams are pictured as internal colored edges between boundary vertices.}\label{fig15}
\end{center}
\noindent
Non-vacuum LO diagrams with $N>4$ may be obtained in the same way. However, the structure of the effective vertices becomes difficult to build explicitly. For this reason we will use the \textit{Ward identities} for effective vertices with valence higher than $2$. \\

\subsubsection{Full non-branching sector} 

We now move on the full non-branching sector, including $3$-valent non-branching interaction bubbles. As recalled in the previous section, the LO graphs may be obtained from the contraction of some $0$-dipoles, corresponding to the connected sum of two quartic pseudo-melons. In the HS representation, the  $0$-dipoles are the arcs on vertices having more than one colored or bicolored edges. Moreover, in the non-branching sector, the contracted $0$-dipoles have to be the arcs between two bicolored edges with the same couple of colors. See Figure \eqref{fig16}a. The contraction of the arc ($e$) between the two bicolored edges generates a new type of bicolored edges that we call \textit{breaking edges}, whose breaking point corresponds to the point of contact with the vertex in the HS representation. See Figure \eqref{fig16}b. 

\begin{center}
\includegraphics[scale=1.3]{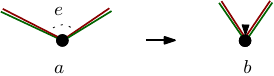} 
\captionof{figure}{Contraction of a $0$-dipole forming an arc ($e$) between two identical bicolored edges (a) and the resulting diagram (b). The $3$-valent resulting from the interactions corresponds to a pair of bicolored edges with the same color hooked to the same point on the vertex. They form \textit{breaking edges}, making  a contact on a vertex with \textit{the breaking point}, marked with a black arrow.} \label{fig16}
\end{center}

As a result, from proposition \eqref{LOvaccum1}, we deduce the following:
\begin{proposition}
Leading order vacuum Feynman graphs are trees in the HS representation, whose edges may be simple colored or bi colored edges as well as  breaking bicolored edges.  The power counting remains the same as for purely quartic sector: $\omega=5$, meaning that $\rho(\mathcal{G})=-2$ for LO vacuum graphs. \label{LOvaccum2}
\end{proposition}
The last condition on the invariance of the power counting may be easily checked.

\subsection{Structure equations for effective vertices}

In this section we investigate the structure of the LO effective vertices with $2$, $4$, $6$ and $8$ points. We will use the method discussed on \cite{Lahoche:2018vun}, by using the structure of the LO graphs established in the previous section. The aim is to get a closed set around just-renormalizable interactions, using them to parametrize the entirety of the renormalization group flow in the sector including melons and pseudo-melons. As we will see in the next section, the momentum dependence for four-point effective vertices plays a crucial role for the computation of the anomalous dimension, providing a contribution, which  not included in standard crude truncations. Moreover, six and eight-point effective vertices are require in order to close the infinite hierarchical system coming from Wetterich equation \eqref{Wett}. From this equation, it is clear that the flow for $\Gamma^{(n)}$ involves $\Gamma^{(n=2)}$. Then, in the melonic sector, the system will be closed if we compute the six-point effective vertices having  a three-valent melon graph as boundary graph. In the same way, to close the non-branching pseudo-melonic sector around marginal interactions requires eight-point functions with non-branching pseudo-melonic graph as boundary graph.  \\

\subsubsection{ Two and four-point effective vertices}

\noindent
Let us start with the two-point function. We denote as $\Sigma$ the leading order self-energy, so that the LO effective propagators $G$ can be  decomposed as:
\begin{equation}
G=\left(-\Delta+m^2+r_s-\Sigma\right)^{-1}=C_s+C_s\Sigma C_s+C_s\Sigma C_s\Sigma C_s+\cdots\,.\label{effectivetwo}
\end{equation}
where $C_s$ is the bare propagator given in \eqref{propar}. We get the following closed equation:
\begin{proposition}
The leading order self-energy $\Sigma$ satisfies the following closed equation:
\begin{align}
\nonumber \Sigma(\vec{p}\,)=-2\lambda_{4,1}&\,\sum_{i=1}^d \sum_{\vec{q}\in\mathbb{Z}^d} \delta_{p_iq_i} \,G(\vec{q}\,)-2\lambda_{4,2}\,\sum_{j<i}^d \sum_{\vec{q}\in\mathbb{Z}^d} \delta_{p_iq_i}\delta_{p_jq_j} \,G(\vec{q}\,)\\
&-6\lambda_{6,1}\, \sum_{j<i}^d \sum_{\vec{q},\vec{q}\,^\prime\in\mathbb{Z}^d} \delta_{p_iq_i}\delta_{p_jq_j}\delta_{p_iq_i^\prime}\delta_{p_jq_j^\prime} \,G(\vec{q}\,)\,G(\vec{q}\,^\prime)\,.\label{closed1}
\end{align}
\end{proposition}
\textit{Proof.}
From the previous section, we know that LO 1PI two-point graphs may be obtained from the cutting of an internal tadpole edge. They correspond to leafs on the HS representation, and the final vertex on the leaf can be hooked to a single colored edge, to a bicolored edge or to a breaking bicolored edges. As a result, the opening tadpole may be localized on a quartic melonic vertex, either on a quartic pseudo melonic or on a $3$-valent pseudo-melonic vertex. Opening a melonic tadpole on a LO graph,  means that there are two half dotted edges hooked to this vertex, and connecting it to the rest of the diagram. But it is easy to check that the remaining part of the diagram hooked to this vertex is nothing but the LO two-point function ${G}$. The same result holds when a pseudo-melonic tadpole is deleted. Finally, when the deleted tadpole is on a $3$-valent vertex, it is easy to check that the only $3$-dipole contractible configuration corresponds to the contraction of this vertex with two effective two-point functions in order to form two effective $3$-dipoles. Graphically, all these configurations correspond to the equation:
\begin{equation}
\Sigma=\sum_i\,\vcenter{\hbox{\includegraphics[scale=0.9]{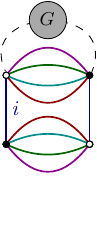} }}+\sum_{i>j}\,\vcenter{\hbox{\includegraphics[scale=0.9]{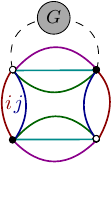} }}+\sum_{i>j}\,\vcenter{\hbox{\includegraphics[scale=0.9]{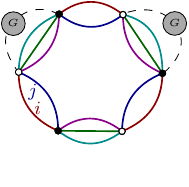} }}\,.
\end{equation}
By expressing these diagrams as an equation and taking into account all the symmetry factors, we obtain the closed equation \eqref{closed1}.
\begin{flushright}
$\square$
\end{flushright}

The structure of 1PI four-point function may be obtained from the same strategy in terms of elementary essential or marginal couplings, as well as effective two-point functions. All the configurations for boundary vertices are pictured on Figure \eqref{fig15}. Note that, for the quartic pseudo-melonic boundaries, we have to add the ones coming from $3$-valent pseudo melonic vertices. 
The interior of the diagram can be determined for each configuration with the two following statements:
\begin{enumerate}
\item The structure of the graph is a tree in the HS representation.
\item  The connectivity of the external faces between the boundary vertices has to be ensured following their respective nature.
\end{enumerate}
For the rest of this paper we focus our attention on the renormalization group flow for local interactions, corresponding to purely tensorial invariants.  We focus on the zero-momenta effective vertices. Even with this simplification, the computation of the zero-momenta four-point function remains difficult, especially when the boundary graph is a quartic melon, due to the large number of configurations. Then, to simplify the proofs, we split the computation into some partial results, corresponding the elementary ‘‘building block'' configurations. \\

\noindent
First, let us consider  a LO four-point graph having quartic melonic boundary, such that the external edges are fully connected to melonic vertices. Such a configuration corresponds to Figure \eqref{fig15}a. Let us call ‘‘red'' the color of the external faces running through the interior of the diagram between the ends vertices. In the HS representation, such a graph corresponds to a two-ciliated tree and the two cilia are joined together with a red path made of a succession of colored or bicolored edges where one of them is red. Figure \eqref{fig17}  provides  an example of this configuration. We call this path the \textit{skeleton}, and the length of the skeleton the number of edges building with  him.  Moreover, we call \textit{purely melonic} the LO two-ciliated trees whose skeleton edges have a single color. 
\begin{center}
\includegraphics[scale=1.2]{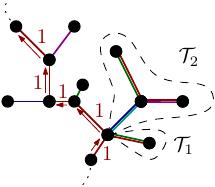} 
\captionof{figure}{A tree with two cilia contributing to the LO four-point function with melonic boundary. The red arrows follow the path of the skeleton corresponding to the color $1$, joining together the boundary vertices.}\label{fig17}
\end{center}
Let $\Gamma_{a}^{(4)}(\vec{p}_1,\vec{p}_2,\vec{p}_3,\vec{p}_4)$ be  the LO $4$-point function obtained as a sum of Feynman graphs having melonic boundary, with nonzero external momenta. The skeleton being labeled with a single color, taking into account the structure of the ends vertices, $\Gamma_{a}^{(4)}(\vec{p}_1,\vec{p}_2,\vec{p}_3,\vec{p}_4)$ have to be decomposed as:
\begin{equation}
\Gamma_{a}^{(4)}(\vec{p}_1,\vec{p}_2,\vec{p}_3,\vec{p}_4)=: \sum_{i=1}^d \Gamma_{a}^{(4)\,,i}(\vec{p}_1,\vec{p}_2,\vec{p}_3,\vec{p}_4)\,,\label{decompa}
\end{equation}
where:
\begin{equation}
\Gamma_{a}^{(4)\,,i}(\vec{p}_1,\vec{p}_2,\vec{p}_3,\vec{p}_4)=: \Pi_a^{(2)}(p_{1i},p_{3i})\, \sym \mathcal{V}^{(4,1)\,i}_{\vec{p}_1,\vec{p}_2,\vec{p}_3,\vec{p}_4}\,,\label{pidef}
\end{equation}
in accordance with the perturbative expansion \eqref{pertexpmelo}. In the rest of this paper, we will refer to $\Pi_a^{(2)}(p_{1i},p_{3i})$ as \textit{effective skeleton function}, and we  denote by $\Pi_a^{(2)}\equiv \Pi_a^{(2)}(0,0)$ its zero momenta value. Note that the upper index $2$ indicates the number of cilia on the graphs contributing to it. From their structure, obviously, $\Pi_a^{(2)}$ split into purely melonic contributions, whose skeletons are chain of melons, and mixing contributions, whose skeletons includes bicolored edges insertions. We will denote them respectively as $\Pi_{a,0}^{(2)}$ and $\Pi_{a,1}^{(2)}$, such that:
\begin{equation}
\Pi_a^{(2)}=\Pi_{a,0}^{(2)}+\Pi_{a,1}^{(2)}\,.
\end{equation}

\noindent
The decomposition \eqref{decompa} can be generalized for all the contributions pictured on Figure \eqref{fig15}. From connectivity of the boundary graphs, it is clear that the configurations (a), (b) and (c) contribute to the effective vertex function having melonic boundary, while the last one (d) contributes to the effective vertex function having pseudo-melonic boundary. From the  configurations (a), (b) and (c), the corresponding trees in HS representation have skeletons labeled with a single color, while the skeleton of the configuration (d) has a bicolored skeleton, labeled by a pair of different colors. We then have the decomposition:
\begin{equation}
\Gamma_{\text{melo}}^{(4)}(\vec{p}_1,\vec{p}_2,\vec{p}_3,\vec{p}_4)=: \sum_{i=1}^d \Gamma_{\text{melo}}^{(4)\,,i}(\vec{p}_1,\vec{p}_2,\vec{p}_3,\vec{p}_4)\,,\label{decompmelo}
\end{equation}
\begin{equation}
\Gamma_{\text{pseudo--melo}}^{(4)}(\vec{p}_1,\vec{p}_2,\vec{p}_3,\vec{p}_4)=: \sum_{j<i}^d \Gamma_{\text{pseudo--melo}}^{(4)\,,ij}(\vec{p}_1,\vec{p}_2,\vec{p}_3,\vec{p}_4)\,,\label{decomppseudomelo}
\end{equation}
where:
\begin{equation}
\Gamma_{\text{melo}}^{(4)\,,i}(\vec{p}_1,\vec{p}_2,\vec{p}_3,\vec{p}_4)= \Pi_1^{(2)}(p_{1i},p_{3i})\, \sym \mathcal{V}^{(4,1)\,i}_{\vec{p}_1,\vec{p}_2,\vec{p}_3,\vec{p}_4}\,,
\end{equation}
with
\begin{equation}
\Pi_1^{(2)}(p_{1i},p_{3i}):=\Pi_a^{(2)}(p_{1i},p_{3i})+\Pi_b^{(2)}(p_{1i},p_{3i})+\Pi_c^{(2)}(p_{1i},p_{3i})\,,
\end{equation}
and:
\begin{equation}
\Gamma_{\text{pseudo--melo}}^{(4)\,,ij}(\vec{p}_1,\vec{p}_2,\vec{p}_3,\vec{p}_4)=\Pi_{2,ij}^{(2)}(p_{1i},p_{1j};p_{3i},p_{3j})\sym \mathcal{V}^{(4,2)\,ij}_{\vec{p}_1,\vec{p}_2,\vec{p}_3,\vec{p}_4}
\end{equation}
with
\begin{equation}
\Pi_{2,ij}^{(2)}(p_{1i},p_{1j};p_{3i},p_{3j}):=\Pi_{d,ij}^{(2)}(p_{1i},p_{1j};p_{3i},p_{3j})\,,
\end{equation}
$\Pi_{d,ij}^{(2)}$ being the skeleton function corresponding to the configuration (d) on Figure \eqref{fig15}. The following result holds: 

\begin{lemma}\label{lemmaa0}
The zero-momenta effective skeleton function for purely melonic graphs, $\Pi_{a,0}^{(2)}$ writes as:
\begin{equation}
\Pi_{a,0}^{(2)}=\frac{2\lambda_{4,1}}{1+2\lambda_{4,1}\mathcal{A}_{4;2}}\label{eqpia}
\end{equation}
where :
\begin{equation}
\mathcal{A}_{m;n}:=\sum_{\vec{p}\in\mathbb{Z}^{m}} \,G^n(\vec{p}\,)\,.
\end{equation}
\end{lemma}

\noindent
\textit{Proof.}
Let us consider a LO tree having purely melonic red skeleton. To each vertex in the way of this red path are hooked some connected components $\mathcal{T}_1,\mathcal{T}_2,\cdots$. They are 1PI two-point graphs from construction, and it is easy to check that all of them provides a contribution to the effective self-energy $\Sigma$:
\begin{equation}
\vcenter{\hbox{\includegraphics[scale=1.1]{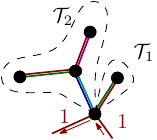} }}\equiv\vcenter{\hbox{\includegraphics[scale=0.9]{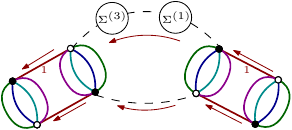} }}\,, \label{vertexHS1}
\end{equation}
where $\Sigma^{(n)}$ denote a contribution of order $n$ for the self-energy. Then, the two skeleton edges hooked to a vertex in the HS representation split him into two corners, to which some connected trees can be hooked. Summing over all possible trees we then reconstruct the effective two-point function \eqref{effectivetwo} to each corners between skeleton edges:
\begin{equation}
\vcenter{\hbox{\includegraphics[scale=1]{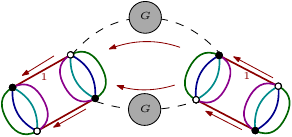} }}\,.\label{effsum1}
\end{equation}
The same structure may be repeated along the length of the skeleton. Then, summing over lengths, we get the first contribution to the 1PI four-point function with melonic vertices on its boundaries: 
\begin{align}
\nonumber&\vcenter{\hbox{\includegraphics[scale=0.6]{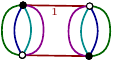} }}+\vcenter{\hbox{\includegraphics[scale=0.6]{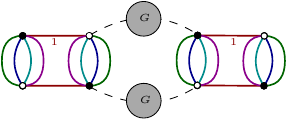} }}+\vcenter{\hbox{\includegraphics[scale=0.6]{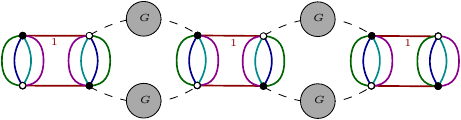} }}+\cdots\\
&\,\,\,= \vcenter{\hbox{\includegraphics[scale=0.6]{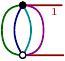} }}\left[\sum_n\left(\vcenter{\hbox{\includegraphics[scale=0.6]{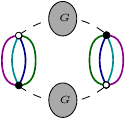} }}\right)^n\,\right]\vcenter{\hbox{\includegraphics[scale=0.6]{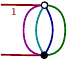} }}=\frac{1}{1-\vcenter{\hbox{\includegraphics[scale=0.6]{bound3.pdf} }}}\times \vcenter{\hbox{\includegraphics[scale=0.6]{order1.pdf} }} \label{gamma4melon}
\end{align}
This is formally the structure of the four-point vertex. The effective loop of length two can be easily computed, and we get, once again in accordance with \eqref{pertexpmelo}:
\begin{equation}
\vcenter{\hbox{\includegraphics[scale=0.6]{bound3.pdf} }}=-2\lambda_{4,1}\mathcal{A}_{4,2}\,.
\end{equation}
Note that the factor $2$ in front of this expression comes to the symmetry of the melonic vertices insertion. Figure \eqref{fig18moins} below provides us an illustration. 

\begin{center}
\includegraphics[scale=0.8]{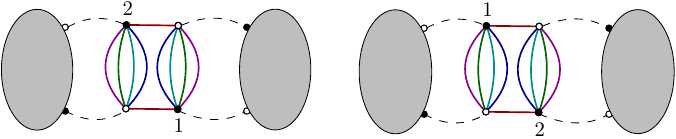} 
\captionof{figure}{The two configurations for a melonic insertion. The black node labeled by $1$ can be contracted on the right or on the left.}\label{fig18moins}
\end{center}

\noindent
The translation of the final diagram \eqref{gamma4melon} into equation is then straightforward, and we get:
\begin{equation}
\frac{1}{1-\vcenter{\hbox{\includegraphics[scale=0.6]{bound3.pdf} }}}\times \vcenter{\hbox{\includegraphics[scale=0.6]{order1.pdf} }}=\frac{1}{1+2\lambda_{4,1}\mathcal{A}_{4,2}} \times 4\lambda_{4,1}=2\Pi_{a,0}^{(2)}\,,
\end{equation}
the last equality, coming from definition \eqref{pidef} completing the proof. 

\begin{flushright}
$\square$ 
\end{flushright}

\begin{center}
\includegraphics[scale=1.2]{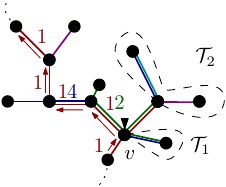} 
\captionof{figure}{A skeleton with two pseudo-melonic insertion. The red arrows follow the path of the skeleton corresponding to the color $1$, joining together the boundary melonic vertices.}\label{fig18}
\end{center}

\noindent
In addition to purely melonic configurations, we have to take into account bicolored edge insertions along the skeleton. Moreover from Figure \eqref{fig15}a, it is clear that all these bicolored insertions have to contain the color red in the pair. An example is pictured on Figure\eqref{fig18}. Note that this constraint on the existence of a common color for the edges along the skeleton as well as the linear topology of the skeleton will be a precious help for enumeration of the different configurations. \\

\noindent
Taking into account bicolored insertions introduces another difficulty. Quartic pseudo melons are not the only source of bicolored edge. We have to take into account a new type of edge, coming from breaking edges. To understand how they occur in the way of the skeleton, let us consider the example pictured in Figure \eqref{fig18}. We have one breaking edge, with colors red and green (1 and 2), and following the same argument leading to equation \eqref{vertexHS1} the contribution of the vertex labeled with $v$ writes explicitly as:
\begin{equation}
\vcenter{\hbox{\includegraphics[scale=1]{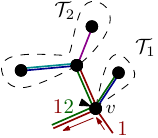} }}\equiv\, \vcenter{\hbox{\includegraphics[scale=0.8]{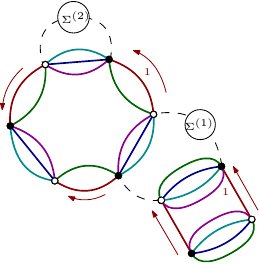} }}\,.\label{vertexHS2}
\end{equation}

\noindent
As a result, summing over all trees having the same skeleton, we get, as for equation \eqref{effsum1}:
\begin{equation}
\vcenter{\hbox{\includegraphics[scale=0.8]{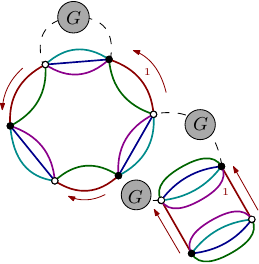} }}\,.
\end{equation}
We then observe that we have to distinguish two types of quartic pseudo-melons for the counting. The purely quartic pseudo-melonic interactions, which we call \textit{type 1}, and the effective quartic pseudo-melonic interactions, arising from the contraction of a six-point pseudo melon with an effective two-point function, which we call \textit{type 2} quartic pseudo-melon. We denote them by:
\begin{equation}
\vcenter{\hbox{\includegraphics[scale=0.9]{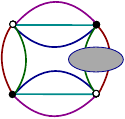} }}\,:=\,\vcenter{\hbox{\includegraphics[scale=0.8]{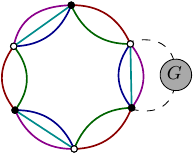} }}\,.
\end{equation}
Type 2 vertices behave like Type-1 ones. They have the same canonical dimension and the same boundary graph, the only difference comes from their respective weight. Nevertheless, in order to investigate the combinatorial structure of the skeleton, we have to sum over all repetitions of the same (type 1 or type 2) vertex, as for the melonic chain in equation \eqref{gamma4melon}. To this end, we introduce the zero momenta purely pseudo melonic effective skeleton functions $\Pi_{d,1i}^{(2)}$, with $i\neq 1$, so that $\Pi_{d,1i}^{(2)}$ is a chain of pseudo melons of type 1 or 2 with intermediate colors $1i$. Obviously, such a configuration corresponds to the structure pictured in Figure \eqref{fig15}d, this is why we labeled these effective function with a lower index $d$. Figure \eqref{fig20} provides an example of tree with two cilia and a purely bicolored skeleton.

\begin{center}
\includegraphics[scale=1]{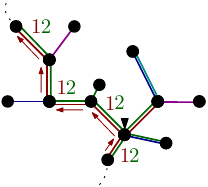} 
\captionof{figure}{A typical contribution for $\Gamma^{(4)}_{d}$} \label{fig20}
\end{center}

\noindent
Therefore we have an analogous  lemma to lemma \eqref{lemmaa0}:

\begin{lemma}
The zero momenta purely pseudo-melonic function $\Pi_{d,1i}$ has the following expression:
\begin{equation}
\Pi_{d,12}^{(2)}:=\frac{\pi_{d,1}+\pi_{d,2}+\pi_{d,1}\pi_{d,2}}{1-\pi_{d,1}\pi_{d,2}}\,,\label{eqpid}
\end{equation}
where the partial effective purely pseudo-melonic skeleton functions $\pi_{d,1}$ and $\pi_{d,2}$ are defined as:
\begin{equation}
\pi_{d,1}=\frac{2\lambda_{4,2}}{1+2\lambda_{4,2} \mathcal{A}_{3,2}}\,\quad \pi_{d,2}=\frac{6\lambda_{6,1} \mathrm{b} }{1+6\lambda_{6,1} \mathrm{b}  \,\mathcal{A}_{3,2}}\,,
\end{equation}
with
\begin{equation}
\mathrm{b} :=\sum_{\vec{q}\in\mathbb{Z}^3}\,G(\vec{q}\,)\,.\label{littleB}
\end{equation}
\end{lemma}

\noindent
\textit{Proof.}
The graphs structure is very reminiscent of the pure melonic case. The essential difference comes from the fact that we have to distinguish two elementary building block configurations, respectively made of chains of type-1 and type--2 pseudo melons. For the first case, when elementary quartic type--1 pseudo-melons form a chain with zero momenta running throughout the boundaries of the external faces, we have the formal sum :
\begin{align}
\nonumber&\vcenter{\hbox{\includegraphics[scale=0.7]{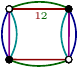} }}+\vcenter{\hbox{\includegraphics[scale=0.7]{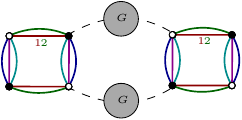} }}+\vcenter{\hbox{\includegraphics[scale=0.7]{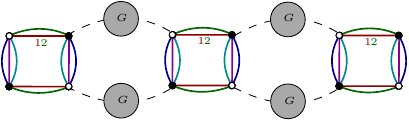} }}+\cdots\\
&=\vcenter{\hbox{\includegraphics[scale=0.7]{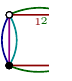} }}\left[\sum_{n=0}^\infty \left(\vcenter{\hbox{\includegraphics[scale=0.7]{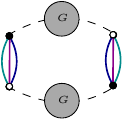} }}\right)^n\,\right]\vcenter{\hbox{\includegraphics[scale=0.7]{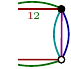} }}=\frac{1}{1-\vcenter{\hbox{\includegraphics[scale=0.7]{order2neckcenter.pdf} }}}\times\vcenter{\hbox{\includegraphics[scale=0.7]{order1neck.pdf} }}\,.\label{kerneck1}
\end{align}
The second case is when type-2 pseudo-melonic interactions, coming from $3$-valent vertices build a chain:
\begin{align}
\nonumber &\vcenter{\hbox{\includegraphics[scale=0.7]{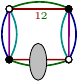} }}+\vcenter{\hbox{\includegraphics[scale=0.7]{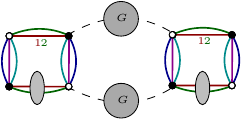} }}+\vcenter{\hbox{\includegraphics[scale=0.7]{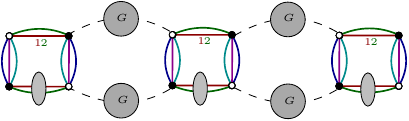} }}+\cdots\\
&=\vcenter{\hbox{\includegraphics[scale=0.7]{order1neckleft.pdf} }}\left[\sum_{n=0}^\infty \left(\vcenter{\hbox{\includegraphics[scale=0.6]{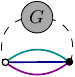} }}\times\vcenter{\hbox{\includegraphics[scale=0.7]{order2neckcenter.pdf} }}\right)^n\,\right]\vcenter{\hbox{\includegraphics[scale=0.7]{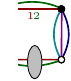} }}=\frac{1}{1-\vcenter{\hbox{\includegraphics[scale=0.6]{pti.pdf} }}\times\vcenter{\hbox{\includegraphics[scale=0.7]{order2neckcenter.pdf} }}}\times\vcenter{\hbox{\includegraphics[scale=0.7]{order1neck2.pdf} }}\,.\label{kerneck2}
\end{align}
The two kernels in brackets must be computed explicitly. Introducing the function $\mathrm{b} $ it is easy to check:
\begin{equation}
 \vcenter{\hbox{\includegraphics[scale=0.7]{order2neckcenter.pdf} }}=-2\lambda_{4,2} \mathcal{A}_{3,2}\,, \qquad \vcenter{\hbox{\includegraphics[scale=0.7]{pti.pdf} }}=3\mathrm{b} \,,
\end{equation}
so that the formal equations \eqref{kerneck1} and \eqref{kerneck2} lead to the two partial effective skeleton functions $\pi_{d,1}$ and $\pi_{d,2}$. Note the factor $2$ occurring in the denominator of the formula has the same origin as the factor $2$ occurring in expression \eqref{gamma4melon}: It comes from the symmetry exchange for the pseudo-melonic vertices. Moreover, the factor $3$ in front of $\mathrm{b} $ comes from the rotational symmetry of the $3$-valent vertices : there are $3$ different ways to build an effective four-dipole on a $3$-valent vertex to obtain a type-2 quartic pseudo melon. \\

\noindent
To obtain the complete four-point kernel $\Pi_{d,1i}$, we then have to sum over all possible partitions of type-1 and type-2 domains. We have to distinguish three cases:\\
\begin{itemize}
\item The boundary vertices are both of type--1,
\item The boundary vertices are both of type--2,
\item One boundary vertex is of type--1, the second of type--2.
\end{itemize}
For the two first  cases, we have the sequences:
\begin{equation}
\pi_{d,1}+\pi_{d,1}\pi_{d,2}\pi_{d,1}+\pi_{d,1}\pi_{d,2}\pi_{d,1}\pi_{d,2}\pi_{d,1}+\cdots=\pi_{d,1}\sum_{n=0}^\infty\,(\pi_{d,2}\pi_{d,1})^n\,,
\end{equation}
for type--1 boundary vertices, and 
\begin{equation}
\pi_{d,2}+\pi_{d,2}\pi_{d,1}\pi_{d,2}+\pi_{d,2}\pi_{d,1}\pi_{d,2}\pi_{d,1}\pi_{d,2}+\cdots=\pi_{d,2}\sum_{n=0}^\infty\,(\pi_{d,2}\pi_{d,1})^n\,,
\end{equation}
for type--2 boundaries. For the heteroclite boundaries however, fixing the end vertices we get:
\begin{equation}
\pi_{d,1}\pi_{d,2}+\pi_{d,1}\pi_{d,2}\pi_{d,1}\pi_{d,2}+\cdots=\sum_{n=1}^\infty (\pi_{d,1}\pi_{d,2})^n\,.
\end{equation}
All the sums can be easily computed as geometric progressions. Computing them, we  finally get  the effective skeleton function $\Pi_{d,12}$.
\begin{flushright}
$\square$
\end{flushright}

\noindent
As we will see, a large part of contributions to the effective function $\Pi_{a,1}^{(2)}$ require the knowledge of the momentum dependence of the effective pseudo-melonic function $\Pi_{d,12}^{(2)}(p_1,p_2;p_1^\prime,p_2^\prime)$. More precisely, because we are only interested for zero momenta four-point functions, we consider the restriction $p_1=p_1^\prime=0$, $p_2=p_2^\prime=p$:
\begin{equation}
\tilde{\Pi}_{d,12}^{(2)}(p):=\Pi_{d,12}^{(2)}(0,p;0,p)\,,
\end{equation}
which  can be easily deduced from the previous proof:
\begin{corollary}
\begin{equation}
\tilde{\Pi}_{d,12}^{(2)}(p)=\frac{\pi_{d,1}(p)+\pi_{d,2}(p)+\pi_{d,1}(p)\pi_{d,2}(p)}{1-\pi_{d,1}(p)\pi_{d,2}(p)}\,,
\end{equation}
where the momentum-dependents effective vertex functions $\pi_{d,1}(p)$ and $\pi_{d,2}(p)$ are defined as:
\begin{equation}
\pi_{d,1}(p)=\frac{2\lambda_{4,2}}{1+2\lambda_{4,2} \mathcal{A}_{3,2}(p)}\,\quad \pi_{d,2}=\frac{6\lambda_{6,1} \mathrm{b}(p) }{1+6\lambda_{6,1} \mathrm{b}(p)  \,\mathcal{A}_{3,2}(p)}\,,
\end{equation}
and:
\begin{equation}
\mathcal{A}_{3,2}(p)=\sum_{\vec{q}\in\mathbb{Z}^4}\,\delta_{q_2,p}\,G^2(\vec{q}\,)\,,\qquad \mathrm{b} :=\sum_{\vec{q}\in\mathbb{Z}^4}\,\delta_{q_2,p}\,G(\vec{q}\,)\,.\label{littleB2}
\end{equation}
\end{corollary}

\noindent
From these elementary ‘‘pure'' building block functions, we can easily deduce all the allowed configurations for each configuration in Figure \eqref{fig15}a, b and c, and the complete skeleton functions may be summarized with the following statement:
\begin{proposition}
The complete zero momenta skeleton functions $\Pi_1^{(2)}$ and $\Pi_{2,12}^{(2)}$ are given in terms of the essential and marginal couplings $\lambda_{4,1}$, $\lambda_{4,2}$, $\lambda_{6,1}$ and effective two-point function $G(\vec{p}\,)$ as:

\begin{equation}
\Pi_1^{(2)}=\Pi_{a,0}^{(2)}+\frac{\Pi_{a,0}\,\mathcal{B}\,\Pi_{a,0}}{1-\Pi_{a,0}\,\mathcal{B}}+(1+\bar{\mathcal{B}})\frac{\Pi_{a,0}\,\bar{\mathcal{B}}}{1-\Pi_{a,0}{\mathcal{B}}}+{\mathcal{D}}\,,
\end{equation}
\begin{equation}
\Pi_{2,12}^{(2)}=\frac{\pi_{d,1}+\pi_{d,2}+\pi_{d,1}\pi_{d,2}}{1-\pi_{d,1}\pi_{d,2}}\,,
\end{equation}
where $\mathcal{B}$, $\bar{\mathcal{B}}$ and $\mathcal{D}$ are respectively given by formulas \eqref{formulacool1}, \eqref{formulacool2}, \eqref{formulacool3}.

\end{proposition}

\noindent
\textit{Proof.}
The components $\Pi_{d,12}^{(2)}$ and $\Pi_{a,0}^{(2)}$ have been computed and are given equations \eqref{eqpia} and \eqref{eqpid}. To complete the proof , we then have to compute the components $\Pi_{a,1}^{(2)}$, $\Pi_{b}^{(2)}$ and $\Pi_{c}^{(2)}$. To this end, we will proceed step by step, computing each component separately. \\

\noindent
$\bullet$ \textit{Computation of $\Pi_{a,1}^{(2)}$.} \\
\noindent
The skeletons of trees contributing to $\Pi_{a,1}^{(2)}$ have melonic boundaries and common color on their edges, with at least one bicolored edge. The most general configuration corresponds to a succession of colored and bicolored edges; and to take into account all possible configurations, we introduce a new elementary building block replacing the purely melonic pattern \eqref{effsum1}. This elementary building block is itself a sum of blocks, made of a succession of pure pseudo-melonic blocks between melonic boundaries. The first  of these elementary building blocks are then:
\begin{equation}
\mathcal{A}_{1i}:=\vcenter{\hbox{\includegraphics[scale=0.8]{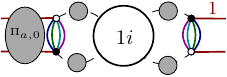} }}\,,
\end{equation}
\begin{equation}
\mathcal{A}_{1ij}:=\vcenter{\hbox{\includegraphics[scale=0.8]{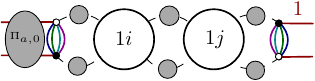} }}\,,
\end{equation}
\begin{equation}
\mathcal{A}_{1ijk}:=\vcenter{\hbox{\includegraphics[scale=0.8]{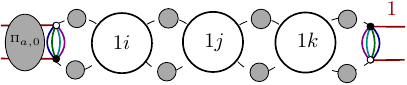} }}\,,
\end{equation}
\begin{equation}
\mathcal{A}_{1ijkl}:=\vcenter{\hbox{\includegraphics[scale=0.8]{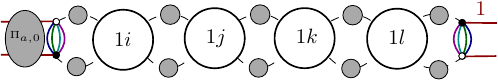} }}\,.
\end{equation}
\begin{equation*}
\cdots
\end{equation*}
where the grey bubble represents the insertion of an effective purely melonic pattern $\Pi_{a,0}^{(2)}$, and the white bubbles represent the insertion of a pure bicolored pattern $\Pi_{d,1i}^{(2)}$ i.e.  the color $1$ (red on the figures) being common on melonic and pseudo-melonic insertions. The indices $i$, $j$, $k$ and $l$ are such that consecutive indices are different. There are then $d-1=4$ different ways to choose the first index $i$, and $d-2=3$ ways for all successive indices $j$, $k$, $l$, ...\\
\noindent
The value of each diagram being independent to the choice of the selected values for these indices, the complete elementary pattern including pseudo-melonic contributions may be written as:
\begin{equation}
\vcenter{\hbox{\includegraphics[scale=0.8]{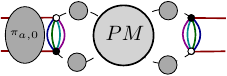} }}\equiv\Pi_{a,0}\mathcal{B}\,,
\end{equation}
with:
\begin{equation}
\Pi_{a,0}\mathcal{B}:=4(\mathcal{A}_{12}+3\mathcal{A}_{123}+3^2\mathcal{A}_{1232}+3^3\mathcal{A}_{12323}+\cdots)\,.\label{temperature}
\end{equation}
where the big grey bubble that we called ‘‘PM'' denotes the sum of contributions coming from chains of pseudo-melonic interactions. As for the elementary purely melonic patterns \eqref{effsum1}, the same scheme can be repeated 	
up to infinity, as in \eqref{gamma4melon}:
\begin{align}
\vcenter{\hbox{\includegraphics[scale=0.8]{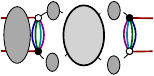}}}&+\vcenter{\hbox{\includegraphics[scale=0.8]{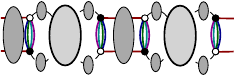}}}+\cdots=\sum_{n=1}^\infty \left(\vcenter{\hbox{\includegraphics[scale=0.8]{effneckprime.pdf}}}\right)^n\,,
\end{align}
and the formal sum can be explicitly computed as a geometric progression, leading to:
\begin{equation}
\sum_{n=1}^\infty \left(\vcenter{\hbox{\includegraphics[scale=0.8]{effneckprime.pdf}}}\right)^n=\frac{\Pi_{a,0}\,\mathcal{B}}{1-\Pi_{a,0}\,\mathcal{B}}\,.
\end{equation}
Finally  the complete effective vertex function $\Gamma^{(4),1}_{a,1}$ having skeleton function $\Pi_{a,1}^{(2)}$ and completing \eqref{gamma4melon}  is decomposed as:
\begin{equation}
\Gamma^{(4),1}_{a,1}=\vcenter{\hbox{\includegraphics[scale=0.8]{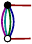}}}\left[\sum_{n=1}^\infty \left(\vcenter{\hbox{\includegraphics[scale=0.8]{effneckprime.pdf}}}\right)^n\right]\vcenter{\hbox{\includegraphics[scale=0.8]{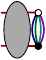}}}\,.
\end{equation}
Explicitly:
\begin{align}
\Gamma^{(4),1}_{a,1}(\vec{0},\vec{0},\vec{0},\vec{0})=&2\times\left[\frac{\Pi_{a,0}\,\mathcal{B}\Pi_{a,0}}{1-\Pi_{a,0}\,\mathcal{B}}\right]\,.
\end{align}
Note that all the effective functions commute. We preserved their order to keep the structure of the diagrams. Moreover, note that the skeleton with length one has been included on the purely melonic contribution $\Pi_{a,0}^{(2)}$. \\

\noindent
$\bullet$ \textit{Computation of $\Pi_{b}^{(2)}$ and $\Pi_{c}^{(2)}$ .}\\
\noindent
The next contributions coming from Figure \eqref{fig15}b and c, and whose effective skeleton functions are respectively denoted as $\Pi_{b}^{(2)}$ and $\Pi_{c}^{(2)}$ can be computed in the same way.  In Figure       \eqref{fig19} we give a picture of typical trees contributing to each of these two functions.
\begin{center}
\includegraphics[scale=1]{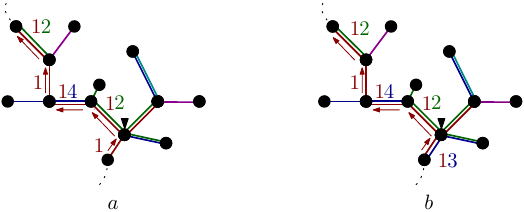} 
\captionof{figure}{Typical contributions to $\Gamma^{(4)}_{b}$ (a) and $\Gamma^{(4)}_{c}$ (b).}\label{fig19}
\end{center}
Let us start with the computation of $\Pi_{b}^{(2)}$, whose building trees have a melon and a pseudo melon on their boundaries as on Figure \eqref{fig19}a. Obviously, the first edge has to be monocolored, and the ending edge has to be bicolored. The effective skeleton function then has to begin with an  effective melon and end with an effective pseudo-melonic function, connecting together with an history involving all possible configurations of edges having a common color. The elementary pattern then connects together the effective boundaries. For the next term, the same pattern repeats, and we get the structure:
\begin{equation}
\sum_{i\neq 1}\,\vcenter{\hbox{\includegraphics[scale=0.8]{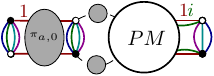} }}+ \sum_{i,j\neq 1}\,\vcenter{\hbox{\includegraphics[scale=0.8]{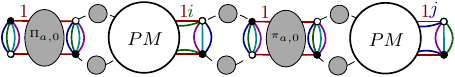} }}+\cdots\,,\label{seriesterm1}
\end{equation}
where the effective white pseudo-melonic vertex denoted as ‘‘PM'' is nothing but that we called ‘‘PM'' with a gray bubble, where we have extracted a boundary pseudo-melonic vertex. Completing the series in \eqref{seriesterm1}, we then get:
\begin{equation}
\vcenter{\hbox{\includegraphics[scale=0.8]{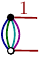} }}\left[\sum_{n=0}\left(\sum_{i\neq 1}\,\vcenter{\hbox{\includegraphics[scale=0.8]{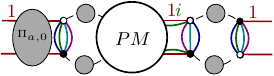} }}\right)^n\right]\sum_{i\neq 1}\vcenter{\hbox{\includegraphics[scale=0.8]{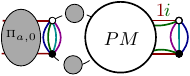} }}\,.
\end{equation}
Obviously,  the elementary building block in the  bracket is nothing that we called $\Pi_{a,0}\mathcal{B}$ in equation \eqref{temperature}. Moreover, the diagram on the right is nothing but this elementary building block amputated from its right external loop. We denote it as $\Pi_{a,0}\bar{\mathcal{B}}$:
\begin{equation}
\Pi_{a,0}\bar{\mathcal{B}}:=\sum_{i\neq 1}\vcenter{\hbox{\includegraphics[scale=0.8]{conf23right.pdf} }}\,,
\end{equation}
and we get:
\begin{equation}
\Pi_b^{(2)}:=\frac{\Pi_{a,0}\bar{\mathcal{B}}}{1-\Pi_{a,0}{\mathcal{B}}}\,.
\end{equation}

\noindent
Finally, for the computation of $\Pi_c^{(2)}$, we have to distinguish the pairs of colors of the boundary pseudo-melonic vertices. We will denote them as  $(1i)$ and $(1j)$, we have to treat separately the case $i=j$ and the case $i\neq j$. For the case $i\neq j$, the elementary pattern must be the following, connecting together two effective pseudo-melonic functions of type $(1i)$ and $(1j)$:
\begin{equation}
\sum_{i\neq j}\vcenter{\hbox{\includegraphics[scale=1]{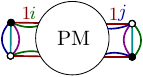} }}=:2\mathcal{D}\,,
\end{equation}
the white bubble being a sum over skeleton with pseudo-melonic interactions only. The factor $2$ comes from $\sym \mathcal{V}^{(4,1)\,i}_{\vec{0},\vec{0},\vec{0},\vec{0}}\,,=2$  for zero external momenta.\\
\noindent
This elementary pattern has to be completed with melonic insertions between the two effective boundaries, leading to the contribution:
\begin{equation}
\sum_{i,j}\vcenter{\hbox{\includegraphics[scale=1]{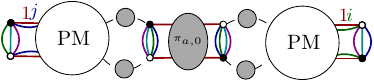} }}\,.
\end{equation}
Note that, because of the melonic insertions, the external faces of colors $i$ and $j$ do not cross one in the other, and the constraint $i\neq j$ can be removed. The same structure can be repeated up to  infinity and  leading to the sum:
\begin{equation}
\sum_{i,j}\vcenter{\hbox{\includegraphics[scale=0.9]{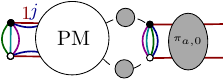} }}\left[\sum_{n=0}\left(\vcenter{\hbox{\includegraphics[scale=0.9]{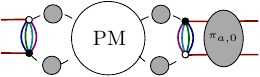} }}\right)^n\right]\vcenter{\hbox{\includegraphics[scale=0.9]{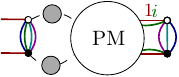} }}\,.
\end{equation}
Then,  the kernel $\Pi_c^{(2)}$ is written as:
\begin{equation}
\Pi_c=\mathcal{D}+\bar{\mathcal{B}}\,\Pi_{b}^{(2)}\,.
\end{equation}
\noindent
$\bullet$ \textit{Computation of $\mathcal{B}$, $\bar{\mathcal{B}}$ and $\mathcal{D}$.}\\

\noindent
To complete the proof we have to compute the kernels $\mathcal{B}$, $\bar{\mathcal{B}}$ and $\mathcal{D}$. Let us start with $\mathcal{B}$ defined in \eqref{temperature}. The successive contributions in this equation i.e.  $\mathcal{A}_{12}$, $\mathcal{A}_{123}$, ... can be easily expressed clearly. First of all, we get:
\begin{equation}
\mathcal{A}_{12}=\Pi_{a,0}^{(2)}\,\sum_{p} \mathcal{A}_{3,2}(p) \Pi_{2,12}^{(2)}(p)\mathcal{A}_{3,2}(p)\,.
\end{equation}
Now, let $\mathcal{A}_{123}$ be a quantity  obtained from gluing $\Pi_{d,12}$ and $\Pi_{d,13}$. In addition to the ‘‘short" faces of length two,  coming from gluing of the two effective vertices, there are two long faces of effective length four, for colors $2$ and $3$, such that:
\begin{align}
\nonumber \mathcal{A}_{123}=\Pi_{a,0}^{(2)}\,\sum_{p,p^\prime}& \mathcal{A}_{3,2}(p) \Pi_{2,12}^{(2)}(p) \mathcal{C}(p,p^\prime)\Pi_{2,13}^{(2)} (p^\prime\,)  \mathcal{A}_{3,2}(p^\prime)\,,
\end{align}
where we defined:
\begin{equation}
\mathcal{C}(p,p^\prime):=\sum_{\vec{q}\,\vec{q}\,^\prime\in\mathbb{Z}^{d-1}} G^2(\vec{q}\,)\delta_{q_2p}\delta_{q_3 p^\prime} \,.
\end{equation}
Note that we have distinguished the indices on $\Pi_{2,12}^{(2)}(p)$ and $\Pi_{2,13}^{(2)}(p)$ for convenience; the structure of $\mathcal{C}$ took into account the strand structure of the diagram. Then, we may use the indice of  $\Pi_{2,12}^{(2)}(p)$ only. Then, in the same way, for $\mathcal{A}_{1232}$, we get:
\begin{align}
\mathcal{A}_{1232}=\Pi_{a,0}^{(2)}\sum_{p,p^{\prime},p^{\prime\prime}}\mathcal{A}_{3,2}(p) \Pi_{2,12}^{(2)}(p) \mathcal{C}(p,p^\prime)\Pi_{2,12}^{(2)} (p^\prime)\mathcal{C}(p^\prime,p^{\prime\prime})\Pi_{2,12}^{(2)} (p^{\prime\prime})\mathcal{A}_{3,2}(p^{\prime\prime})\,.
\end{align}
In order to write conveniently the general term, we introduce the compact matrix notation:
\begin{align}
\mathcal{A}_{1232}=\Pi_{a,0}^{(2)}\sum_{p,p^\prime}\mathcal{A}_{3,2}(p) \left[\Pi_{2,12}^{(2)}\,\mathcal{C}\,\Pi_{2,12}^{(2)}\, \mathcal{C}\,\Pi_{2,12}^{(2)}\right](p,p^\prime)\mathcal{A}_{3,2}(p^{\prime\prime})\,.
\end{align}
The Equation \eqref{temperature} becomes therefore :
\begin{equation}
\mathcal{B}=4\Pi_{a,0}^{(2)}\sum_{p,p^\prime}\mathcal{A}_{3,2}(p)\mathcal{K}(p,p^\prime)\mathcal{A}_{3,2}(p^\prime)\,,\label{formulacool1}
\end{equation}
with
\begin{equation}
\mathcal{K}:=\Pi_{2,12}^{(2)}\left(1+3\mathcal{C}\,\Pi_{2,12}^{(2)}+(3\mathcal{C}\,\Pi_{2,12}^{(2)})^2+\cdots\right)=\Pi_{2,12}^{(2)}(1-3\mathcal{C}\,\Pi_{2,12}^{(2)})^{-1}\,.\label{formulacool}
\end{equation}

\noindent
In the same way, we defined the function $\bar{\mathcal{B}}$ as ${\mathcal{B}}$ amputated to the right external loop as:
\begin{equation}
\bar{\mathcal{B}}:=4(\bar{\mathcal{A}}_{12}+3\bar{\mathcal{A}}_{123}+3^2\bar{\mathcal{A}}_{1232}+3^3\bar{\mathcal{A}}_{12323}+\cdots)\,.\label{temperature2}
\end{equation}
with:
\begin{equation}
\bar{\mathcal{A}}_{12}=\Pi_{a,0}^{(2)}\, \mathcal{A}_{3,2}\Pi_{2,12}^{(2)}\,,
\end{equation}
\begin{align}
\nonumber \bar{\mathcal{A}}_{123}=\Pi_{a,0}^{(2)}\,\sum_{p}& \mathcal{A}_{3,2}(p) \Pi_{2,12}^{(2)}(p) \mathcal{C}(p,0)\Pi_{d,13}  \,,
\end{align}
such that, in a compact matrix form:
\begin{equation}
\bar{\mathcal{B}}=4\Pi_{a,0}^{(2)}\sum_{p,p^\prime}\mathcal{A}_{3,2}(p)\mathcal{K}(p,0)\,.\label{formulacool2}
\end{equation}

\noindent
Finally, the last function that we have to compute is $\mathcal{D}$, built of chains of pseudo-melons of type-1 or type-2, such that the  boundaries are of type $(1i)$ and $(1j)$ with $i\neq j$. It can be easily deduced from decomposition \eqref{temperature}, keeping only the contributions having such a boundary. We get:
\begin{equation}
\mathcal{D}=4\times 3\,\left[\Pi_{2,12}^{(2)}\mathcal{C}\,\Pi_{2,12}^{(2)}\,\left(1+ 2\,(\mathcal{C}\,\Pi_{2,12}^{(2)})+(2\,\mathcal{C}\,\Pi_{2,12}^{(2)})^2+\cdots\right)\right](0,0)\,,
\end{equation}
which can be computed as formula \eqref{formulacool}:
\begin{equation}
\mathcal{D}=12\,\left[\Pi_{2,12}^{(2)}\mathcal{C}\,\Pi_{2,12}^{(2)}\,\left(1- 2\,(\mathcal{C}\,\Pi_{2,12}^{(2)})\right)^{-1}\right](0,0)\,. \label{formulacool3}
\end{equation}
\begin{flushright}
$\square$
\end{flushright}

\noindent
The \textit{renormalization conditions} impose the values of zero-momenta couplings. They define the asymptotic couplings in the IR, which is given  for $s\to -\infty$, so that:
\bea
\Pi_1^{(2)}(s\to-\infty)=2\lambda_{4,1}^r\,,\label{rencond1}
\\
\Pi_{2,12}^{(2)}(s\to-\infty)=2\lambda_{4,2}^r\,,\label{rencond2}
\eea
the upper index $r$ for ‘‘renormalized'' referring to the finiteness of the corresponding quantities in the continuum limit $\Lambda\to\infty$. Note that this is not obvious, because some quantities like $\mathrm{b}$ and $\mathcal{A}_{mn}$ are divergent. The finiteness of the limit is guaranteed from renormalizability of the theory. In practice, counter-terms may be defined in order to ensure the finiteness of the effective skeleton function for arbitrary $s$, from deep UV. In this intermediate regime, these functions define \textit{effective couplings} at scale $s$:

\begin{definition}\label{defrenconds}
The effective essential and marginal couplings at scale $s$, $\lambda_{4,1}(s)$ and $\lambda_{4,2}(s)$, respectively associated to the quartic melonic and quartic pseudo-melonic interactions are defined as
\end{definition}
\begin{equation}
\Pi_1^{(2)}(s)=2\lambda_{4,1}(s)\,,\label{rencond12}
\end{equation}
and:
\begin{equation}
\Pi_{2,12}^{(2)}(s)=2\lambda_{4,2}(s)\,.\label{rencond22}
\end{equation}

\noindent
Note that in these equations, we introduce the explicit dependence on $s$ for effective vertex functions. This dependence comes from their definition, and have been left without all confusion in the hope to simplify the notations. 

\subsubsection{Six and eight point effective vertices}\label{68point}

The equations for LO four-point function is  obtained above. We have to compute the same equations for six and eight-point effective vertices in this section. We will investigate successively the effective six and eight-points vertices having melonic, pseudo-melonic or intertwining boundary graphs. Indeed, we will see that, in addition to the melonic and pseudo-melonic boundaries, we get mixing boundaries, having intermediate canonical dimensions between melons and pseudo-melons as mentioned in our Introduction. These mixing boundaries correspond to connected sums of elementary quartic melons and/or pseudo-melons, with colors respecting the LO tadpole deletions leading to six and eight points function from LO four-point functions. We will detail all of them for each case. However before starting  this investigation, let make a remark about the existence of mixing configurations. An elementary example, corresponding to the connected sum of a quartic melon and a quartic pseudo-melon is given in Figure \eqref{fig21} b, and it is easy to see that it comes from the contraction of a $0$-dipole between two vertices, as pictured in Figure \eqref{fig20bis}. 
\begin{center}
$\vcenter{\hbox{\includegraphics[scale=0.9]{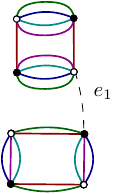} }}\to \vcenter{\hbox{\includegraphics[scale=1]{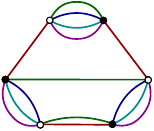} }}$
\captionof{figure}{Contraction of a zero dipole $e_1$ between a melonic vertex and a pseudo-melonic vertex. The contraction leads to the intertwining diagram on the right.}\label{fig20bis}
\end{center}
Then, even if the initial theory space is a direct sum $\mathcal{S}_M\oplus\mathcal{S}_{PM}$, the theory space does not remain along this subspace under the renormalization group flow. A new type of graph arises in the theory space, mixing melons and pseudo-melons, and having intermediate power counting between them. Schematically:
\begin{equation}
‘‘\frac{d}{ds}\left( \mathcal{S}_M\oplus\mathcal{S}_{PM}\right) = \mathcal{S}_M\oplus\mathcal{S}_{PM}\oplus \left\{\vcenter{\hbox{\includegraphics[scale=0.6]{3intermelonbis.pdf} }}+\,\cdots\right\}\,"
\end{equation}
\noindent
As for four-point functions, the LO  six-point functions are obtained from deletion of   the melonic or pseudo-melonic tadpole along the boundary of an external face from a LO four-point graphs. Then, the allowed boundaries for LO graphs may be all deduced from the four leading order configurations pictured in Figure \eqref{fig15}, and it is easy to see from the deletion procedure that there are only four allowed boundaries for LO graphs, see  Figure \eqref{fig21}. \\ 

\noindent
The first graph of  Figure \eqref{fig21} denoted by (a) is an element of the non-branching melonic family, obtained as a connected sum of two identical quartic melons\footnote{As for pseudo-melons, non-branching melons are obtain as connected sums of the same elementary quartic melon. These graphs are described in \cite{Carrozza:2016tih},\cite{Lahoche:2018vun}. }. The second one (b) is a branching melon graph, obtained from Figure \eqref{fig15}b opening a melonic tadpole on the boundary of the green external face, whose end points are hooked on the pseudo-melonic opened boundary vertex. The last diagram (d) corresponds to the non-branching pseudo-melonics, obtained as connected sum of two identical quartic pseudo-melons. Finally, the boundary graph of type (c) corresponds to the intertwining graphs, obtained from Figure \eqref{fig15}d. 
\begin{center}
\includegraphics[scale=1]{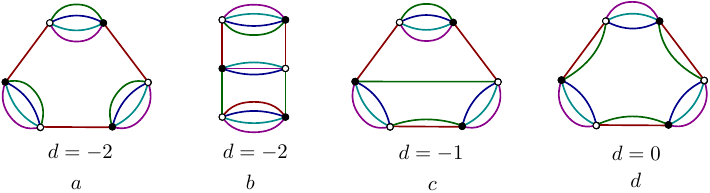} 
\captionof{figure}{The four possible LO boundary graphs for melonic, pseudo-melonic and interwining graphs. A non-branching $3$-valent melon (a), a $3$-valent branching melon (b), a non-branching pseudo-melon (d) and an intertwiner diagram (c). The  canonical dimensions are also  indicated.}\label{fig21}
\end{center}
However, from deletion of three melonic and/or pseudo-melonic tadpoles, it is easy to check that the connectivity of the external faces allows three configurations and  all of these are pictured in Figure \eqref{fig21}. The first one (a) corresponds to a non-branching melon. The last one (c) corresponds to a non-branching pseudo-melon. The intermediate one (b) however, corresponds to the connected sum of a quartic melon and a quartic pseudo-melon, and intertwine between the two sectors. It is clear from the canonical dimensions, indicated on Figure \eqref{fig21}. Note that we have to take into account the intermediate diagram for the computation of the the flow equations, due to the fact that the leading contraction building from  a four-dipole has a quartic pseudo-melon as boundary graph.\\

\noindent
The intertwining configurations are relevant for our analysis. Indeed, we will use two kinds of equations in the next section: The flow equation \eqref{Wett} in order to compute the evolution of the essential and marginal couplings; and the Ward identities, to compute the first derivative with respect to the external momenta of the effective vertex functions, involved in the computation of the anomalous dimension. In any case, all our computations involve a single effective loop, and our restriction in the deep UV sector requires to consider only the LO contractions  i.e.   having the maximal degree of divergences, as relevant contributions for the flow of the couplings and the value of the anomalous dimension. Then, even is its canonical dimension is smaller than the one of the pseudo-melon in Figure \eqref{fig21} d, the graph in Figure \eqref{fig21} c will contribute to the flow of the quartic pseudo-melonic coupling, and have to be retained as a relevant contribution on the Ward identity. Indeed, it compensates its lack of canonical dimension with a melonic "pole", allowing to create a four-dipole. For instance, the following graph
\begin{equation}
\vcenter{\hbox{\includegraphics[scale=0.9]{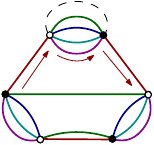} }}\,,
\end{equation}
has power counting $\omega=-2+4-1=1$, and it is easy to check that its boundary graph is a quartic pseudo-melon. The same behavior will be arise for eight-point functions, for which we will only retain those LO   six-points non-branching pseudo-melon after one-loop contraction. \\

\noindent
Let us start with non-branching melonic six-point effective vertices. As for four-point functions, the six-point effective vertex has to be a sum indexed with the color of its external face, being as well the color of the skeleton in the corresponding tree in HS representation. 
\begin{equation}
\Gamma^{(6)}_{s\,,\text{a, melo}}(\{\vec{p}_j\})=\sum_{i=1}^d\Gamma^{(6)\,i}_{s\,,\text{a, melo}}(\{\vec{p}_j\})\,,
\end{equation}

\noindent
the index $a$ referring to  Figure \eqref{fig21} a.   For the four-point functions, we see that the LO graphs are trees organized around a skeleton having the topology of a line. For the  six-point functions, the skeleton will be a \textit{tripod}, with three \textit{arms} hooked to a common vertex. More precisely, investigating the structure of $3$-ciliated trees with skeleton of color $1$, we get two types of tripods, distinguished both on Figure \eqref{fig22}. In the first one case (a), the three arms of color $1$ hooked to the three opened tadpoles and  are fully  hooked on the same vertex $v$ with non-breaking edges. In the second case (b), the three arms are hooked to the same vertex $v$ from a common breaking edge.   As for four-point functions, we can fix the size of the tripod skeleton and sum over trees having such a skeleton, relaxing in a second time the length of the different arms.  Precisely  as for  the four-point functions, we define the zero-momenta melonic \textit{effective tripod function} $\Pi_{1,a}^{(3)}$ as:
\begin{equation}
\Gamma^{(6)\,i}_{s\,,\text{a,melo}}(\vec{0},\vec{0},\vec{0},\vec{0},\vec{0},\vec{0})=:3!\,\Pi_{1,a}^{(3)}\,,\label{defeff1}
\end{equation}

\begin{center}
\includegraphics[scale=1]{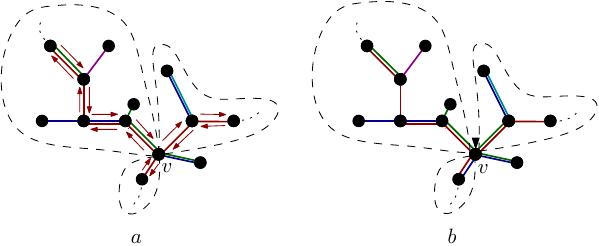} 
\captionof{figure}{Typical tree configurations contributing to the LO $6$-point functions with melonic boundaries. In (a), the three arms of the red skeleton are hooked on the same vertex $v$ from non-breaking edges. In (b), they are hooked to $v$ with a common breaking edge.}\label{fig22}
\end{center}

\noindent
Similarly, effective vertex having boundaries of type (b) and (c) on Figure \eqref{fig21} may be denoted by $\Gamma^{(6)\,ij}_{s\,,\text{b,melo}}$ and $\Gamma^{(6)\,ij}_{s\,,\text{c,inter}}$ and  both of these contributions  depend on a pair of indices. We define the corresponding effective skeleton functions as:
\begin{equation}
\Gamma^{(6)\,ij}_{s\,,\text{b,melo}}(\vec{0},\vec{0},\vec{0},\vec{0},\vec{0},\vec{0})=:3!\,\Pi_{1,b}^{(3)}\,,
\end{equation}
\begin{equation}
\Gamma^{(6)\,ij}_{s\,,\text{c,inter}}(\vec{0},\vec{0},\vec{0},\vec{0},\vec{0},\vec{0})=:3!\,\Pi_{1,c}^{(3)}\,.
\end{equation}

\noindent
and we have the following statement:
\begin{proposition}\label{propositionphi61}
The effective zero-momenta tripod functions $\Pi_{1,a}^{(3)}$, $\Pi_{1,b}^{(3)}$ and $\Pi_{1,b}^{(3)}$ are expressed in terms of the essential and marginal couplings as\footnote{Note that, as for four-point functions, the upper index $3$ refers to the number of cilia.}:
\begin{align}
\nonumber\Pi_{1,a}^{(3)}&=16\,\bigg\{  \bigg(\lambda_{4,1}^3(s)\mathcal{A}_{4,3}+12\lambda_{4,1}^2(s)\lambda_{4,2}(s)\mathcal{A}_{3,3}+36\lambda_{4,2}^2(s)\lambda_{4,1}(s)\mathcal{A}_{2,3}\\&+24\lambda_{4,2}^3(s)\mathcal{A}_{1,3}\bigg)
-12\lambda_{4,1}^3(s)\sum_{p\in\mathbb{Z}} \lambda_{6,1}(p,s)(\mathcal{A}_{3,2}(p))^3+12\lambda_{6,1}(s)\lambda_{4,1}^2(s)(\mathcal{A}_{3,2})^2\bigg\}\,,\label{phi61}
\end{align}

\begin{align}
\Pi_{1,b}^{(3)}=3!\times 16\,\left\{2\lambda_{4,1}^2(s)\lambda_{4,2}(s) \mathcal{A}_{3,3}+3\lambda_{6,1}(s)\lambda_{4,1}^2(s)(\mathcal{A}_{3,2})^2\right\}\,,\label{effskel61}
\end{align}

\begin{align}
\Pi_{1,c}^{(3)}=3! \times 2\times \left\{\,8\lambda_{4,1}(s)\lambda_{4,2}^2(s) \mathcal{A}_{3,3}+24\lambda_{4,2}^3(s)\mathcal{A}_{2,3}-6\lambda_{4,1}(s)\lambda_{6,1}(s)\mathcal{A}_{3,2})\right\}\,.\label{effskel62}
\end{align}

where $\lambda_{4,1}(s)$, $\lambda_{4,2}(s)$ and $\lambda_{6,1}(s)$ are effective couplings at scale $s$ and $\lambda_{6,1}(p,s)$ is the $p$ dependent effective coupling at the momentum $p$ given by:
\bea
\lambda_{6,1}(p,s)&=&\Bigg(\frac{\lambda_{6,1}(s)-\frac{8}{3}\lambda_{4,2}^3(s)\mathcal{A}_{3,3}}{1-3!\lambda_{4,2}(s)\mathcal{A}_{3,2}+12\lambda_{4,2}^2(s)\mathcal{A}_{3,2}^2-16\lambda_{4,2}^3(s)\mathcal{A}_{3,2}^3}\Bigg)\Big(1-3!\lambda_{4,2}(s)\mathcal{A}_{3,2}(p)\cr
&+&12\lambda_{4,2}^2(s)\mathcal{A}^2_{3,2}(p)-16\lambda_{4,2}^3(s) \mathcal{A}^3_{3,2}(p)\Big)+\frac{8}{3}\lambda^3_{4,2}(s)\mathcal{A}_{3,3}(p)
\eea
\end{proposition}

\noindent
In order to prove this statement, we have to consider the intermediate result about effective non-branching pseudo-melonic functions:
\begin{lemma}\label{lemmaphi62}
The bare coupling $\lambda_{6,1}$ may be expressed in terms of the effective essential and marginal effective pseudo-melonic couplings at scale $s$ as:
\begin{equation}
\lambda_{6,1}=\frac{\lambda_{6,1}(s)-\frac{8}{3}\lambda_{4,2}^3(s)\mathcal{A}_{3,3}}{1-3!\lambda_{4,2}(s)\mathcal{A}_{3,2}+12\lambda_{4,2}^2(s)\mathcal{A}_{3,2}^2-16\lambda_{4,2}^3(s)\mathcal{A}_{3,2}^3}\,.
\end{equation}
\end{lemma}

\noindent
\textit{Proof.}
As for the  trees having melonic boundary, the LO trees contributing to the effective tripod function $\Pi_{2,ij}^{(3)}$ split into two families: the trees whose three arms are hooked to a common vertex, and the trees whose arms are hooked to  a common breaking edge corresponding to the boundary graph of the global tree. Listing all the possible LO configurations, we get :
\begin{align}
\nonumber\vcenter{\hbox{\includegraphics[scale=1]{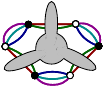} }} &=\vcenter{\hbox{\includegraphics[scale=0.7]{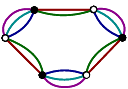} }} +\vcenter{\hbox{\includegraphics[scale=0.8]{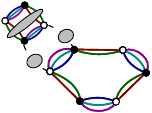} }} +\vcenter{\hbox{\includegraphics[scale=0.8]{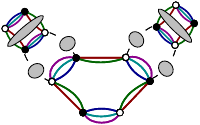} }}\\ &+\vcenter{\hbox{\includegraphics[scale=0.7]{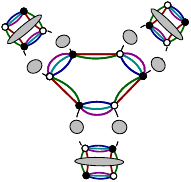} }}
+\vcenter{\hbox{\includegraphics[scale=0.7]{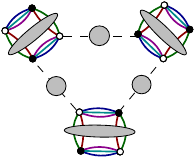} }}\,, \label{sixpointdecomp}
\end{align}
where the first term with a grey bubble represents the effective six-point vertex having pseudo-melonic boundary with fixed bicolored edges (red and green on the figure):
\begin{equation}
\Gamma^{(6)\,ij}_{s\,,\text{pseudo-melo}}(\vec{0},\vec{0},\vec{0},\vec{0},\vec{0},\vec{0})\equiv \vcenter{\hbox{\includegraphics[scale=1]{phi60.pdf} }} \,,
\end{equation}
the function $\Gamma^{(6)\,ij}_{s\,,\text{melo}}$ being defined such that the effective vertex function for pseudo-melonic boundaries is written  as $\Gamma^{(6)}_{s\,,\text{pseudo-melo}}=\sum_{j<i}\Gamma^{(6)\,ij}_{s\,,\text{pseudo-melo}}$. The function $\Gamma^{(6)\,ij}_{s\,,\text{pseudo-melo}}$ may be then computed directly from renormalization conditions for four-point functions \eqref{rencond2}. Taking into account the  symmetry factors, we get:
\begin{align}
\nonumber\Gamma^{(6)\,ij}_{s\,,\text{pseudo-melo}}=(3!)^2 \lambda_{6,1}&\left[1-3!\lambda_{4,2}(s)\mathcal{A}_{3,2}+2\times 3!\lambda_{4,2}^2(s)\mathcal{A}_{3,2}^2-2^4\lambda_{4,2}^3(s)\mathcal{A}_{3,2}^3\right]\\
&+3!2^4\lambda_{4,2}^3(s)\mathcal{A}_{3,3}\,.
\end{align}
Note that the symmetry factors as well as relative signs may be easily understood and computed considering the first terms of the perturbative expansion, and identifying them with the corresponding terms in this expansion. Moreover, note that we reintroduced the variable $s$ to make difference between effective and bare couplings. We define the effective tripod skeleton function $\Pi_{2,ij}^{(3)}$ as:
\begin{equation}
\Gamma^{(6)\,ij}_{s\,,\text{pseudo-melo}}=(3!)^2\Pi_{2,ij}^{(3)}.
\end{equation}
With the renormalization conditions, we provide new definition, which completes Definition \eqref{defrenconds}:
\begin{definition}\label{defrenconds2}
The effective marginal coupling $\lambda_{6,1}(s)$ at scale $s$ is defined as:
\begin{equation}
\Pi_{2,ij}^{(3)}=\lambda_{6,1}(s)\,,\qquad \Pi_{2,ij}^{(3)}(s\to-\infty)=\lambda_{6,1}^r \label{rencond3}\,,
\end{equation}
\end{definition}
\noindent
we get the equality:
\bea
\lambda_{6,1}(s)=\lambda_{6,1}\left[1-3!\lambda_{4,2}(s)\mathcal{A}_{3,2}+2\times 3!\lambda_{4,2}^2(s)\mathcal{A}_{3,2}^2-2^4\lambda_{4,2}^3(s)\mathcal{A}_{3,2}^3\right]+\frac{8}{3}\lambda_{4,2}^3(s)\mathcal{A}_{3,3}\,.
\eea
which ends the proof. 

\begin{flushright}
$\square$
\end{flushright}
\noindent

However, using  the  Lemma \eqref{lemmaphi62}, we are able to  provide the proof of  Proposition \eqref{propositionphi61}.\\

\noindent
\textit{Proof of proposition \eqref{propositionphi61}.}

\noindent
We will proceed as for the proof of the previous lemma, listing all the allowed configurations compatible with the corresponding boundary diagram, pictured in Figure \eqref{fig21}a.
 As explained before, we have to distinguish the case
 when the three arms are hooked to the same vertex, and the case when the three arms are hooked to a common breaking edge, having a common color with the three arms hooked to him. These two configurations are pictured in Figure \eqref{fig22}.
 However, the classification has to be refined from Lemma \eqref{lemmaphi62}. Indeed, among the configurations of type \eqref{fig22}a, some of them are made with a tripod kernel  having a pseudo-melonic boundary graph, and allows  one  to build an effective breaking edge to which are hooked three arms. 
This kernels are nothing that the fourth contribution in the equation \eqref{sixpointdecomp}. Adding it to the other ones having a common breaking edge for the three arms, we build an effective six-point vertex having pseudo-melonic boundary, with melonic arms hooked to him, corresponding to effective melonic four-point functions. It is easy to check that all allowed configurations are those pictured in Figure \eqref{fig23}, the configurations (e) and (f) having a kernel corresponding to an effective pseudo-melonic function. Taking into account the respective canonical dimensions for each effective insertions,  it is easy to see that all these configurations have the same power-counting, $\omega=-2$, in accordance with the expected canonical dimension for $3$-valent melonic bubbles given equation \eqref{dimensiongeneral} . From renormalization conditions \eqref{rencond1}, \eqref{rencond2} and \eqref{rencond3}; and taking into account relative signs and symmetry factors considering the first terms of the perturbative expansion, we get:
\begin{align}
\nonumber&\Gamma^{(6)\,i}_{s\,,\text{a, melo}}=3!\times 16\,\bigg\{  \bigg(\lambda_{4,1}^3(s)\mathcal{A}_{4,3}+12\lambda_{4,1}^2(s)\lambda_{4,2}(s)\mathcal{A}_{3,3}+36\lambda_{4,2}^2(s)\lambda_{4,1}(s)\mathcal{A}_{2,3}\\\quad&+24\lambda_{4,2}^3(s)\mathcal{A}_{1,3}\bigg)
-12\lambda_{4,1}^3(s)\sum_{p\in\mathbb{Z}} \lambda_{6,1}(p,s)(\mathcal{A}_{3,2}(p))^3+12\lambda_{6,1}(s)\lambda_{4,1}^2(s)(\mathcal{A}_{3,2})^2\bigg\}\,.
\end{align}
We then deduce the expression of $\Pi_{1,a}^{(3)}$ for definition \eqref{defeff1}. 

\begin{center}
\includegraphics[scale=0.8]{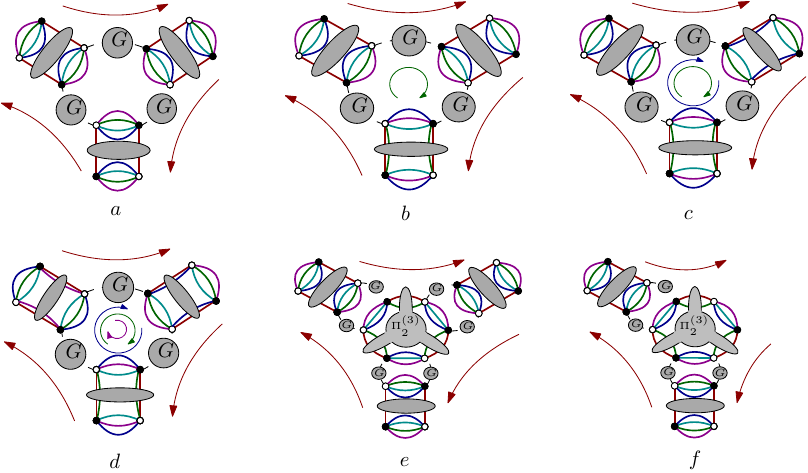} 
\captionof{figure}{All the possible configurations having a non-branching melon as boundary graph. The grey bubbles denote effective vertices whose boundary graphs are explained. The trajectories of the external faces running in the interior of the diagrams are pictured with colored arrows.}\label{fig23}
\end{center}

\noindent
In the same way, the configurations contributing to effective six-point vertex functions having branching melonic and intertwining boundaries can be easily obtained from a list of all allowed configurations. As for the previous case, some of these contributions can be resumed as effective pseudo-melonic effective vertices, such that the relevant LO contributions for each case are pictured in Figure \eqref{fig23bis}  and Figure \eqref{fig24} below.

\begin{center}
\includegraphics[scale=0.8]{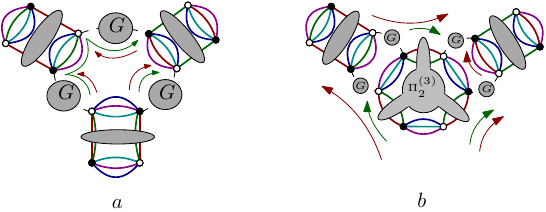} 
\captionof{figure}{Two configurations contributing to the six-point effective vertices having branching melonic boundaries.}\label{fig23bis}
\end{center}
\noindent
Expressing this relationship in detail, we get for zero external momenta effective functions:
 \begin{equation}
\Gamma^{(6)\,ij}_{s\,,\text{b,melo}}=(3!)^2\times 8\,\left\{4\lambda_{4,1}^2(s)\lambda_{4,2}(s) \mathcal{A}_{3,3}+3\lambda_{6,1}(s)\lambda_{4,1}^2(s)(\mathcal{A}_{3,2})^2\right\}\,,
\end{equation}
 for branching melonic boundaries, and :
 \begin{equation}
\Gamma^{(6)\,ij}_{s\,,\text{c,inter}}=(3!)^2 \times  \left\{\,16\lambda_{4,1}(s)\lambda_{4,2}^2(s) \mathcal{A}_{3,3}+48\lambda_{4,2}^3(s)\mathcal{A}_{2,3}-6\lambda_{4,1}(s)\lambda_{6,1}(s)\mathcal{A}_{3,2})\right\}\,,
\end{equation}
for intertwining boundaries.

\begin{center}
\includegraphics[scale=0.8]{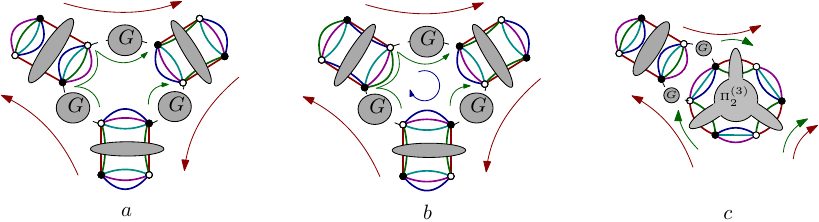} 
\captionof{figure}{Three configurations contributing to the intertwining $6$-point effective vertices.}\label{fig24}
\end{center}

 \begin{flushright}
 $\square$
 \end{flushright}

\noindent

Let us  now focus on  the leading order eight-point functions. They can be obtained as for four and six point functions from deleting a new tadpole in the boundary of an opened face running through the interior of the diagrams. Once again all the eight-point function can be classified from their power-counting and their boundary graphs. Moreover, in the flow equations, the melonic interactions become closed from six-point functions. Then, the eight-point functions are relevant only to close the pseudo-melonic interactions. As a result, only the leading order functions having a boundary graph whose one-loop contraction leads to a pseudo-melonic boundary will contribute in our flow equations. Figure \eqref{fig25} provides us an example. 

\begin{center}
$\vcenter{\hbox{\includegraphics[scale=1.1]{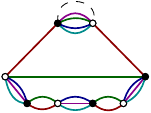} }}\,\to\,\vcenter{\hbox{\includegraphics[scale=0.9]{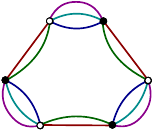} }}$
\captionof{figure}{One-loop contraction on the melonic ‘‘pole" in a eight-point boundary graph. The corresponding eight-point function has divergent degree bounded as $\omega=-2$. The created four-dipole then increases it of $2$; so that the global scaling becomes $\omega+2=0$.}\label{fig25}
\end{center}

\noindent
From these considerations, it is easy to check that there are only two relevant boundaries for leading order eight-points graphs, pictured in Figure \eqref{fig25} below. 
\begin{center}
\includegraphics[scale=1.2]{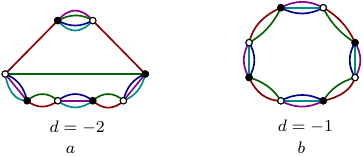} 
\captionof{figure}{Allowed boundaries for relevant eight-point functions with their respective canonical dimensions. As for six-point boundaries, type (a) intertwines between melons and pseudo-melons and has a single melonic ‘‘pole", whereas type (b) is a purely non-branching eight-point graph.}\label{fig25}
\end{center}

\noindent
In both cases, the corresponding eight-point functions are labeled with a pair of indices:
\begin{equation}
\Gamma_{s\,,\text{pseudo-melo}}^{(8)}=\sum_{j<i} \Gamma_{s\,,\text{pseudo-melo}}^{(8)\,,ij}\,,\quad \Gamma_{s\,,\text{inter}}^{(8)}=\sum_{j<i} \Gamma_{s\,,\text{inter}}^{(8)\,,ij}\,,
\end{equation}
and we define their respective zero momenta skeleton functions as:
\begin{equation}
\Gamma_{s\,,\text{inter}}^{(8)\,,ij}=:4!\Pi_{1}^{(4)}\,,\quad \Gamma_{s\,,\text{pseudo-melo}}^{(8)\,,ij}=:4!\Pi_{2}^{(4)}\,.\label{defgamma8}
\end{equation}
As for six-point functions, there are essentially two different configurations for the four arms of the skeleton:
\begin{enumerate}
\item The four arms hooked to external deleted pseudo-melonic tadpole can be hooked to the same vertex.

\item The four arms can be hooked to form effective tripods, hooked together with a common bicolored path.
\end{enumerate}

\begin{center}
\includegraphics[scale=1.3]{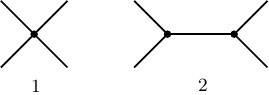} 
\captionof{figure}{Two allowed topology for configurations of the four arms of the leading order $8$-point functions. In the case (1), the four arms are hooked to the same vertex. In the case (2), the arms form effective tripods, hooked together with a bicolored path.}
\end{center}

\noindent 
Figure \eqref{fig25bis} provides some examples of trees corresponding to these two configurations for the pseudo-melonic boundary. The first one (a) correspond to the configuration (1): The four arms of the bicolored skeleton are hooked to the same vertex. The second (b) and third (c) both correspond to the configuration (2): The four arms are organized as two $3$-valent tripods and are hooked together with a common bicolored path.
\begin{center}
\includegraphics[scale=0.8]{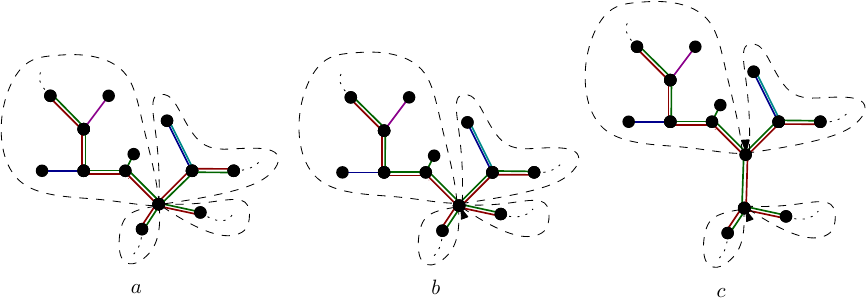} 
\captionof{figure}{Three configurations for the arms of the four skeletons. In (a), the four arms are hooked to the same vertex. In (b) two arms are hooked to the same breaking edge. In (c) the two remaining edges are hooked to another common breaking edge. }\label{fig25bis}
\end{center}
We have the following statement:
\begin{proposition}
The zero-momenta skeleton functions $\Pi_{1}^{(4)}$ and $\Pi_{2}^{(4)}$ are given in terms of essential and marginal effective couplings at scale $s$, $\lambda_{4,1}(s)$, $\lambda_{4,2}(s)$ and $\lambda_{6,1}(s)$ as:
\begin{align}
\nonumber\Pi_{1}^{(4)}=&-\frac{4!2^4 3!}{3!} \lambda_{4,2}^3\lambda_{4,1} \mathcal{A}_{3,4} -\frac{ (4!)^2}{3!}\Pi_{1,c}^{(3)} \lambda_{4,2}^2(s) \mathcal{A}_{3,3}+ \lambda_{6,1}(s)\lambda_{4,2}(s)\lambda_{4,1}(s) \mathcal{A}_{3,3}\\\nonumber
&+\frac{4!}{2}\Pi_{1,c}^{(3)}\, \frac{\lambda_{6,1}(s)-\frac{8}{3}\lambda_{4,2}^3(s)\mathcal{A}_{3,3}}{1-3!\lambda_{4,2}(s)\mathcal{A}_{3,2}+12\lambda_{4,2}^2(s)\mathcal{A}_{3,2}^2-16\lambda_{4,2}^3(s)\mathcal{A}_{3,2}^3}\\
&\quad\times\bigg(1-4\lambda_{4,2}(s)\mathcal{A}_{3,2}+4\lambda_{4,2}^2(s)\mathcal{A}_{3,2}^2 \bigg)\,,
\end{align}
\begin{align}
\nonumber&\Pi_{2}^{(4)}=-\frac{9\times 4!}{2}\lambda_{6,1}(s) \frac{\lambda_{6,1}(s)-\frac{8}{3}\lambda_{4,2}^3(s)\mathcal{A}_{3,3}}{1-3!\lambda_{4,2}(s)\mathcal{A}_{3,2}+12\lambda_{4,2}^2(s)\mathcal{A}_{3,2}^2-16\lambda_{4,2}^3(s)\mathcal{A}_{3,2}^3} \\
&\quad\times\left(1-4\lambda_{4,2}(s)\mathcal{A}_{3,2}+4\lambda_{4,2}^2(s)\mathcal{A}_{3,2}^2 \right)-2^4 3! \lambda_{4,2}^4(s) \mathcal{A}_{3,4}+\frac{4!}{2}\lambda_{6,1}(s)\lambda_{4,2}^2 \mathcal{A}_{3,3}\,.
\end{align}
\end{proposition}

\noindent
\textit{Proof.}

\noindent
Like for four and six-point functions, we have to list all leading order configurations and, for each of them, compute their respective boundary graphs and their divergent degrees. From this analysis, and taking into account effective summations like for six-points graphs, we easily check that there are only five configurations whose complete set is given in Figure \eqref{fig26}, having  a pseudo-melonic boundary graph.
\begin{center}
\includegraphics[scale=0.8]{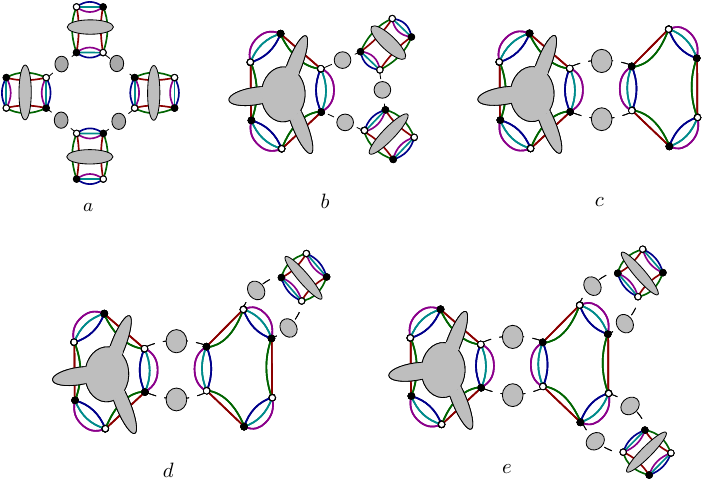} 
\captionof{figure}{The three leading order configurations for eight-point functions having non-branching pseudo-melonic boundaries.}\label{fig26}
\end{center}
The two first  configurations (a) and (b) can be easily computed from renormalization conditions \eqref{rencond1}, \eqref{rencond2} and \eqref{rencond3}, and symmetry factors as well as the relative signs may be fixed from comparison with lowest orders of the perturbative expansion. The three remaining configurations (c), (d) and (e) require to be carefully checked on the counting procedure. We get explicitly, using Lemma \eqref{lemmaphi62}:
\begin{align}
\nonumber &\Gamma_{s\,,\text{pseudo-melo}}^{8\,,ij}=-\frac{9\times(4!)^2}{2}\lambda_{6,1}(s) \frac{\lambda_{6,1}(s)-\frac{8}{3}\lambda_{4,2}^3(s)\mathcal{A}_{3,3}}{1-3!\lambda_{4,2}(s)\mathcal{A}_{3,2}+12\lambda_{4,2}^2(s)\mathcal{A}_{3,2}^2-16\lambda_{4,2}^3(s)\mathcal{A}_{3,2}^3} \\
&\quad\left(1-4\lambda_{4,2}(s)\mathcal{A}_{3,2}+4\lambda_{4,2}^2(s)\mathcal{A}_{3,2}^2 \right)-4!2^4 3! \lambda_{4,2}^4(s) \mathcal{A}_{3,4}+\frac{(4!)^2}{2}\lambda_{6,1}(s)\lambda_{4,2}^2 \mathcal{A}_{3,3}\,.
\end{align}
Then  the expression of  $\Pi_2^{(4)}$ can be derived easily. \\

\noindent
Moving on to the configurations having intertwining boundaries, it is easy to check that the only ones allowed such  boundaries are pictured in Figure \eqref{fig27} below; and the corresponding zero momenta eight-point function writes as:
\begin{align}
\nonumber\Gamma^{(8)\,ij}_{s\,,\text{inter}}=&-\frac{(4!)^22^4 3!
}{3!} \lambda_{4,2}^3\lambda_{4,1} \mathcal{A}_{3,4} -\frac{3(4!)^22^3}{3!}\Pi_{1,c}^{(3)} \lambda_{4,2}^2(s) \mathcal{A}_{3,3}+4! \lambda_{6,1}(s)\lambda_{4,2}(s)\lambda_{4,1}(s) \mathcal{A}_{3,3}\\\nonumber
&+\frac{(4!)^2}{2}\Pi_{1,c}^{(3)}\, \frac{\lambda_{6,1}(s)-\frac{8}{3}\lambda_{4,2}^3(s)\mathcal{A}_{3,3}}{1-3!\lambda_{4,2}(s)\mathcal{A}_{3,2}+12\lambda_{4,2}^2(s)\mathcal{A}_{3,2}^2-16\lambda_{4,2}^3(s)\mathcal{A}_{3,2}^3}\\
&\quad\times\bigg(1-4\lambda_{4,2}(s)\mathcal{A}_{3,2}+4\lambda_{4,2}^2(s)\mathcal{A}_{3,2}^2 \bigg)\,,
\end{align}
where the $1/3!$ in front of the effective skeleton function $\Pi_{1,c}^{(3)}$ comes from its definition \eqref{effskel62} -- it suppresses the counting of the external contractions on black (or white) nodes. 

\begin{center}
\includegraphics[scale=0.9]{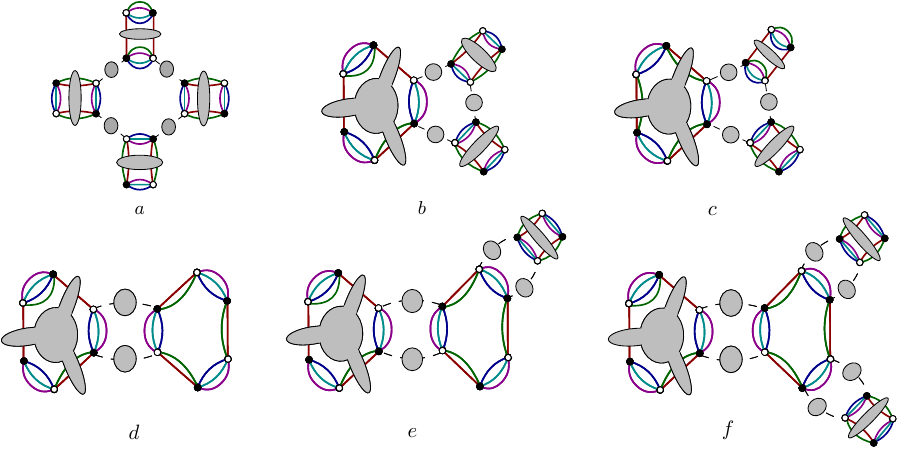} 
\captionof{figure}{The four leading order configurations for eight-point functions having intertwining boundaries.}\label{fig27}
\end{center}

\begin{flushright}
$\square$
\end{flushright}

\section{Flow equations and their fixed points}\label{sec5}

The leading effective  vertex being obtained, let us now focus on  the building of the leading order renormalization group flow corresponding to the initial conditions \eqref{initialaction}, from effective interactions generated from essential and marginal couplings at scale $s$. Once again, we limit our investigation on the UV regime $\Lambda\gg e^s\gg 1$. 

\subsection{Flow for local couplings}

From Definition \eqref{definitionTomTom}, local interactions correspond to pure connected tensorial invariants. Therefore, we have defined the effective couplings at scale $s$ as the zero-momenta $n$-point functions corresponding to a given boundary graph. Then, we can compute the successive functional derivatives of the Wetterich equation \eqref{Wett}, and identify in both sides the terms corresponding, at leading order to the same boundary graph when the external momenta are setting to zero. \\

\noindent
The effective couplings at scale $s$ have been defined from equations \eqref{rencond12}, \eqref{rencond22} and \eqref{rencond3}. To complete the Definitions  \eqref{defrenconds} and \eqref{defrenconds2}, we have to define the \textit{dynamical mass} $m^2(s)$; which is a relevant coupling with canonical dimension equal to $2$:
\begin{definition}\label{defrenconds3}
The effective mass at scale $s$ is defined as:
\begin{equation}
m^2(s):={\Gamma}_s^{(2)}(\vec p=\vec{0}\,)\,.
\end{equation}
\end{definition}
We have the following proposition:
\begin{proposition}\label{flow1}
In the deep UV sector ($\Lambda\gg e^s\gg 1$) and in the symmetric phase, the exact flow equations for essential and marginal coupling in the sector mixing melons and non-branching pseudo-melons are given by:
\begin{align}
&\dot{m}^2=-10\,\lambda_{4,1} \mathcal{I}_{4,2}-20\lambda_{4,2}\mathcal{I}_{3,2}\,,\\
&\dot{\lambda}_{4,1}=-\left(2\Pi_{1,a}^{(3)}+\frac{2}{3}\Pi_{1,b}^{(3)}\right)\mathcal{I}_{4,2}+4\lambda_{4,1}^2\mathcal{I}_{4,3}+16\lambda_{4,1}\lambda_{4,2}\mathcal{I}_{3,3} \,,\\
&\dot{\lambda}_{4,2}=-\left(3\lambda_{6,1}+\frac{1}{6}\Pi_{1,c}^{(3)}\right)\mathcal{I}_{3,2}+4\lambda_{4,2}^2\mathcal{I}_{3,3}\,,\\
&\dot{\lambda}_{6,1}=-\left(\frac{1}{6}\Pi_{2}^{(4)}+\frac{1}{4!}\Pi_1^{(4)}\right)\mathcal{I}_{3,2}+12\lambda_{4,2}\lambda_{6,1}\mathcal{I}_{3,3}-8\lambda_{4,2}^3\mathcal{I}_{3,4}\,.
\end{align}
where the dot means derivative with respect to $s$, and:
\begin{equation}
\mathcal{I}_{m,n}:= \sum_{\vec{p}\in\mathbb{Z}^{m}} \dot{r}_s(\vec{p}\,)\,G^n(\vec{p}\,)\,.
\end{equation}
\end{proposition}

\noindent
\textit{Proof.} 
Deriving the exact flow equation \eqref{Wett} with respect to $M$ and $\bar M$, we deduce an equation describing the flow of the two-point function $\Gamma_s^{(2)}$ in terms of $\Gamma^{(4)}_s$ only, in the symmetric phase:
\begin{equation}
\dot{\Gamma}_s^{(2)}(\vec p\,)=-\sum_{\vec q}\Gamma^{(4)}_{s}(\vec p,\vec q;\vec p,\vec q\,)\,\frac{\dot r_s(\vec q\,)}{[\Gamma_s^{(2)}(\vec q\,)+r_s(\vec q\,)]^2}\,,\label{Gamma2}
\end{equation}
 where $\partial_s A=:\dot A$. First, $\Gamma^{(4)}_{s,\vec p\vec q;\vec p,\vec q}$ split into melonic and pseudo-melonic contributions. We just considered  only the boundary graphs of the effective functions to simplify the notations and then  the previous equation  is written graphically as:
\begin{equation}
\dot{\Gamma}_s^{(2)}(\vec p\,)=-\sum_{i=1}^d\left(\vcenter{\hbox{\includegraphics[scale=0.7]{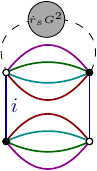} }}+\vcenter{\hbox{\includegraphics[scale=0.7]{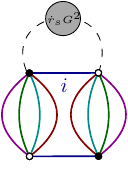} }}\right)-\sum_{i<j}^d\left(\vcenter{\hbox{\includegraphics[scale=0.7]{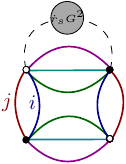} }}+\vcenter{\hbox{\includegraphics[scale=0.7]{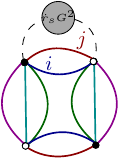} }}\right)\,.
\end{equation}
In the deep UV limit, the second and fourth contributions may be discarded. Indeed, with respect to the first one, the second diagram lacks three faces, and its proper power counting is $\omega=-1$. In the same way, the fourth diagram scales as $\omega=1$, while the first and third diagrams scale as $\omega=2$. The relevant contributions are then:
\begin{equation}
\dot{\Gamma}_s^{(2)}(\vec p\,)=-2\sum_{i=1}^d\left(\vcenter{\hbox{\includegraphics[scale=0.7]{Wett3.pdf} }}\right)-2\sum_{i<j}^d\left(\vcenter{\hbox{\includegraphics[scale=0.7]{Wett1.pdf} }}\right)\,.
\end{equation}
The dynamic mass $m^2(s)$ has been defined in Definition \eqref{defrenconds3}. Then, setting $\vec{p}=\vec{0}$ in the previous equation, we get:
\begin{equation}
\dot{m}^2=-2d\,\lambda_{4,1} \sum_{\vec{p}\in\mathbb{Z}^{d-1}} \dot{r}_s(\vec{p}\,)\,G^2(\vec{p}\,)-2\frac{d(d-1)}{2}\lambda_{4,2} \sum_{\vec{p}\in\mathbb{Z}^{d-2}} \dot{r}_s(\vec{p}\,)\,G^2(\vec{p}\,)\,.
\end{equation}

\noindent
In the same way, for  the four-point functions we have to derive once again one time with respect to $M$ and $\bar{M}$; and keep only the even $N$-point functions, having the same number of derivatives with respect to the two mean fields.   We get:
\begin{align}
\nonumber\dot{\Gamma}^{(4)}_{s}=-\sum_{\vec p}\dot r_s(\vec p\,) G^2_s(\vec p\,)&\Big[\Gamma^{(6)}_{s}(\vec p,\vec 0,\vec 0,\vec p,\vec 0,\vec 0\,)-2\sum_{\vec p\,'}\Gamma^{(4)}_{s}(\vec p,\vec 0,\vec p\,',\vec 0\,)G_s(\vec p\,')\Gamma^{(4)}_{s}(\vec p\,',\vec 0,\vec p,\vec 0\,)\Big]\\
&+2\sum_{\vec p}\dot r_s(\vec p\,) G^3_s(\vec p\,)[\Gamma^{(4)}_{s}(\vec p,\vec 0,\vec p,\vec 0\,)]^2\,. \label{flowfour}
\end{align}
Graphically, we get the contributions (we left the external momenta to simplify the notations):
\begin{align}
\nonumber\dot{\Gamma}_{s}^{(4)}=&-4\sum_{i=1}^d\left(3\times\vcenter{\hbox{\includegraphics[scale=0.6]{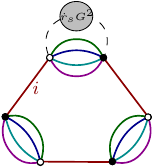} }}\cdots\right)-4\sum_{i<j}^d\left(\,\vcenter{\hbox{\includegraphics[scale=0.65]{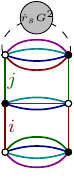} }}+\,\vcenter{\hbox{\includegraphics[scale=0.65]{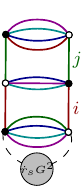} }}+\cdots\right)\\\nonumber
&-4\sum_{i<j}^d\left(3\times\vcenter{\hbox{\includegraphics[scale=0.6]{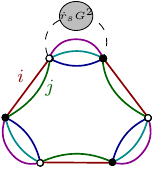} }}+\vcenter{\hbox{\includegraphics[scale=0.6]{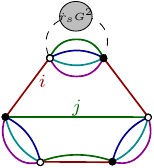} }}+2\times\vcenter{\hbox{\includegraphics[scale=0.6]{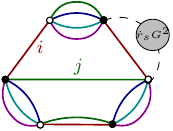} }}+\cdots\right)\\\nonumber
&+\sum_{i,j}^d\left(4\times\vcenter{\hbox{\includegraphics[scale=0.7]{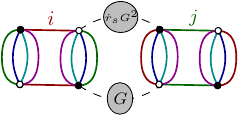} }}+\cdots\right)+\sum_{i,j,k}^d\left(4\times \vcenter{\hbox{\includegraphics[scale=0.7]{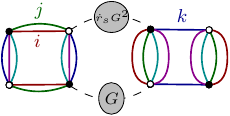} }}+\cdots\right)\\
&+\sum_{i<j;k<l}^d\left(4\times\vcenter{\hbox{\includegraphics[scale=0.7]{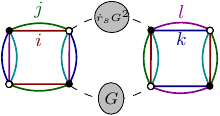} }}+\cdots\right)\,,\label{Wett4}
\end{align}
where once again we only drawn the boundary graphs of  effective functions. The factors involved in this relation  count  the number of contractions leading to the corresponding  diagram, including permutations of the external points. This is the origin of the factor $4=(2!)^2$ in front of the six-point contractions. Moreover, the same factor $4$  in the front of the four-point contributions  comes from the derivation itself : A first factor $2$ comes from the derivative of $G^2$ in \eqref{Gamma2}; and a second factor $2$ comes from a discarded term in the first derivation, involving three-point functions. The absence of the factor coming from permutation of external points arises because of the definition of the corresponding skeleton functions.\\

\noindent
However, among these contractions, some of them contribute to the melonic sector, and some others contribute to the pseudo-melonic one; and we have to classify them with respect to their boundary graphs. For instance, in the first line, the two first diagrams have melonic boundaries, while the two first diagrams in the second line have pseudo-melonic boundaries. The diagrams in the third and fourth lines require to be careful analyzed because their boundary graphs depend on the selected indices $i,j,k,l$. For instance, for $i=j$, the first diagram in the third line has pseudo-melonic boundary, while it has disconnected boundary for $i\neq j$:
\begin{equation}
\vcenter{\hbox{\includegraphics[scale=1]{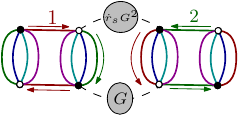} }} \sim \vcenter{\hbox{\includegraphics[scale=1]{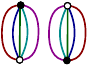} }} \,.
\end{equation}
This diagram will not be taken into account, for melonic and pseudo-melonic flows. Moreover, it is completely discarded in the UV limit, its proper power counting begins $\omega=-2$; while the one of the case $i=j$ is $\omega=0$, as expected for a relevant contribution to the flow of the marginal coupling $\lambda_{4,1}$\footnote{Disconnected contributions can be taken into account; and mixing sectors involving their contributions have be investigated in \cite{BenGeloun:2018ekd}, from a truncation approach.}. In the same way, the second contribution in the third line is melonic for $k=j$ or $k=i$, and completely disconnected for $i\neq j\neq k$. Therefore, taking into account only the relevant contributions for each sectors we get:
\begin{align}
\nonumber\dot{\Gamma}_{s\,,\text{melo}}^{(4)}=-4&\sum_{i=1}^d\left(3\times\vcenter{\hbox{\includegraphics[scale=0.6]{Wett41.pdf} }}\,\right)-4\sum_{i<j}^d\left(\,\vcenter{\hbox{\includegraphics[scale=0.65]{Wett4pioupiou.pdf} }} +\,\vcenter{\hbox{\includegraphics[scale=0.65]{Wett4pioupiou2.pdf} }}\,\right)\\
&+\sum_{i}^d\left(4\times\vcenter{\hbox{\includegraphics[scale=0.7]{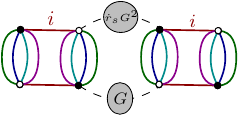} }}\right)+\sum_{i}^d\left(4\times\vcenter{\hbox{\includegraphics[scale=0.7]{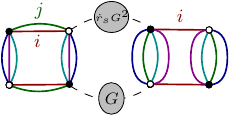} }}\right)\,,
\end{align}
\begin{align}
\nonumber\dot{\Gamma}_{s\,,\text{pseudo-melo}}^{(4)}&=-4\sum_{i<j}^d\left(3\times\vcenter{\hbox{\includegraphics[scale=0.6]{Wett42.pdf} }}+\vcenter{\hbox{\includegraphics[scale=0.6]{Wett43.pdf} }}\right)+\sum_{i<j}^d\left(4\times\vcenter{\hbox{\includegraphics[scale=0.7]{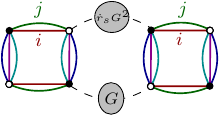} }}\right)\,.
\end{align}

\noindent
Finally, using renormalization conditions \eqref{rencond12}, \eqref{rencond22} and \eqref{rencond3} to identify the relevant couplings on both sides as ${\Gamma}_{s\,,\text{melo}}^{(4)}=4\lambda_{4,1}$ and ${\Gamma}_{s\,,\text{pseudo-melo}}^{(4)}=4\lambda_{4,2}$, we get:
\begin{align}
\nonumber\dot{\lambda}_{4,1}=-&\left(2\Pi_{1,a}^{(3)}+\frac{2}{3}\Pi_{1,b}^{(3)}\right)\sum_{\vec{p}\in\mathbb{Z}^{d-1}} \dot{r}_s(\vec{p}\,)\,G^2(\vec{p}\,)+4\lambda_{4,1}^2\sum_{\vec{p}\in\mathbb{Z}^{d-1}} \dot{r}_s(\vec{p}\,)\,G^3(\vec{p}\,)\\
&+16\lambda_{4,1}\lambda_{4,2} \sum_{\vec{p}\in\mathbb{Z}^{d-2}} \dot{r}_s(\vec{p}\,)\,G^3(\vec{p}\,)
\end{align}
and :
\begin{equation}
\dot{\lambda}_{4,2}=-\left(3\lambda_{6,1}+\frac{1}{6}\Pi_{1,c}^{(3)}\right)\sum_{\vec{p}\in\mathbb{Z}^{d-2}} \dot{r}_s(\vec{p}\,)\,G^2(\vec{p}\,)+4\lambda_{4,2}^2\sum_{\vec{p}\in\mathbb{Z}^{d-2}} \dot{r}_s(\vec{p}\,)\,G^3(\vec{p}\,)\,,
\end{equation}

\noindent
Finally, following the same strategy, we get for the six-point function:
\begin{align}
\dot{\Gamma}^{(6)}_{s}&=-\sum_{\vec p}\dot{r}_s(\vec p\,)\Big[\Gamma^{(8)}_{s}(\vec p,\vec 0,\vec 0,\vec 0,\vec p,\vec0,\vec 0,\vec 0)G^2(\vec p\,)-18\Gamma^{(4)}_{s}(\vec p,\vec 0,\vec p,\vec0)  \Gamma^{(6)}_{s}(\vec p,\vec 0,\vec 0,\vec p,\vec 0,\vec 0)G^3(\vec p)\cr\nonumber
&\qquad \qquad+36(\Gamma^{(4)}_{s}(\vec p,\vec 0,\vec p,\vec0)\,)^3G^4(\vec p)\Big]\\\nonumber
&=-(3!)^2\sum_{i<j}^d\left(4\times\vcenter{\hbox{\includegraphics[scale=0.8]{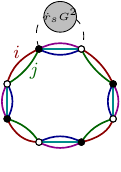} }}+\,4\times \vcenter{\hbox{\includegraphics[scale=0.8]{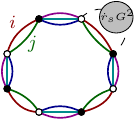} }}+\cdots\right)-(3!)^2\sum_{i<j}^d\left(\vcenter{\hbox{\includegraphics[scale=0.8]{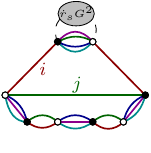} }}+\cdots\right)\\\nonumber
&\qquad+18\times 4\times \sum_{j<i}^d\left(3\times\vcenter{\hbox{\includegraphics[scale=0.75]{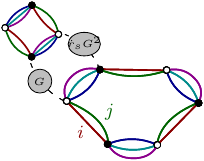} }}\cdots\right)-36\sum_{i<j}^d\left(\vcenter{\hbox{\includegraphics[scale=0.75]{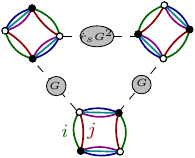} }}+\cdots\right)\,,
\end{align}
where the additional factor $4$ in front of the first term in line four comes from the counting of the permutation of the external points hooked to the six-point effective vertex. Then, from the renormalization condition \eqref{rencond3}, we have $\Gamma^{(6)}_{s\,,\text{pseudo-melo}}=(3!)^2\lambda_{6,1}$, and keeping only the relevant contributions in the UV having pseudo-melonic boundaries, we get finally:

\begin{equation}
\dot{\lambda}_{6,1}=-\left(4\times\vcenter{\hbox{\includegraphics[scale=0.8]{Wett81.pdf} }}+\vcenter{\hbox{\includegraphics[scale=0.8]{Wett83.pdf} }}\right)+216\times\left(\vcenter{\hbox{\includegraphics[scale=0.75]{Wett84.pdf} }}\right)-36\times\left(\vcenter{\hbox{\includegraphics[scale=0.75]{Wett86.pdf} }}\right)\,.
\end{equation}
equivalently we get
\begin{align}
\nonumber\dot{\lambda}_{6,1}=&-\left(\frac{1}{6}\Pi_{2}^{(4)}+\frac{1}{4!}\Pi_1^{(4)}\right)\sum_{\vec{p}\in\mathbb{Z}^{d-2}} \dot{r}_s(\vec{p}\,)\,G^2(\vec{p}\,)+12\lambda_{4,2}\lambda_{6,1}\sum_{\vec{p}\in\mathbb{Z}^{d-2}} \dot{r}_s(\vec{p}\,)\,G^3(\vec{p}\,)\\
&-8\lambda_{4,2}^3 \sum_{\vec{p}\in\mathbb{Z}^{d-2}} \dot{r}_s(\vec{p}\,)\,G^4(\vec{p}\,)\,,
\end{align}
which ends  the proof.
\begin{flushright}
$\square$
\end{flushright}

\subsection{Anomalous dimension}

Until now we have no explicitly introduced counter-terms. However, fields and couplings in the original action \eqref{initialaction} have to be completed with some counter-terms so that the quantum corrections are finite. For all couplings, including mass, these counter-terms have to be understood included in the definition of the bare couplings themselves; the renormalization conditions in the deep IR fixing the renormalization prescription. However, we need for our discussion to appear explicitly the wave-function counter-terms, which we call $Z_{-\infty}$ following the notations used in \cite{Lahoche:2018vun} -- the subscript $-\infty$ referring to the deep IR limit $s\to -\infty$, in which the asymptotic renormalization prescription are fixed. The bare kinetic kernel $C^{-1}(\vec{p}\,)$ is then replaced by:
\begin{equation}
C^{-1}(\vec{p}\,)=(Z_{-\infty} \vec{p}\,^2+m^2)\Theta_a^{-1}(\Lambda^2-\vec{p}\,^2)\,,\label{Cregularized}
\end{equation}
where $\Theta_a^{-1}(\Lambda^2-\vec{p}\,^2)$ is a smooth invertible distribution depending on a real parameter $a$, such that $\lim\limits_{a \to 0}\Theta_a(x)=\Theta(x)$. As example we write:
\begin{equation}
\Theta_a(x):=\frac{1}{a\sqrt{\pi}}\int_{-\infty}^x \,e^{-y^2/a^2}dy  \,.
\end{equation}
Moreover, because $\Lambda>e^s$ the regulator is unaffected from the modification $r_s\to r_s/\Theta_a$ in the limit $a\to 0$; so that the regularized propagator $C_s$ can be written as
\begin{equation}
C_s^{-1}=(Z_{-\infty} \vec{p}\,^2+m^2+r_s(\vec{p}\,))\Theta_a^{-1}(\Lambda^2-\vec{p}\,^2)\,.\label{Cregularized2}
\end{equation}
Our renormalization prescriptions, defining the asymptotic mass $m_r^2$ in the same time, is  such that (see  \cite{Lahoche:2018vun}):
\begin{equation}
\Gamma_{s\to-\infty}(\vec{p}\,) =m^2_r+\vec{p}\,^2+\mathcal{O}(\vec{p}\,^2)\,.
\end{equation}
Note that this definition matches only if the trajectory does not cross a non-Gaussian fixed point.  A more general definition, including this case, fixes the renormalization conditions to a finite scale $s_0$ so that $e^{s_0}\ll e^{s}$. For arbitrary $s$, we conventionally define the wave function renormalization and the anomalous dimension such that :
\begin{definition} 
\textbf{Anomalous dimension.}
The wave function renormalization $Z(s)$ and the anomalous dimension $\eta(s)$ -- both depending on $s$ are defined as:
\begin{equation}
Z(s):=\frac{\partial }{\partial p_1^2}\Gamma^{(2)}_s(\vec{p}\,)\bigg\vert_{\vec{p}=\vec{0}}\,,\quad \eta(s):=\frac{\dot{Z}}{Z}\,. \label{anomalous}
\end{equation}
\end{definition}
The flow equations deduced in the previous section are ‘‘exact''  i.e. up to the limit of our approximation scheme. However, the anomalous dimension, formally given from Equation \eqref{anomalous} remains unknown. To exploit our flow equations, we then have to complete our approximation scheme. We have expressed all the LO effective vertices in terms of essential and marginal couplings, as well as effective two-point functions $\Gamma^{(2)}_s$. Formally, these functions are fixed from closed equation \eqref{closed1}; however, solving exactly this equation remains an open challenge in tensor field theories \cite{Samary:2014tja}-\cite{Samary:2014oya}. Following the method explained in \cite{Lahoche:2018vun}, in  order to extract available equations for renormalization group flow, especially in view of a numerical analysis, we have to fix the form of $\Gamma^{(2)}_s$; and we adopt the following  definition:
\begin{definition}\label{defZ}
In the symmetric phase, and in the range of momenta contributing significantly in the domain defined by the distribution $\dot{r}_s$; the effective two-point function $\Gamma^{(2)}_s$ is assumed to be truncated around the two first terms in derivative expansion
\begin{equation}
\Gamma^{(2)}_s(\vec{p}\,):=Z(s)\vec{p}\,^2+m^2(s)\,,
\end{equation}
so that $Z(s)$ matches with the definition \eqref{anomalous}. 
\end{definition}
Note that this definition is compatible with usual truncations in the symmetric phase, that is with derivative and mean field expansion considered in the literature \cite{Blaizot:2006vr}-\cite{Defenu:2014jfa}. Moreover, in non-symmetric phase, a dependence of $Z(s)$ on the mean fields $M$ and $\bar{M}$ is expected.  Finally, the domain of  the momentas defined by the distribution $\dot{r}_s$ is important.  It corresponds to the relevant integration domain involved in the flow equations \eqref{Wett}; and may be viewed in a sense as an approximation ‘‘slice by slice'' along the flow line. However, it cannot be used globally in the full range of momenta, especially in the deep UV, for $\vec{p}\,^2\sim \Lambda^2$, without dangerous contradictions with Ward-identity, as we will briefly discuss at the end of this section. \\

\noindent
In addition to this definition, we recall the usual definition for \textit{renormalized and dimensionless} couplings:
\begin{definition} \label{rencouplings}
In the deep UV, the dimensionless and renormalized couplings, $\bar{m}^2$, $\bar{\lambda}_{4,1}$, $\bar{\lambda}_{4,2}$ and $\bar{\lambda}_{6,1}$ are defined as:
\begin{equation}
m^2=Ze^{2s}\bar{m}^2\,,\quad {\lambda}_{4,1}=Z^2\bar{\lambda}_{4,1}\,,\quad {\lambda}_{4,2}=Z^2e^s\bar{\lambda}_{4,2}\,,\quad {\lambda}_{6,1}=Z^3\bar{\lambda}_{6,1}\,.
\end{equation}
Moreover, we denote respectively by $\beta_m$, $\beta_{4,1}$, $\beta_{4,2}$ and $\beta_{6,1}$ their  derivative with respect to $s$. 
\end{definition}

\noindent
Now, let us give the set of  WT-identities, which will help to extract the anomalous dimension $\eta$.

\begin{proposition} 
\textbf{(UV-regularized Ward-Takahashi identity)}\\
\noindent
Let $Z_s(J,\bar{J})$ be  the one-parameter generating functional for the theory with microscopic action \eqref{initialaction}:
\begin{equation}
Z_s(J,\bar{J}):=\int d\mu_{C_s}[T,\bar{T}] e^{-S_{\text{int}}(T,\bar{T})+\bar{J}T+\bar{T}J}\,,\quad d\mu_{C_s}[T,\bar{T}] :=dTd\bar{T} \,e^{-\bar{T}C^{-1}_sT}\,.\label{defZmu}
\end{equation}
The free energy ${W}_s(J,\bar{J}):=\ln Z(J,\bar{J})$ satisfies the functional differential equation:
\begin{equation}
\sum_{\vec{p}_\bot,\vec{p}\,^\prime_\bot} \prod_{j\neq1} \delta_{p_jp_j^\prime}\bigg\{ \left(C^{-1}_s(\vec{p}\,)-C^{-1}_s(\vec{p}\,^\prime)\right)\left[\frac{\partial^2 W_s}{\partial \bar{J}_{\vec{p}\,^\prime}\,\partial {J}_{\vec{p}}}+\bar{M}_{\vec{p}}M_{\vec{p}\,^\prime}\right]-\bar{J} _{\vec{p}}\,M_{\vec{p}\,^\prime}+{J} _{\vec{p}\,^\prime}\bar{M}_{\vec{p}}\bigg\}=0\,,\label{Ward}
\end{equation}
where the mean fields $M$ and $\bar{M}$ have been defined in \eqref{means}, and $\vec{p}_\bot:=(0,p_2,\cdots,p_d)\in\mathbb{Z}^{d-1}$. 
\end{proposition}

\noindent
\textit{Proof (Sketched).}
The Ward--Takahashi identity was extensively discussed in the literature \cite{Lahoche:2018vun}. We provide only the sketch of this proof, to take into account the little modifications coming from the sharp UV-regulation; and the new boundary graphs coming from pseudo-melonic sector.  The given proofs follow the steps detailed in \cite{Lahoche:2018vun}. \\

\noindent
Let us consider the unitary transformations $\textbf{U}\in\mathcal{U}^{\times d}$ acting independently over each component of the tensors $T$ and $\bar{T}$. $\textbf{U}$ is a $d$-dimensional vector $\textbf{U}=(U_1,U_2,\cdots,U_d)$ whose components $U_i$ are unitary matrices acting on the indices of color $i$. The action of $\textbf{U}$ on the two tensors is defined as (we sum over repeated indices):
\begin{align}
\textbf{U}[T]_{p_1,p_2,\cdots,p_d}&:=[U_1]_{p_1q_1}[U_2]_{p_2q_2}\cdots[U_d]_{p_dq_d}T_{q_1,q_2,\cdots,q_d}\\
\textbf{U}[\bar{T}]_{p_1,p_2,\cdots,p_d}&:=[U_1^*]_{p_1q_1}[U_2^*]_{p_2q_2}\cdots[U_d^*]_{p_dq_d}\bar{T}_{q_1,q_2,\cdots,q_d}\,,\label{defvar}
\end{align}
where the star exponent $*$ means the complex conjugation. Obviously, $\sum_{\vec{p}}\bar{T}_{\vec{p}}T_{\vec{p}}$ and any higher valence tensorial interactions are invariant under any such transformations. Then:
\begin{equation}
\textbf{U}[S_{\text{int}}]=S_{\text{int}}\,.\label{invariance}
\end{equation}
In contrast, it is not the case for the kinetic and source terms, due to the nontrivial propagator and sources $J$ and $\bar{J}$, which  breaks   the unitary invariance. However, the functional integral defining the generating functional $Z_s(J,\bar{J})$ has to be insensitive on the bad transformation of the kinetic term because of the formal translation invariance of the Lebesgue integration measure. Then, it has to be invariant under any unitary transformation; and as a direct consequence, the two-point function $\langle\bar{T}_{\vec{p}}T_{\vec{q}}\rangle$ and any higher functions transform like a  trivial representation of $\textbf{U}\otimes \textbf{U}^*$. \\

\noindent
We can translate this conclusion as follows: In the vicinity of the unity we can write $U=\mathbb{I}+i\epsilon$, where $\epsilon=\epsilon^\dagger$ is a Hermitian matrix and $\mathbb{I}$ the identity matrix. Then, at the first order
\begin{equation}
\textbf{U}=\textbf{I}+\sum_i\vec{\epsilon}_i\,,
\end{equation}
where $\textbf{I}:=\mathbb{I}^{\otimes d}$ and $\vec{\epsilon}_i=\mathbb{I}^{\otimes (i-1)}\otimes\epsilon_i\otimes\mathbb{I}^{d-i+1}$. In this infinitesimal prescription, the invariance of the generating functional simply means $\vec{\epsilon}_i [Z_s(J,\bar{J})]=0$. Explaining each terms at the first order in ${\epsilon}_i$, we then get:
\begin{equation}
\int \vec{\epsilon}_i[d\mu_{C_s}] e^{-S_{\text{int}}+\bar{J}T+\bar{T}J} +\int d\mu_{C_s} \left\{\vec{\epsilon}_i[S_{\text{int}}]+\bar{J}\,\vec{\epsilon}_i[T]+\vec{\epsilon}_i[\bar{T}]J\right\}e^{-S_{\text{int}}+\bar{J}T+\bar{T}J}=0\,.\label{variation}
\end{equation}
Obviously, this equation has to be true for all $i$. From equation \eqref{invariance}, it follows that $\vec{\epsilon}_i[S_{\text{int}}]=0$. Moreover, from definition \eqref{defvar}, it is easy to deduce the infinitesimal variations for single tensor fields:
\begin{equation}
\vec{\epsilon}_i[T]_{\vec{p}}=\sum_{\vec{p}\,^\prime}{\epsilon}_{i\,p_ip_i^\prime}\,\prod_{j\neq i} \delta_{p_jp_j^\prime}\,T_{\vec{p}\,^\prime}\,.\label{vartensor}
\end{equation}
Finally, from definition \eqref{defZmu}; the variation of the Gaussian measure $d\mu_{C_s}$ can be easily computed. In matrix notation:
\begin{equation}
 \vec{\epsilon}_i[d\mu_{C_s}] =-( \vec{\epsilon}_i[\bar{T}]C^{-1}_s T+\bar{T}C^{-1}_s  \vec{\epsilon}_i[{T}])d\mu_{C_s}\,.
\end{equation}
Because of the  Hermiticity of the matrix $\epsilon_i$, the two terms in bracket have opposite relative signs. Then  consider the equation  \eqref{variation}, and let us   use  the explicit variation \eqref{vartensor} and finally  rewriting each tensor field $T$ and $\bar{T}$ respectively as $\partial/\partial \bar{J}$ and $\partial/\partial J$, we deduce the formula \eqref{Ward} setting $i=1$. 
\begin{flushright}
$\square$
\end{flushright}
In practice, in the symmetric phase, the Ward identity allows to link up $\Gamma_s^{(n+2)}$ to the $\Gamma_s^{(n)}$ and their derivative. In this paper, we only keep the two first  relations, involving only essential and marginal sectors  and write the following  corollary:

\begin{corollary}
\textbf{First and second LO zero-momenta WT-identities}
In the symmetric phase, the zero momenta four and six-point functions satisfy:
\begin{align}
&Z_{-\infty} \left(\frac{1}{2}\Pi_1^{(2)}\mathcal{L}_1+2\Pi_2^{(2)}\mathcal{L}_2\right)=-\frac{\partial}{\partial p_1^2}\,\left[\Gamma_s^{(2)}(\vec{p}\,)- C^{-1}(\vec{p}\,) \right]\bigg\vert_{\vec{p}=\vec{0}}\,, \label{Wardprime}\\
&Z_{-\infty} \left(\mathcal{L}_1\left(\Pi_{1,a}^{(3)}+\frac{1}{3}\Pi_{1,b}^{(3)}\right)-8\lambda_{4,1}^2\,\mathcal{U}_1-32\lambda_{4,1}\lambda_{4,2}\mathcal{U}_2\right)=-\frac{\partial}{\partial p_1^2} \, \Pi^{(2)}_{1}(p_1,0)\bigg\vert_{p_1=0}\,,\label{Wardsecond1}\\
&Z_{-\infty} \left(\mathcal{L}_2\left(6\lambda_{6,1}+\frac{1}{3}\Pi_{1,c}^{(3)}\right)-8\lambda_{4,2}^2\,\mathcal{U}_2\right)=-\frac{\partial}{\partial p_1^2} \, \Pi^{(2)}_{2}(p_1,0)\bigg\vert_{p_1=0}\,,\label{Wardsecond2}
\end{align}
where:
\begin{equation}
Z_{-\infty}\mathcal{L}_i:=\sum_{\vec{p}_\bot\in\mathbb{Z}^{d-i}} \frac{\partial C^{-1}_s(\vec{p}_\bot)}{\partial p_1^2}\,G^{2}(\vec{p}_\bot)\,,\quad Z_{-\infty}\mathcal{U}_i:=\sum_{\vec{p}_\bot\in\mathbb{Z}^{d-i}} \frac{\partial C^{-1}_s(\vec{p}_\bot)}{\partial p_1^2}\,G^{3}(\vec{p}_\bot)\,,
\end{equation}
\end{corollary}
\textit{Proof (Sketched).}
Assuming we are in the symmetric phase and taking the derivative of the equation \eqref{Ward} with respect to $M_{\vec{q}\,^\prime}$ and $\bar{M}_{\vec{q}}$; and vanishing the sources $J=\bar{J}=0$, we get:
\begin{align}
\nonumber\sum_{\vec{p}_\bot, \vec{p}_\bot\,^{\prime}} \prod_{j\neq 1}  \delta_{p_jp_j^\prime}  \bigg[\big[C_s(\vec{p}\,^{2})-C_s(\vec{p}\,^{\prime\,{2}})\big]&\bigg[\frac{\partial^2 G_{\vec{p},\vec{p}\,^\prime}}{\partial M_{\vec{q}\,^\prime}\,\partial \bar{M}_{\vec{q}}}+\delta_{\vec{p}\vec{q}}\,\delta_{\vec{p}\,^\prime,\vec{q}\,^\prime}\bigg]-\Gamma^{(2)}_{s\,,\vec{q}\vec{p}}\,\delta_{\vec{q}\,^\prime\vec{p}\,^\prime}+\Gamma^{(2)}_{s\,,\vec{q}\,^\prime\vec{p}\,^\prime}\delta_{\vec{p}\vec{q}}\\
&-r_s(\vec{p}\,^2)\delta_{\vec{q}\vec{p}}\,\delta_{\vec{q}\,^\prime\vec{p}\,^\prime}+r_s(\vec{p}^{\,\prime\,2})\delta_{\vec{q}\,^\prime\vec{p}\,^\prime}\delta_{\vec{p}\vec{q}}\bigg] =0\,,\label{Ward2}
\end{align}
where we used the fact that following \eqref{Legendre} the inverse of the two-point function $G=\partial^2 W_s/\partial J\partial \bar{J}$ is $\Gamma^{(2)}_s+r_s$, implying : $\partial \Gamma_s/\partial M=\bar{J}-r_s \bar{M}$.  Then  it is easy to check that:
\begin{equation}
\frac{\partial^2 G_{\vec{p},\vec{p}\,^\prime}}{\partial M_{\vec{q}\,^\prime}\,\partial \bar{M}_{\vec{q}}}=-G(\vec{p}\,){\Gamma}^{(4)}_{s}(\vec{p}_\bot,\vec{p}_\bot, \vec{0},\vec{0})G(\vec{p}\,^\prime)\,.
\end{equation}
 Inserting this relation in the equation \eqref{Ward2}, and taking the limit $p_1\to p_1^\prime\to 0$ and $\vec{q}=\vec{q}\,^\prime=\vec{0}$, we get:
\begin{equation}
\sum_{\vec{p}_\bot} \, G^2(\vec{p}_{\bot})\,{\Gamma}^{(4)}_{s}(\vec{p}_\bot,\vec{p}_\bot, \vec{0},\vec{0})=-\frac{\partial}{\partial p_1^2}\,\left[\Gamma_s^{(2)}(\vec{p}\,)- C^{-1}(\vec{p}\,) \right]\big\vert_{\vec{p}=\vec{0}}
\end{equation}
Recall  that $C^{-1}:=C_s^{-1}-r_s$. In fact, it is easy to see that  the LO contractions of $G^2$ are those building a $4$-dipole in the melonic contribution, and a $3$-dipole in the pseudo-melonic one. In the first case, this implies that we have to keep only the contribution of $\Gamma_{s\,\text{melo}}^{(4),1}$. In the second case,  we keep only the contributions of $\Gamma_{s\,\text{pseudo-melo}}^{(4),1i}$ for $i\neq 1$. Finally, because the $\sym$ is the decomposition of the four-point functions, only half of them contribute at LO. The Formula \eqref{Wardprime} follows. \\

\noindent
To prove   formulas \eqref{Wardsecond1} and \eqref{Wardsecond2}, we have to derive two times with respect to $M$, and two times with respect to $\bar{M}$. Straightforwardly we get:
\begin{equation}
\sum_{\vec p_\bot,\vec p\,'_\bot}\delta_{\vec p_\bot\vec p\,'_\bot}\Delta C_s(\vec p,\vec p\,')\,\tilde{\Gamma}^{(6)}_{s,\vec p_4\vec p_2\vec p\,';\vec p_3\vec p_1\vec p}\,G_s(\vec p\,)G_s(\vec p\,')+\sum_{\vec p_\bot,\vec p\,'_\bot}\delta_{\vec p_\bot\vec p\,'_\bot}X_{\vec p_1,\vec p_2,\vec p_3,\vec p_4;\vec p\vec{p}\,^\prime}=0\,,\label{sixward}
\end{equation}
where we used the following definitions:
\begin{equation}
\tilde{\Gamma}^{(6)}_{s,\vec p_4\vec p_2\vec p\,';\vec p_3\vec p_1\vec p}:=\Gamma^{(6)}_{s,\vec p_4\vec p_2\vec p\,';\vec p_3\vec p_1\vec p}-4\Gamma^{(4)}_{s,\vec p\vec p_4;\vec p_1\vec p\,''}G_s(\vec p\,'')\Gamma^{(4)}_{s,\vec p\,''\vec p_2;\vec p_2\vec p\,'}\,,
\end{equation}
and
\begin{equation}
X_{\vec p_1,\vec p_2,\vec p_3,\vec p_4;\vec p\vec{p}\,^\prime}:=-\delta_{\vec p_3 \vec p\,'}\Gamma^{(4)}_{s,\vec p_2\vec p_4;\vec p\vec p_1}
-\delta_{\vec p_1\vec p\,'}\Gamma^{(4)}_{s,\vec p_2\vec p_4;\vec p\vec p_3}
+\delta_{\vec p_4 \vec p}\,\Gamma^{(4)}_{s,\vec p\,'\vec p_2;\vec p_1\vec p_3}
+\delta_{\vec p_2 \vec p}\,\Gamma^{(4)}_{s,\vec p\,'\vec p_4;\vec p_1\vec p_3}
\end{equation}
and finally  we introduced the notation $\Delta C_s(\vec p,\vec p\,'):=C_s(\vec{p}\,^{2})-C_s(\vec{p}\,^{\prime\,{2}})$. There is an important difference with the previous Ward identity. While the boundary graph of any connected two-point graph is the elementary melon, denoted as $\gamma_1$ in the Section \eqref{Pseudomelonic}; the boundaries of connected four-point graphs can be melonics or pseudo-melonics. In both sides of the equation \eqref{sixward}, we then have to identify the contributions having the same boundary graphs; exactly like for the flow equations in the previous section, equation \eqref{Wett4} for instance. Moreover, the intertwining sector has to be taken into account in the computation of the pseudo-melonic contribution. Once again, picturing only the boundary structure of the effective vertices, the left-hand side of the equation \eqref{sixward} is written as:
\begin{align}
\nonumber&\left\{\,\vcenter{\hbox{\includegraphics[scale=0.6]{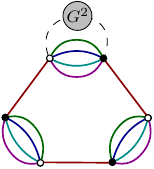} }}+\vcenter{\hbox{\includegraphics[scale=0.6]{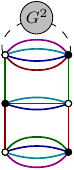} }}+\vcenter{\hbox{\includegraphics[scale=0.6]{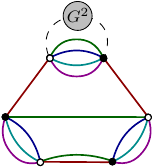} }}+\vcenter{\hbox{\includegraphics[scale=0.6]{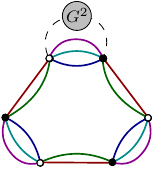} }}+\vcenter{\hbox{\includegraphics[scale=0.6]{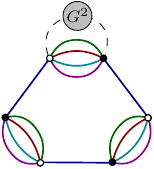} }}+\vcenter{\hbox{\includegraphics[scale=0.6]{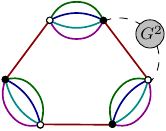} }}+\vcenter{\hbox{\includegraphics[scale=0.6]{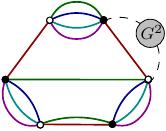} }}+\cdots\right\}\\
&\qquad\qquad-2\left\{\vcenter{\hbox{\includegraphics[scale=0.6]{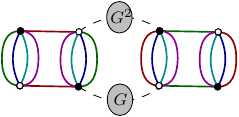} }}+\vcenter{\hbox{\includegraphics[scale=0.6]{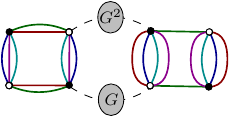} }}+\vcenter{\hbox{\includegraphics[scale=0.6]{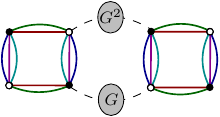} }}+\cdots\right\}\,,
\end{align}
and it is not hard to select, among these contractions, the LO ones, having respectively melonic or pseudo-melonic boundaries. For instance, the first one in the upper line has quartic melonic boundary, while the second ones have pseudo-melonic boundaries. We then have to split the series of allowed contractions as:
\begin{align}
\nonumber&\left\{\,\vcenter{\hbox{\includegraphics[scale=0.6]{ward1.pdf} }}+\vcenter{\hbox{\includegraphics[scale=0.6]{Wardnew.pdf} }}\right\}-2\left\{\vcenter{\hbox{\includegraphics[scale=0.6]{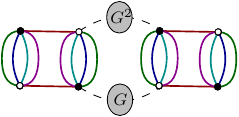} }}+\vcenter{\hbox{\includegraphics[scale=0.6]{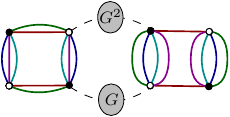} }}\right\}\,,
\end{align}
for relevant melonic, and
\begin{align}
\left\{\,\vcenter{\hbox{\includegraphics[scale=0.6]{Ward2.pdf} }}+\vcenter{\hbox{\includegraphics[scale=0.6]{Ward3.pdf} }}\right\}-2\left\{\vcenter{\hbox{\includegraphics[scale=0.6]{Ward43.pdf} }}\right\}\,,
\end{align}
for relevant pseudo-melonic.  Then, setting $\vec{p}_3=\vec{p}_4=\vec{0}$ in a first time, $\vec{p}_1=(p_1^\prime,\vec{0}_{\bot})$ , in a second time $\vec{p}_2=(p_1,\vec{0}_{\bot})$, and finally $p_1\to p_1^\prime\to 0$, we get the two formula \eqref{Wardsecond1} and \eqref{Wardsecond2}; taking into account the renormalization conditions \eqref{rencond12}, \eqref{rencond22} and \eqref{rencond3}. 

\begin{flushright}
$\square$
\end{flushright}
To complete this proposition, we may compute the derivative in both sizes, for $C^{-1}$ and $C^{-1}_s=C^{-1}+r_s$, using the regularized version \eqref{Cregularized} and \eqref{Cregularized2} of these kinetic kernels, we get:
\begin{equation}
\frac{\partial C^{-1}}{\partial p_1^2}(\vec{p}\,)=Z_{-\infty}\Theta_a^{-1}(\Lambda^2-\vec{p}\,^2)-(Z_{-\infty}\vec{p}\,^2+m^2)\frac{\Theta^\prime_a}{\Theta_a^2}(\Lambda^2-\vec{p}\,^2)\,.
\end{equation}
Then, taking the limit $a\to 0$, and due to the fact that $\Theta^\prime=\delta$, we get, for the right-hand side:
\begin{equation}
\frac{\partial C^{-1}}{\partial p_1^2}(\vec{p}=\vec{0}\,)=Z_{-\infty}\,.\label{Cprime}
\end{equation}
In the same way:
\begin{equation}
\frac{\partial C^{-1}_s}{\partial p_1^2}(\vec{p}\,)=\left(Z_{-\infty}+\frac{\partial r_s}{\partial p_1^2}(\vec{p}\,)\right)\Theta_a^{-1}(\Lambda^2-\vec{p}\,^2)-(Z_{-\infty}\vec{p}\,^2+m^2)\frac{\Theta^\prime_a}{\Theta_a^2}(\Lambda^2-\vec{p}\,^2)\,.
\end{equation}
Moreover, we have to take into account the factors $\Theta_a^2$ of $\Theta_a^3$ coming from $G^2$ of $G^3$ in the definitions of the functions $\mathcal{L}_i$ and $\mathcal{U}_i$. Indeed:
\begin{equation}
G=\frac{C_s}{1-C_s\Sigma(\vec{p}\,)}\propto \Theta(\Lambda^2-\vec{p}\,^2)\,,
\end{equation}
which leads to, in the limit $a\to 0$:
\begin{align}
&Z_{-\infty}\,\mathcal{L}_i:=\sum_{\vec{p}_\bot\in\mathbb{Z}^{d-i}} \left(Z_{-\infty}+\frac{\partial r_s}{\partial p_1^2}(\vec{p}\,)\right)\,G^{2}(\vec{p}_\bot)-\frac{5-i}{2}(Z_{-\infty}\Lambda^2+m^2)G^2(\Lambda)\Omega_{d-i} \Lambda^{3-i}\,,\label{newL}\\ &Z_{-\infty}\,\mathcal{U}_i:=\sum_{\vec{p}_\bot\in\mathbb{Z}^{d-i}} \left(Z_{-\infty}+\frac{\partial r_s}{\partial p_1^2}(\vec{p}\,)\right)\,G^{3}(\vec{p}_\bot)-\frac{5-i}{2}(Z_{-\infty}\Lambda^2+m^2)G^3(\Lambda)\Omega_{d-i} \Lambda^{3-i}\,,
\end{align}
where for the ‘‘boundary terms'' we used  the fact that $G(\vec{p}\,)$ only depends on $\vec{p}\,^2$, from which we introduced the notation: $G(\vec{p}\,)\delta(\Lambda^2-\vec{p}\,^2)=:G(\Lambda)\delta(\Lambda^2-\vec{p}\,^2)$. Moreover, we computed the following sum using integral approximation:
\begin{equation}
\sum_{\vec{p}_\bot} \delta(\Lambda^2-\vec{p}\,^2_{\bot}) \sim \int d^{d-i}x\, \delta(\Lambda^2-\vec{x}\,^2)=\frac{1}{2} (5-i)\Omega_{d-i} \Lambda^{3-i}\,.
\end{equation}
In the continuum limit, i.e., for $\Lambda \to \infty$, the boundary terms in equations (201) and (202) are entirely suppressed for $\mathcal{U}_i$ ($i=1,2$) and $\mathcal{L}_2$. For $\mathcal{L}_1$, however, the boundary contribution remains of order $\mathcal{O}(1)$. Within this contribution, the piece proportional to $Z_{-\infty}\Lambda^2$ can be safely discarded, as it is subleading compared to the logarithmic divergence of the first integral. Accounting for the proper scaling of the mass (including counter-terms), the second piece is also of order $\mathcal{O}(1)$. A technical subtlety arises because this term lacks the overall $Z_{-\infty}$ factor, making a direct comparison difficult. Nevertheless, since our analysis focuses on the deep infrared regime ($k = e^s \ll \Lambda$), this boundary term (which is strictly fixed by microscopic initial conditions at the scale $\Lambda$) does not compromise our IR conclusions. To circumvent these spurious cutoff effects entirely, one can adopt alternative regularization schemes, such as dimensional regularization. By analytically continuing the dimension of the internal group manifold from $U(1)$ to $U(1)^D$, boundary terms are structurally avoided, and divergences manifest simply as poles in $\epsilon = 1 - D$. Our physical conclusions remain robust and can be consistently derived in either scheme, provided we introduce the following definition:
\begin{definition}
In the dimensional regularization, the continuum limit corresponds to $D\to 1$.
\end{definition}

From this definition, the irrelevance of these sharp-cutoff boundary terms can be further established through a more rigorous decoupling argument. As previously noted, the boundary term generated at $\vec{p}\,^2 = \Lambda^2$ is superficially logarithmically divergent. To explicitly isolate its potential feedback on the infrared dynamics, we can perform a Taylor expansion with respect to the infrared scale $k = e^s$. Expanding in powers of the dimensionless ratio $k/\Lambda$ yields the generic form:
\begin{equation}
\text{Boundary Term}(k, \Lambda) = \mathcal{C}_0(\Lambda) + \mathcal{C}_1(\Lambda) \frac{k}{\Lambda} + \mathcal{O}\left( \left(\frac{k}{\Lambda}\right)^2 \right)\,.
\end{equation}
In the strict continuum limit ($\Lambda \to \infty$), all terms carrying a genuine dynamical dependence on the flowing scale $k$ are trivially suppressed and vanish. The only surviving contribution is the zero-th order term $\mathcal{C}_0(\Lambda)$. This residual term depends exclusively on the microscopic initial conditions.

Furthermore, the universality of the renormalization group dictates that the macroscopic flow constraints evaluated at scale $s$ must be insensitive to the chosen regularization scheme. Since such scale-free boundary artifacts are structurally absent in dimensional regularization, consistency requires that the sharp-cutoff residual $\mathcal{C}_0(\Lambda)$ must either be exactly zero or a non-dynamical, scheme-dependent artifact entirely absorbed into the initial bare parameters. In either case, its scale derivative vanishes identically ($\partial_s \mathcal{C}_0(\Lambda) = 0$), ensuring that these boundary terms completely decouple from the local Wetterich flow equations and the local consequence (at scale $s$) associated Ward-Takahashi identities.

The Ward identity allows to compute the derivative of the four-point functions at zero momenta, or in other words, to keep  additional information coming from the momentum dependence of the effective vertex. Note that such a dependence does not appear in the standard crude truncations; as it can be easily checked from Appendix \eqref{AppA} where the truncation method around just-renormalizable interactions has been briefly recalled to compare with the flow equations deduced from effective vertex method. The knowledge of these derivatives allows to compute the anomalous dimension from its definition \eqref{anomalous} like any other couplings. However, the Ward identities \eqref{Wardprime}, \eqref{Wardsecond1}, and \eqref{Wardsecond2} are not directly practicable because of the presence of $Z_{-\infty}$ and divergent quantities like $\mathcal{L}_1$. These quantities refer to the global history of the flow, but to be usable, the local flow equation \eqref{Wett} requires local information. Then, we have to deal with these divergent terms to extract local information from Ward identities. To this end, let us first  consider the case of the pure melonic sector, setting $\lambda_{4,2}=\lambda_{6,1}=0$. Note that this sector is stable, in the sense that relevant pseudo-melonic contributions cannot be generated from the melonic sector itself in the UV. In a second time, we will  consider the mixing sector. \\

$\bullet$ \textbf{Melonic sector}\\
\noindent
Setting $\lambda_{4,2}=\lambda_{6,1}=0$, only the melonic sector remains in the theory, and the relevant theory space in the  UV sector contains only melonic functions, having melonic graphs as boundary. The effective marginal coupling $\lambda_{4,1}(s)\equiv \lambda(s)$ remains the only relevant parameter to control the large  behavior of the flow. Then, all correlation functions may be expressed in terms of this parameter as well as effective two-point functions. The corresponding six-point function may be obtained from our results of the section \eqref{sec4} vanishing the pseudo-melonic couplings (see \cite{Lahoche:2018vun} for more detail). As mentioned before, the interest of this discussion allows only to understand the way to deal with nonlocal terms in the Ward identity, a point which in not discussed in the previous work. For convenience, and because it has to be precise, we limit our study to the standard \textit{modified Litim regulator} \cite{Tetradis:1995br}-\cite{Litim:2000ci}:
\begin{equation}
r_s(\vec{p}\,):=Z(s)(e^{2s}-\vec{p}\,^2)\Theta(e^{2s}-\vec{p}\,^2)\,,\label{regulatorlitim}
\end{equation}

\noindent
In the melonic sector, and using \eqref{newL}, and \eqref{Cprime}, the Ward identities \eqref{Wardprime}, \eqref{Wardsecond1} are reduced to\footnote{We discard the boundary terms for this discussion.} :
\begin{align}
&\qquad \quad 2Z_{-\infty} \lambda\,\mathcal{L}=-Z+Z_{-\infty}\,,\label{W1}\\
&Z_{-\infty} \Pi^{(3)}\,\mathcal{L}-8Z_{-\infty} \lambda^2\, \mathcal{U} =-\frac{d}{dp^2} \Pi^{(2)} \label{W2}\,,
\end{align}
where to simplify the notations we left the irrelevant lower index $1$ for $\mathcal{L}$ and $\mathcal{U}$, and we denote as $\Pi^{(3)}$ and $\Pi^{(2)}$ the quantities respectively called $\Pi^{(3)}_{1,a}$ and $\Pi^{(2)}_1$. Defining $\bar{Z}:=Z/Z_{-\infty}$, we get from the first equation \eqref{W1}:
\begin{equation}
\bar{Z}=1-2\lambda \mathcal{L}\,.\label{Zbar}
\end{equation}
Note that this equation has been derived in \cite{Lahoche:2018vun} directly from closed equation \eqref{closed1}. A first important relation between beta functions and anomalous dimension can be deduced from the first relation \eqref{W1}. This is  useful as a consistency ingredient for our incoming approximation as well as a test for other approximation schemes. This is the case  like the ones considered in the reference paper \cite{Lahoche:2018vun} for the computation of the effective loop integrals involved in the EVE method, and for other methods like truncation, considered in Appendix \eqref{AppA}. From the definition \eqref{anomalous}, $\eta=\dot{\bar{Z}}/\bar{Z}$, and from definition \eqref{Zbar}:
\begin{equation}
\dot{\bar{Z}}=-2(\dot{\lambda} \mathcal{L}+\lambda \dot{\mathcal{L}})=\frac{\dot{\lambda}}{\lambda} (\bar{Z}-1)-2\lambda (\dot{\mathcal{A}}_{4,2}+\dot{\Delta})\,.\label{eqmagic}
\end{equation}
The difference $\Delta:=\mathcal{L}-\mathcal{A}_{4,2}$ involves the derivative of the regulator function: $\Delta=\frac{1}{Z_{-\infty}}\sum_{\vec{p}\in\mathbb{Z}^{d-1}} G^2(\vec{p}\,)r_s^{\prime}(\vec{p}\,)$ , the ‘‘prime'' meaning derivative with respect to the variable $p_1^2$, setting equal to zero. For the Litim regulator:
\begin{equation}
r_s^{\prime}(\vec{p}\,)=-Z(s)\Theta(e^{2s}-\vec{p}\,^2)\to \dot{r}_s(\vec{p}\,)=-[\eta (e^{2s}-\vec{p}\,^2)+2e^{2s}]r_s^{\prime}(\vec{p}\,)\,.
\end{equation}
The two distributions $r_s^{\prime}$ and $\dot{r}_s$ are proportionals. Thus, the approximation \eqref{defZ} used for computation of the sums $\mathcal{I}_{n,m}$ in the flow equations must be used for computation of $\Delta$ and $\dot{\Delta}$. However, as explained before Definition \eqref{defZ}, this approximation becomes totally wrong for the computation of $\dot{\mathcal{A}}_{2,4}$, and all divergent quantities, for which the deep UV terms survive. We will discuss this point carefully. From now, let us consider the structure equations for melonic four-point functions. They only involve $\Pi_{1,a}^{(2)}\equiv \Pi^{(2)}=2\lambda(s)$, and it is explicitly given from Lemma \eqref{lemmaa0}:
\begin{equation}
\lambda(s)=\frac{\lambda}{1+2\lambda {\mathcal{A}}_{4,2}}\,,
\end{equation}
where, in the right hand side, coupling without explicit $s$ dependence designates the bare coupling. Deriving it with respect to $s$, we get:
\begin{equation}
\dot{\lambda}(s)=-2\lambda^2(s) \dot{\mathcal{A}}_{4,2}\,. 
\end{equation}
Finally, because of the definition of $r_s$, $\Delta$ is proportional to $\bar{Z}/Z^2$. Extracting this factor as $\Delta=:\frac{\bar{Z}}{Z^2} \bar{B}(s)$, and defining the renormalized coupling following Definition \eqref{rencouplings} $\bar{\lambda}:=\lambda/Z^2$, we obtain from equation \eqref{eqmagic}:
\begin{equation}
\eta=\frac{\dot{\lambda}}{\lambda}-2\bar{\lambda} (-\eta \bar{B}(s)+\dot{\bar{B}}(s))\,.
\end{equation}
The contribution $\bar{B}(s)$ can be easily computed using integral approximation, valid for $e^s\gg 1$,
\begin{equation}
\bar{B}(s)= - \frac{\Omega_{d-1}}{(1+\bar{m}^2)^2} \, \to \dot{\bar{B}}(s)=2 \frac{\Omega_{d-1}}{(1+\bar{m}^2)^3}\,\beta_m\,,
\end{equation}
where $\Omega_{d-1}$ denotes the hyper-volume of the unit ball in dimension $d-1$. Inserting this result in equation \eqref{eqmagic}, and using the definition of $\beta_{\lambda}=(\dot{\lambda}-2\eta\lambda)/Z^2$, we get:
\begin{corollary}\label{corconstraint}
In the deep UV and for purely melonic sector, and with approximation \eqref{defZ}, the Ward identity rely the beta functions and anomalous dimension as:
\begin{equation}
\beta_{\lambda}=-\eta \bar{\lambda} \left(1-2\bar{\lambda} \frac{\Omega_{d-1}}{(1+\bar{m}^2)^2}\right)+4\bar{\lambda}^2 \frac{\Omega_{d-1}}{(1+\bar{m}^2)^3}\,\beta_m\,.
\end{equation}
\end{corollary}
This equation is an additional constraint on the flow, and among their consequences, it adds a strong constraint on the fixed points. Indeed, let us define  a fixed point $p=(\bar{m}^2_*,\bar{\lambda}_*)$ such that $\beta_{\lambda}(p)=\beta_m(p)=0$, we must have:
\begin{equation}
\eta_*=0 \,,\quad \text{or}\quad 2\bar{\lambda}_* \frac{\Omega_{d-1}}{(1+\bar{m}^2_*)^2}=1\,.
\end{equation}
The fixed point found in \cite{Lahoche:2018vun}, recovered in section \eqref{sectiondern} and Appendix \eqref{AppA} was such that $p\approx (-0,55, 0,003)$, $\eta\approx 0,7$ and violates the two previous conditions. This seems to be a strong argument in favor of a disappearance of the fixed point, which violates the Ward identities. However, this argument requires to be carefully analyzed. First of all, the Ward identity is written in the symmetric phase. Also, the constraint depends on the choice of the regulator, except for one and two loop approximation as well, in accordance to their universality. Then, such a statement requires a proper analysis on the influence of the regulator. However, the strong gap between the obtained values and the requirement of the constraint seems to indicate that this fixed point is well a spurious consequence of the lack of the Ward identity constraint. We keep this important discussion for a work in preparation. At this stage, the constraint given from corollary \eqref{corconstraint} provides us  important information for the computation of the integrals. Indeed, if we try to compute the integral $\mathcal{A}_{4,2}$ using approximation \eqref{defZ}, introducing a cutoff on the high momenta, we get:
\bea
 \mathcal{A}_{4,2}=(\bar{B}+a\ln(\Lambda))/Z^2.
\eea
 The derivative with respect to $s$ then generates a factor $-2\eta \mathcal{A}_{4,2}$, having the same expected behavior in $\Lambda$ as $\bar{Z}$. As a result, we generate a term $-2\eta \bar{\lambda}$, which exactly compensates the same term coming from $\dot{\lambda}/\lambda$. As a consequence, the leading term in $\bar{\lambda}$ writes as $\beta_{\lambda}^{(2)}=\eta^{(1)} \bar{\lambda}$, where $\eta^{(1)}$ denotes the first order term in the expansion of $\eta$ in power of $\lambda$. From Appendix \eqref{AppA} or from a direct computation, we get $\eta^{(1)}=8\Omega_{d-1}\lambda$, implying $\beta_{\lambda}^{(2)}=8\Omega_{d-1}\lambda^2\geq 0$. This result is in complete violation of universality. Indeed, a direct calculation shows that the theory is asymptotically free and  $\beta_{\lambda}^{(2)}=-\eta^{(1)}\bar{\lambda}$. In the decomposition of $\mathcal{A}_{4,2}$, the problem comes from the big logarithm term $a \ln(\Lambda)/Z^2(s)$, and especially from the dependence on $Z(s)$. The bound of the integration is so far from the allowed windows of momenta ensuring validity of the approximation, and the factor in front of this big logarithm does not have to depend on $s$. With this hypothesis, and assuming that $1/\bar{Z}\to 0$ in the continuum limit $\Lambda\to\infty$, we recover the exact result \eqref{corconstraint}. For this reason, we will use this approximation only for convergent integrals, and more generally whenever  these deep UV contributions are completely discarded. This is the case for the integral approximations used in the reference \cite{Lahoche:2018vun}. \\

\noindent
Even to close this discussion, let us return on the expression \eqref{Zbar}. From renormalization conditions, $Z(s\to-\infty)=1$ and $\lambda(s\to-\infty)=\lambda_r$, we have:
\begin{equation}
Z_{-\infty}^{-1}=1-2\lambda_r A_{-\infty}\,, \label{Zinfmel}
\end{equation}
where $A_{-\infty}:=\mathcal{L}(s\to\infty)$. From the condition $r_{s\to-\infty}=0$, we have explicitly: 
\begin{equation}
A_{-\infty}=\sum_{\vec{p}\in\mathbb{Z}^{d-1}}\frac{1}{\left[\Gamma_{s\to -\infty}(\vec{p}\,)\right]^2}=\sum_{\vec{p}\in\mathbb{Z}^{d-1}} \left[ \frac{1}{Z_{-\infty}\vec{p}\,^2+m^2-\Sigma_{-\infty}(\vec{p}\,)}\right]^2\,.
\end{equation}
The counter-terms subtract all the sub-divergences, except the global one\footnote{We lack the final subtraction in the Zimmerman forest formula.}. Then, we expect that $Z_{-\infty}^{-1}$ diverges logarithmically in the continuum limit $\Lambda\to\infty$ ($D\to 1$), or equivalently ${Z}_{-\infty}\to 0$ like $1/\ln(\Lambda)$. Then ${Z} \approx -2Z_{-\infty}\lambda \mathcal{L}$, without contradiction with  formula \eqref{corconstraint}; as we can show from a straightforward computation. Moreover, in the second Ward identity \eqref{W2}:
\begin{equation}
\frac{1}{2} Z\Pi^{(3)}\,+8Z_{-\infty} \lambda^3\, \mathcal{U} =\lambda \frac{d}{dp^2} \Pi^{(2)}\,.\label{Wardreduce}
\end{equation}
As for $\mathcal{L}$, the quantity $\mathcal{U}$ splits as:
\bea
\mathcal{U}=\mathcal{A}_{4,3}+\frac{\bar{Z}}{Z^3} \bar{C}(s).
\eea However, $\mathcal{A}_{4,3}$ is superficially convergent and all the sub-divergences have been canceled from renormalization. Then, in the continuum limit, we have  $Z_{-\infty}\mathcal{A}_{4,3}\to 0$. Recall that $\Pi^{(3)}$ is given in section \eqref{68point}, proposition \eqref{propositionphi61}. Then, using integral approximation for the computation of the sum, and Definition \eqref{defZ} for the computation of the convergent integral, we get with $\Omega_4=\pi^2/2$:
\begin{equation}
\Pi^{(3)}=16\lambda^3 \mathcal{A}_{4,3} = 16\lambda^3 \frac{1}{2Z^3e^{2s}}\frac{\pi^2}{1+\bar{m}^2}\left(\frac{1}{(1+\bar{m}^2)^2}+\frac{1}{1+\bar{m}^2}+1\right) \,,
\end{equation}
and :
\begin{equation}
\bar{C}(s)=-\frac{1}{2e^{2s}} \frac{\pi^2}{(1+\bar{m}^2)^3}\,,
\end{equation}
so that equation \eqref{Wardreduce} becomes:
\begin{equation}
 \frac{d}{dp^2} \Pi^{(2)}=\frac{4}{Z^2 e^{2s}}\lambda^2\frac{\pi^2}{1+\bar{m}^2}\left(\frac{1}{1+\bar{m}^2}+1\right) \,.\label{Wardreduce2}
\end{equation}
The right-hand side of the equation \eqref{Wardreduce} may be computed from the same approximation. Because of the structure equation for four-point function, it is easy to check that (see \cite{Lahoche:2018vun}):
\begin{equation}
\frac{d}{dp^2} \Pi^{(2)}=-4\lambda^2(s) \frac{d}{dp^2} \mathcal{A}_{4,2}(p=0)\,,
\end{equation}
where $ \mathcal{A}_{4,2}(p):=\sum_{\vec{q}\in\mathbb{Z}^d}\delta_{pq_1}G^2(\vec{q}\,)$. Once again this function can be computed using approximation \eqref{defZ}, because the derivative with respect to $p_1^2$ is insensitive on the deep UV effect depending on the UV cutoff. Explicitly, we get:
\begin{equation}
\frac{d}{dp^2} \mathcal{A}_{4,2}(p=0)=-\frac{1}{Z^2 e^{2s}}\frac{\pi^2}{1+\bar{m}^2}\left(\frac{1}{1+\bar{m}^2}+1\right)\,,
\end{equation}
so that equation \eqref{Wardreduce2} is identically verified. Then, this simple checking ensures at least the coherence of our approximation for the computation of convergent integral with Ward identities. \\

$\bullet$ \textbf{Mixing sector}\\
\noindent
In the mixing sector, and following the Ward identity \eqref{Wardprime}, the formula \eqref{W1} is replaced by:
\begin{equation}
\bar{Z}=1-2\lambda_{4,1} \mathcal{L}_1-8\lambda_{4,2} \mathcal{L}_2\,.
\end{equation}
As for the purely melonic sector, we expect that $\bar{Z}\sim \ln(\Lambda)$ in the continuum limit. Moreover, once again, we have $\bar{Z}(s\to-\infty)=1/Z_{-\infty}$, so that :
\begin{equation}
Z_{-\infty}^{-1}=1-2\lambda_{4,1}^r \mathcal{A}_{4,2}(-\infty)-8\lambda_{4,2}^r \mathcal{A}_{3,2}(-\infty)\,,
\end{equation}
and $Z_{-\infty} \to 0$ like $1/\ln(\Lambda)$ in the continuum limit. As in the melonic sector, this allows to simplify the Ward identities. However, an additional strong simplification comes from the fact that $\mathcal{A}_{3,2}(-\infty)$ is a superficially convergent integral, whose sub-divergences are canceled from renormalization. Then, we have  $Z_{-\infty}^{-1}\approx-2\lambda_{4,1}^r \mathcal{A}_{4,2}(-\infty)$, meaning that $Z_{-\infty}\mathcal{A}_{3,2}\to0$ in the continuum limit. This can be translated as an approximation for $Z$ itself. Indeed, decomposing $\mathcal{L}_i$ as
\bea
 \mathcal{L}_i=\mathcal{A}_{5-i,2}+k^{d-i-4}\bar{Z} \bar{B}_i/Z^2,
\eea
 with
\begin{equation}
\bar{B}_i=-\frac{\Omega_{d-i}}{(1+\bar{m}^2)^2}\,,
\end{equation}
and imposing in the continuum limit that $Z_{-\infty}\mathcal{A}_{3,2}\to 0$, we get: $Z\approx -2Z_{-\infty}\lambda_{4,1} \mathcal{L}_1-8\lambda_{4,2} \bar{B}_2/(kZ)$. Also $\mathcal{U}_i$ is  decomposed as:
\begin{equation}
Z_{-\infty}\mathcal{U}_i=Z_{-\infty}\mathcal{A}_{5-i,3}+k^{d-i-6}\frac{1}{Z} \bar{\mathcal{C}}_i\,, \quad \bar{\mathcal{C}}_i= -\frac{\Omega_{d-i}}{(1+\bar{m}^2)^3}\,.
\end{equation}
Because $\mathcal{A}_{5-i,3}$ is a superficially convergent quantity, we expect that, up to the sub-divergences canceled from renormalization, it is insensitive on the UV cutoff. Therefore, we have to impose $Z_{-\infty}\mathcal{A}_{5-i,3}\to 0$ in the continuum limit and  the Ward identities \eqref{Wardsecond1} and \eqref{Wardsecond2} reduce to\footnote{We recall that $\Omega_{3}=4\pi/3$.}:
\begin{align}
&\frac{\partial \Pi^{(2)}_{1}}{\partial p_1^2} = \frac{Z}{\lambda_{4,1}}\left(\Pi_{1,a}^{(3)}+\frac{1}{3}\Pi_{1,b}^{(3)}\right)\left(1+\frac{32 \lambda_{4,2}\pi}{3Z^2(1+\bar{m}^2)^2}\right)- \frac{4\lambda_{4,1}\pi}{(1+\bar{m}^2)^3}\left(\pi\lambda_{4,1}+ \frac{32}{3} \lambda_{4,2}\right)\,,\label{deriv1}\\
&\frac{\partial \Pi^{(2)}_{2}}{\partial p_1^2}=\frac{4\pi}{3Z}\frac{1}{(1+\bar{m}^2)^2}\left(12\lambda_{6,1}+\frac{2}{3}\Pi^{(3)}_{1,c}\right)-\lambda_{4,2}^2\,\frac{32}{3Z}\frac{\pi}{(1+\bar{m}^2)^3}\,.\label{deriv2}
\end{align}
The knowledge of these derivatives is the last ingredient required for computation of the anomalous dimension.  Moreover, to find nontrivial fixed point requires to use \textit{renormalized couplings} defined in Definition \eqref{rencouplings}. To this end, we introduce the dimensionless renormalized sums $\bar{\mathcal{I}}_{m,n}$ as:
\begin{equation}
{\mathcal{I}}_{m,n}=:Z^{1-n} e^{(m+2-2n)s}\bar{\mathcal{I}}_{m,n}\,.\label{WdimI}
\end{equation}
Generally, we will adopt the convention that any bared quantity $\bar{x}$ is renormalized and dimensionless, in the sense that we have extracted its proper $Z$ and $e^s$ dependence:
\begin{equation}
x=Z^\alpha e^{d_x \,s} \bar{x}\,,
\end{equation}
$d_x$ being the \textit{canonical dimension} of $x$. From these definitions, the two equations \eqref{deriv1} and \eqref{deriv2} admit dimensionless and renormalized versions as:
\begin{align}
&\bar{\Pi}^{(2)\,\prime}_1 = \frac{1}{\bar{\lambda}_{4,1}}\left(\bar{\Pi}_{1,a}^{(3)}+\frac{1}{3}\bar{\Pi}_{1,b}^{(3)}\right)\left(1+\frac{32 \bar{\lambda}_{4,2}\pi}{3(1+\bar{m}^2)^2}\right)- \frac{4\bar{\lambda}_{4,1}\pi}{(1+\bar{m}^2)^3}\left(\pi\bar{\lambda}_{4,1}+ \frac{32}{3} \bar{\lambda}_{4,2}\right)\,,\label{deriv12}\\
&\bar{\Pi}^{(2)\,\prime}_2=\frac{4\pi}{3}\frac{1}{(1+\bar{m}^2)^2}\left(12\bar{\lambda}_{6,1}+\frac{2}{3}\bar{\Pi}^{(3)}_{1,c}\right)-\bar{\lambda}_{4,2}^2\,\frac{32}{3}\frac{\pi}{(1+\bar{m}^2)^3}\,.\label{deriv22}
\end{align}
Moving on the computation of the anomalous dimension, from equation \eqref{Gamma2} and definition \eqref{anomalous}, we get:
\begin{equation}
\dot{Z}=-\frac{d}{dp_1^2}\left[\sum_{\vec q}\Gamma^{(4)}_{s}(\vec p,\vec q;\vec p,\vec q\,)\,\frac{\dot r_s(\vec q\,)}{[\Gamma_s^{(2)}(\vec q\,)+r_s(\vec q\,)]^2}\right]\bigg\vert_{p_1=0}\,.
\end{equation}
Keeping only the LO terms like for the computation of $\beta_m$, we get:
\begin{equation}
\dot{Z}=-\Pi^{(2)\,\prime}_1 \bar{\mathcal{I}}_{4,2}-4\Pi^{(2)\,\prime}_2 \bar{\mathcal{I}}_{3,2}-\Pi^{(2)}_1 \bar{\mathcal{I}}_{4,2}^\prime-4\Pi^{(2)}_2 \bar{\mathcal{I}}_{3,2}^\prime\,.
\end{equation}
Then, from renormalization conditions \eqref{rencond1}, using definitions \eqref{rencouplings} and \eqref{WdimI}, as well as \eqref{anomalous}; and from the flow equation \eqref{flow1}, we deduce the following final statement:
\begin{proposition}\label{flowwithoutdim}
In the symmetric phase and in the continuum limit, the anomalous dimension $\eta(s)$ defined in \eqref{anomalous} satisfies the  equation:
\begin{equation}
\eta(s)=\frac{(4\bar\lambda_{4,1}-\Pi_1^{(2)\prime})\pi^2+(3\bar\lambda_{4,2}-\Pi_2^{(2)\prime})\frac{32\pi}{3}}{(1+\bar m^2)^2+(\Pi_1^{(2)\prime}-6\bar\lambda_{4,1})\frac{\pi^2}{6}+(\Pi_2^{(2)\prime}-5\bar\lambda_{4,2})\frac{32\pi}{15}},
\end{equation}
and the autonomous system for beta functions:
\begin{align}
&\beta_m=-(2+\eta)\bar{m}^2-10\,\bar{\lambda}_{4,1} \bar{\mathcal{I}}_{4,2}-20\bar{\lambda}_{4,2}\bar{\mathcal{I}}_{3,2}\,,\\
&\beta_{4,1}=-2\eta\bar{\lambda}_{4,1}-\left(2\bar{\Pi}_{1,a}^{(3)}+\frac{2}{3}\bar{\Pi}_{1,b}^{(3)}\right)\bar{\mathcal{I}}_{4,2}+4\bar{\lambda}_{4,1}^2\bar{\mathcal{I}}_{4,3}+16\bar{\lambda}_{4,1}\bar{\lambda}_{4,2}\bar{\mathcal{I}}_{3,3} \,,\\
&\beta_{4,2}=-(1+2\eta)\bar{\lambda}_{4,2}-\left(3\bar{\lambda}_{6,1}+\frac{1}{6}\bar{\Pi}_{1,c}^{(3)}\right)\bar{\mathcal{I}}_{3,2}+4\bar{\lambda}_{4,2}^2\bar{\mathcal{I}}_{3,3}\,,\\
&\beta_{6,1}=-3\eta \bar{\lambda}_{6,1}-\left(\frac{1}{6}\bar{\Pi}_{2}^{(4)}+\frac{1}{4!}\bar{\Pi}_1^{(4)}\right)\bar{\mathcal{I}}_{3,2}+12\bar{\lambda}_{4,2}\bar{\lambda}_{6,1}\bar{\mathcal{I}}_{3,3}-8\bar{\lambda}_{4,2}^3\bar{\mathcal{I}}_{3,4}\,,
\end{align}
where we defined:
\begin{equation}\label{Vinaurelie}
\bar{\mathcal{I}}_{m,n}(p):=\frac{1}{Z^{1-n} e^{(m+2-2n)s}} \sum_{\vec{p}\in\mathbb{Z}^{m+1}}\delta_{p,p_1} \dot{r}_s(\vec{p}\,)\,G^n(\vec{p}\,)\,; \qquad \bar{\mathcal{I}}_{m,n}^{\prime}=\frac{\partial \bar{\mathcal{I}}_{m,n}}{\partial p_1^2}(0)\,.
\end{equation}
\end{proposition}

All the dimensionless sums involved in these propositions may be easily computed using integral approximation. The details are given in Appendix \eqref{AppB}, we get:
\begin{equation}
\bar{\mathcal{I}}_{m,n}(p)=\frac{e^{-(m+2)s}\Omega_m}{(1+\bar{m}^2)^n} \left[2e^{2s}+\eta\left(e^{2s}-\left(\frac{m}{m+2}(e^{2s}-p^2)+p^2\right)\right)\right](e^{2s}-p^2)^{m/2}\,.
\end{equation}

Finally, from the approximation $Z\approx -2Z_{-\infty}\lambda_{4,1} \mathcal{L}_1-8\lambda_{4,2} \bar{B}_2/(kZ)$ used before, and from a straightforward computation following the same steps as for the purely melonic sector, we deduce the complementary statement of Corollary \eqref{corconstraint}:
\begin{corollary}
In the symmetric phase, and taking the continuum limit, the beta functions for $\phi^4$ couplings are related to the anomalous dimension as
\begin{align}
\nonumber\eta \bar{\lambda}_{4,1}&=(\beta_{4,1}+2\eta \bar{\lambda}_{4,1})\left(1-\frac{32\pi}{3}\frac{\bar{\lambda}_{4,2} }{(1+\bar{m}^2)^2}\right)-\frac{\pi^2\bar{\lambda}_{4,1}^2}{(1+\bar{m}^2)^2}\left(\eta+\frac{2\beta_m}{1+\bar{m}^2}\right)\\
&+\frac{32}{3}\frac{\pi\bar{\lambda}_{41}}{(1+\bar{m}^2)^2}\left(\beta_{4,2}+2\eta\bar{\lambda}_{4,2}\right)-\frac{32}{3} \frac{\pi\bar{\lambda}_{4,1}\bar{\lambda}_{4,2}}{(1+\bar{m}^2)^2}\left(\eta+\frac{2\beta_m}{1+\bar{m}^2}\right)\,.
\end{align}
\end{corollary}
Once again, these relations which depend only on the choice of the regulator and on the continuum limit can be used to analyze the robustness of the fixed points obtained from flow equation \eqref{Wett}.  Like for the melonic case, we will extend the discussion about this constraint in forthcoming works.

\subsection{Investigations on the phase space structure}\label{sectiondern}

To conclude this section, we will investigate the structure of the phase space, and the existence of nontrivial fixed points. We divide these investigations into two parts. In a first time we will study the vicinity of the Gaussian fixed point, and argue in favor of an asymptotic safety scenario. In a second time, we will move on to the research of nontrivial fixed points in accordance to this scenario. Due to the complicated structure of the flow equations, we use numerical methods for this purpose.   Another important simplification comes from the choice of the regulator $r_s$. Formally, the exact renormalization group flow described from equation \eqref{Wett} does not depend on the choice of the regulator. However, the approximations required to extract a practicable information from this equation generally depend on this choice. This is especially the case for crude truncations whose method is recalled in the Appendix \eqref{AppA} to compare with the EVE method. We expect that the effective vertex method discussed in this paper does not suffer for the same pathology, or at least,  the dependence on the regularization becomes marginal.

Let us remark that the Definition \eqref{defZ} is not supported with a rigorous proof.
 However, despite the fact that we expected  the reliability on the results deduced from our method,  the approximation scheme that we will use to extract information, especially for non-Gaussian fixed points is suspected to increase the dependence on the regulator.  Due to this difficulty,  a rigorous discussion on the reliability of our results, especially on the choice of the regulator function will be considered for a future work, and in this section we only investigate the plot of the  using the Litim regulator \eqref{regulatorlitim}. For the rest of this section, we denote by $p\equiv (\bar{m}^2,\bar\lambda_{4,1},\bar\lambda_{4,2},\bar\lambda_{6,1})$ the point in the four-dimensional phase space. 

\subsubsection{Vicinity of the non-Gaussian fixed point}

The Gaussian point $p_0:=(0,0,0,0)$ is obviously a fixed point of the system \eqref{flowwithoutdim}, with critical exponents $(-2,-1,0,0)$, in accordance with the canonical dimension of the involved couplings. Expanding the flow equations around $p_0$, and keeping only the leading order contributions, we get  :
\begin{align}
&\beta_m\approx -2\bar{m}^2-10\pi\left(\pi \bar{\lambda}_{4,1}+\frac{16}{3}\bar{\lambda}_{4,2}\right)\\
&\beta_{4,1}\approx-4\pi\left(\pi \bar{\lambda}_{4,1}+\frac{16}{3}\bar{\lambda}_{4,2}\right) \bar{\lambda}_{4,1}\\
&\beta_{4,2}\approx -\bar{\lambda}_{4,2}-8\pi \lambda_{6,1}\\
&\beta_{6,1}\approx -12\pi \left(\pi \bar{\lambda}_{4,1}+\frac{16}{3}\bar{\lambda}_{4,2}\right) \bar{\lambda}_{6,1}\,.
\end{align}
These equations are very reminiscent to some  already studied in the literature \cite{Carrozza:2014rba, Lahoche:2016xiq} for $\phi^6$ models. Setting $\bar{\lambda}_{4,2}=\bar{\lambda}_{6,1}=0 $, we recover the well known asymptotic freedom of the quartic melonic models : $\beta_{4,1}\approx-4\pi^2 \bar{\lambda}_{4,1}^2$, ensuring a well behavior in the UV for perturbative theory. Complementary, for $\bar{\lambda}_{4,1}=0$, the expected UV behavior of the flow lines is radically different. Indeed, let us consider the flow equations in the invariant plane $(\bar{\lambda}_{4,1},\bar{\lambda}_{6,1})$:
\begin{align}\label{systnew22}
&\beta_{4,2}=-\bar{\lambda}_{4,2}-8\pi\bar{\lambda}_{6,1}\\
&\beta_{6,1}=-64\pi\bar{\lambda}_{4,2}\bar{\lambda}_{6,1}\,.
\end{align}  
The situation is quite different from the pure marginal quartic case. In contrast to him, the sign of the beta function depends on the sign of the couplings, and this dependence drastically changes the behavior of the flow lines. When the two couplings are both positive, $\bar{\lambda}_{6,1},\bar{\lambda}_{4,2}\,>0$, a region denoted as $\text{I}$ on Figure \eqref{figflow}a, the beta functions are negative as well; and the two couplings decrease in the same time. Nevertheless, so far from $p_0$, the flow line are incoming on the origin. For the rest of the phase diagram, we have to distinguish two regions, respectively labeled $\text{II}$ and $\text{III}$, such that $\beta_{4,2}<0$ in  region $\text{II}$ and $\beta_{4,2}>0$ in the region $\text{III}$; the two regions being separated with the line $\beta_{4,2}=0$. In  region $\text{II}$, $\bar{\lambda}_{4,2}<0$ and $\bar{\lambda}_{6,1}>0$, then, $\beta_{4,2}<0$ and $\beta_{6,1}>0$. As a result, any trajectory starting in this region necessarily goes ultimately far away of the Gaussian fixed point. Finally, in  region $\text{III}$, the situation is still different, $\beta_{4,2}$ and $\beta_{6,1}$ are both positive; and we expect that any trajectory starting in this region is getting closer to the $\bar{\lambda}_{6,1}$-axis and moving away from the $\bar{\lambda}_{4,2}$-axis. To be more precise, let us consider  region $\text{I}$. 

\begin{center}
\begin{equation*}
\vcenter{\hbox{\includegraphics[scale=0.8]{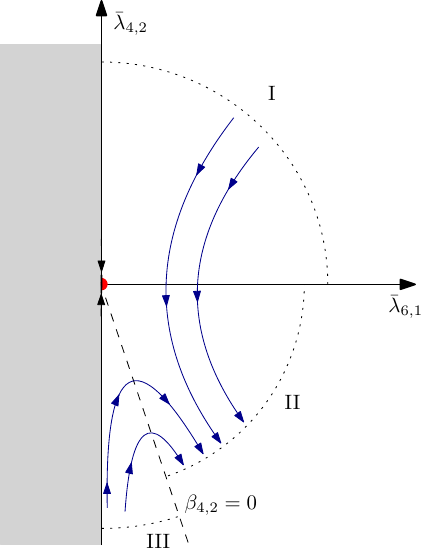} }}\qquad \vcenter{\hbox{\includegraphics[scale=0.45]{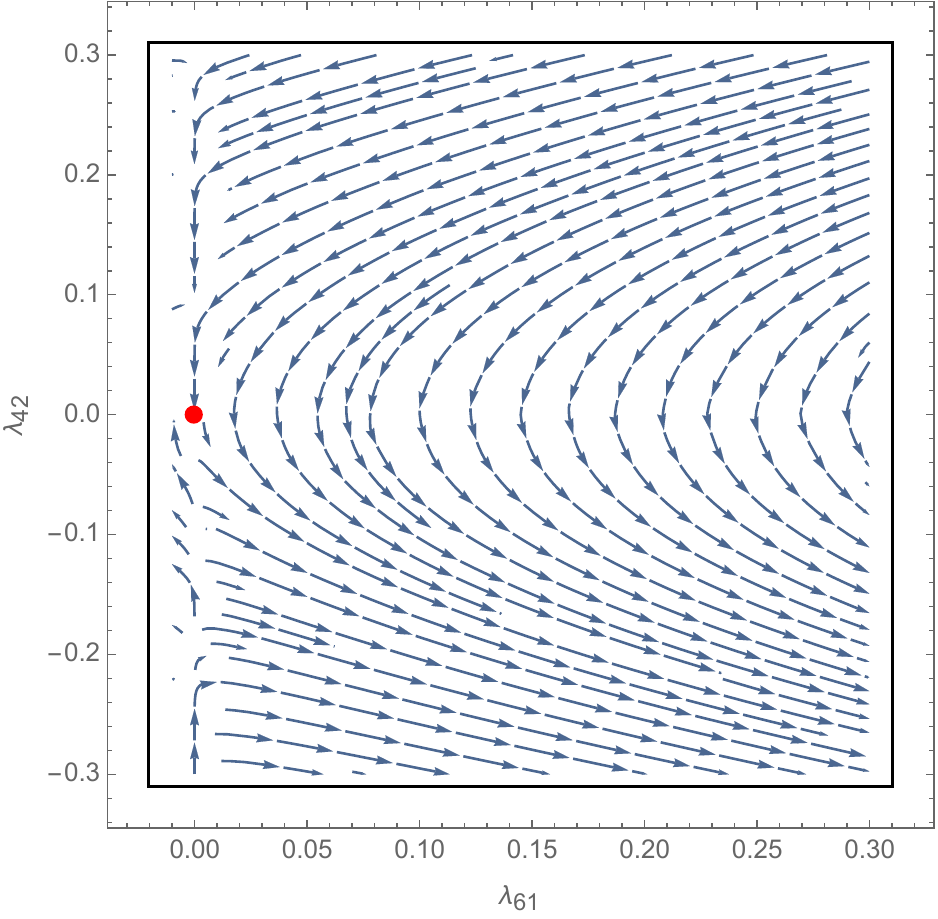}  }}
\end{equation*}
\captionof{figure}{Phase portrait in the plan $(\bar{\lambda}_{4,1},\bar{\lambda}_{6,1})$, in the vicinity of the Gaussian fixed point. On the left, the qualitative phase space, obtained from an analysis of the differential equations describing the flow. On the right, the same region obtained from a numerical analysis. In both cases, the red point corresponds to the Gaussian fixed point.}\label{figflow}
\end{center}
It is easy to show that any trajectory starting in this region (which excludes the  vertical axis) has to cross the $\bar{\lambda}_{6,1}$-axis before reaching the Gaussian fixed point, and then is ultimately repelled from it, to the infinity. Indeed, consider a trajectory whose  points reach the surface of a disk of radius $\epsilon$ around $p_0$. Excluding the axis, the two coordinates are both of order $\epsilon$, and then $\beta_{4,2}=\mathcal{O}(\epsilon)$ while $\beta_{6,1}=\mathcal{O}(\epsilon^2)$. As a consequence, for any variation $\delta s$ of the flow parameter, $\delta \bar{\lambda}_{4,2}\propto \epsilon \delta s$ and $\delta \bar{\lambda}_{6,1}\propto \epsilon^2 \delta s$, implying that $\delta \bar{\lambda}_{4,2}\propto \delta \bar{\lambda}_{6,1}/\epsilon$. For any variation $\delta \bar{\lambda}_{6,1}$ in the direction of the $\bar{\lambda}_{4,2}$-axis, the component of the velocity vector in the direction of the horizontal axis becomes arbitrarily large. The vector field then becomes more and more vertical, and finally goes through the horizontal axis.  It is not so hard to show that around this point corresponds to the minimal distance with the origin of coordinates\footnote{In polar coordinates, denoting by $r(t)$ the distance from the origin, we get easily that $\dot{r}(s)=-r(s)\sin{\theta}(\sin{\theta}+8\pi \cos{\theta}+64r(s)\pi\sin{\theta}\cos{\theta})$, vanishing for $\theta=0$.}. The same phenomena is expected in region $\text{III}$, and the qualitative behavior is pictured on Figure \eqref{figflow}a. Figure \eqref{figflow}b is a numerical integration of the same region, which confirms this expected behavior. As a consequence, it seems that the mixing sector is not asymptotically free, due to the presence of pseudo-melons. Solving this difficulty, and a well understanding of the UV completion of the theory then requires to have more information about the landscape around the Gaussian fixed point, especially concerning the existence of the non-Gaussian UV fixed point to which the outgoing trajectories maybe to ends.

\subsubsection{Non Gaussian fixed points}

To investigate the non-Gaussian fixed points, we have to solve the complete autonomous system given from Proposition \eqref{flowwithoutdim}. However, the complicated structure of these equations requires approximations to  solve them. Due to this difficulty, we limit our investigations on the fixed points which can be reached from a perturbative analysis. Our strategy is the following. Increasing the terms keeping in the perturbative expansion of the exact flow equations, we get a list of fixed points which progressively converge to a single one: 
\begin{equation}
p_1\approx(-0.55, 0.003,0,0)\,,
\end{equation}
having one relevant, one irrelevant and two attractive focal directions. Because all the pseudo-melonic couplings vanish, we call this fixed point ‘‘melonic''. Moreover,it matches with the fixed point already deduced in our previous work \cite{Lahoche:2018vun} in the melonic sector, and with the fixed point called $FP2$ in the Appendix \eqref{AppA} of this paper. This fixed point seems to be a good candidate as UV fixed point, providing an indication in favor of an asymptotic safety scenario. In this respect, the fact that only melonic interactions survives when the trajectories reach this fixed point remains an interesting feature of this theory. However, even if we discard the numerical difficulties, the existence of this fixed point seems to be compromised from the constraint \eqref{corconstraint} coming from Ward identity. Indeed, a simple numerical calculation shows a large deviation at the fixed point $p_1$, indicating a strong violation of the Ward identity for relevant operators. At this stage, we cannot expect that the problem comes from the numerical analysis. Indeed, the same difficulty occurs in the purely melonic sector, where the numerical investigation may be done easily. Moreover, as discussed above, we naively expect that the dependence on the regularization has to be improved with respect to the crude truncations. Nevertheless in both cases the difficulty is the same. A possibility is to consider that this discordance indicates the break down of the validity of the expansion around vanishing mean field, as it is the case in the symmetric phase. More generally, we can expect that the problem comes from a crude reduction of the full theory space.

\section{Conclusion}\label{sec6}
In this paper we have built a version of the non-perturbative renormalization group flow including nontrivial dependence of the effective vertices on the relevant and marginal operators in a sector mixing melonics and pseudo-melonics interactions. This allows to use them to solve the renormalization group flow of the full operators in the UV, and close the infinite hierarchy equation coming from the flow equations. The resulting flow equations seems to indicate the existence of a nontrivial UV attractive fixed point including only purely melonic interactions. However, we showed that Ward identity is strongly violated at this fixed point. As a result, our unique fixed point seem to be unphysical, and the possible existence of other nontrivial fixed points far away of our investigation procedure seems to be necessary. 

The importance to include the constraint coming from Ward identity in the resolution of the flow equation is not a novelty, and is well known in gauge theory, especially in QCD nonperturbative approach. Note that it is not the only limitation of our results. Despite the fact that we do not crudely truncate the flow, we have made some approximations whose consistency have to be supported in forthcoming  works. In particular our investigations have been limited on the symmetric phase, ensuring convergence of any expansion around vanishing means field. Moreover, we have retained only the first terms in the derivative expansion of the two-point function, and only considered the local potential approximation, i.e. potentials which can be expanded as an infinite sum of melonic and pseudo melonic local interactions. Some deviations from ultra locality could be introduce nontrivial effects having the same power counting with ultralocals interactions with higher valences, and then  to contribute with them on the same footing. All these difficulties are not taken into account in our conclusions, and will be discussed in some works in progress.

\begin{appendices}
\section{ Useful formulas}\label{AppB}
In this section we provide the proof of  the useful sums involved in the flow equations. First of all,  let us write 
\begin{equation}
\mathcal{A}_{m,n}:=\sum_{\vec{p}\in\mathbb{Z}^{m}} \,G^n(\vec{p}\,)\,.
\end{equation}
Due to the Heaviside function $\theta(x)$, the above expression can be decomposed into two different contributions depending respectively on $\theta(e^{2s}-p^2)$ and $\theta(p^2-e^{2s})$. The first one corresponds to the integration domain $p^2\leq e^{2s}$ i.e the IR domain. On the other hand the second integration region $p^2> e^{2s}$ corresponds to  the UV domain.  Now integrating over the $m$-ball we get 
\bea\label{sososo}
\mathcal{A}_{m,n}(p)&=&\bigg(\frac{\Omega_m(e^{2s}-p^2)^{\frac{m}{2}}}{Z^n e^{2n s}(\bar m^2+1)^n}+\frac{m\Omega_m}{Z^n}\int_{\sqrt{e^{2s}-p^2}}^\infty\frac{x^{m-1}dx}{(x^2+p^2+e^{2s}\bar m^2)^n}\bigg)\theta(e^{2s}-p^2)\cr
&+&\frac{m\Omega_m}{Z^n}\int_{0}^\infty\frac{x^{m-1}dx}{(x^2+p^2+e^{2s}\bar m^2)^n}\theta(p^2-e^{2s}),
\eea
where $\Omega_m=\frac{\pi^{\frac{m}{2}}}{\Gamma(\frac{m}{2}+1)}$ is the  volume of unit $m$-sphere. The particular case of integers $m$  at $p=0$  are given by the following relations:
\bea
\mathcal{A}_{4,3}&=&\frac{\pi ^2 (\bar m^2+2) e^{-2 s}}{2 (\bar m^2+1)^2 Z^3}+\frac{\pi ^2 e^{-2 s}}{2 (\bar m^2+1)^3 Z^3}
\\
\mathcal{A}_{3,n}&=&\frac{4\pi e^{(3-2n)s}}{3  Z^n(\bar m^2+1)^n} +\frac{4\pi e^{(3-2n)s}}{ Z^n} \int_1^\infty\,\frac{x^2 dx}{(x^2+\bar m^2)^n}
\\
\mathcal{A}_{2,n}&=&\frac{\pi e^{(2-2n)s} }{Z^n(\bar m^2+1)^n } +\frac{2\pi e^{(2-2n)s}}{Z^n} \int_1^\infty\,\frac{x dx}{(x^2+\bar m^2)^n}
\\
\mathcal{A}_{1,n}&=&\frac{2e^{(1-2n)s}}{Z^n(\bar m^2+1)^n }+\frac{e^{(1-2n)s}}{ Z^n}\int_1^\infty\,\frac{dx}{(x^2+\bar m^2)^n}
\eea
Now let us compute the quantities: $\sum_{p\in\mathbb{Z}} (\mathcal{A}_{3,2}(p))^n$ and $\sum_{p\in\mathbb{Z}}\mathcal{A}_{3,3}(p) (\mathcal{A}_{3,2}(p))^3$ using \eqref{sososo}:
\bea
\mathcal{A}_{3,2}(p)&=&\Big[\frac{4\pi(e^{2s}-p^2)^{\frac{3}{2}}}{3Z^2 e^{4s}(\bar m^2+1)^2}+\frac{4\pi }{ e^{2s}Z^2}\int_1^\infty\,\frac{x(e^{2s}x^2-p^2)^{\frac{1}{2}}dx}{(x^2+\bar m^2)^2}\Big]\theta(e^{2s}-p^2)\cr
&+&\frac{4\pi}{Z^2}\int_0^\infty\,\frac{x^2 dx}{(x^2+p^2+e^{2s}\bar m^2)^2}\theta(p^2-e^{2s})
\eea
Note that the support of the distributions $\theta(e^{2s}-p^2)$ and  $\theta(p^2-e^{2s})$  denoted respectively  by $\mathcal D_1$ and $\mathcal D_2$ are such that $\mathcal D_1 \cap \mathcal D_2=\{e^s\}$, which is of null measure in the Lebesgue sens. Then the integration of the functions of the  form $X(p):=(a(p)\theta(e^{2s}-p^2)+b(p)\theta(p^2-e^{2s}))^n$ is therefore:
\bea\label{samovin}
\int dp \,X(p)=\int_{\mathcal{D}_1} \,dp\, a^n(p)\theta(e^{2s}-p^2)+\int_{\mathcal{D}_2} dp\, b^n(p)\theta(p^2-e^{2s}).
\eea
To give more explanation about the above relation, let us remark that  \eqref{samovin}  can be viewed as the following sum
\bea
2\int dp \,X(p)&\equiv&\sum_{p\in \mathcal{D}_1}a^n(p)+\sum_{p\in \mathcal{D}_2}b^n(p)+\sum_{\ell=1}^{n-1}\sum_{p\in\mathcal D_1 \cap \mathcal D_2}\frac{n!}{\ell!(n-\ell)! }a^\ell(p) b^{n-\ell}(p)\cr
&=&\sum_{p\in \mathcal{D}_1}a^n(p)+\sum_{p\in \mathcal{D}_2}b^n(p)+\sum_{\ell=1}^{n-1}\frac{n!}{(n-\ell)! }a^\ell(e^{2s}) b^{n-\ell}(e^{2s}),
\eea
where  in the $UV$ limit i.e.  $s\rightarrow \infty$ the quantity  $a(e^{2s})\equiv e^{-s} Z^\alpha \bar a\rightarrow 0$ and $b(e^{2s})\equiv e^{-s} Z^\alpha \bar b\rightarrow 0$, $\alpha\in \mathbb{Z}$.
Now applying this result to  $\sum_{p\in\mathbb{Z}} (\mathcal{A}_{3;2}(p))^n$ we come to:
\bea
\sum_{p\in\mathbb{Z}} (\mathcal{A}_{3,2}(p))^n&=& 2\Big(\frac{4\pi}{Z^2}\Big)^n e^{(1-n)s}\int_{1}^\infty \,dy\bigg(\int_0^\infty\frac{x^2dx}{(x^2+y^2+\bar m^2)^2}\bigg)^n   \cr
&+&2\Big(\frac{4\pi}{Z^2}\Big)^n e^{(1-n)s} \int_0^1dy\Bigg(\frac{(1-y^2)^{\frac{3}{2}}}{ 3(\bar m^2+1)^2}+ \int_1^\infty\,\frac{x(x^2-y^2)^{\frac{1}{2}}dx}{(x^2+\bar m^2)^2}\Bigg)^n.
\eea
In the same manner
\bea
\sum_{p\in\mathbb{Z}} \mathcal{A}_{3,3}(p)(\mathcal{A}_{3,2}(p))^n&=& 2\Big(\frac{4\pi}{Z^2}\Big)^n\frac{4\pi}{Z^3} e^{-(2+n)s}\int_{1}^\infty \,dy\bigg\{ \int_0^\infty\frac{x^2dx}{(x^2+y^2+\bar m^2)^3}\cr
&\times&\bigg(\int_0^\infty\frac{x^2dx}{(x^2+y^2+\bar m^2)^2}\bigg)^n\bigg\}   \cr
&+&2\Big(\frac{4\pi}{Z^2}\Big)^n \frac{4\pi}{Z^3} e^{-(2+n)s} \int_0^1dy\bigg\{\bigg(\frac{(1-y^2)^{\frac{3}{2}}}{ 3(\bar m^2+1)^3}+ \int_1^\infty\,\frac{x(x^2-y^2)^{\frac{1}{2}}dx}{(x^2+\bar m^2)^3}\bigg)\cr
&\times&\bigg(\frac{(1-y^2)^{\frac{3}{2}}}{ 3(\bar m^2+1)^2}+ \int_1^\infty\,\frac{x(x^2-y^2)^{\frac{1}{2}}dx}{(x^2+\bar m^2)^2}\bigg)^n\bigg\}
\eea

As in the previous paragraph  the sum \eqref{Vinaurelie} can be integrated in the $d-1$ ball as follows:
\bea
\mathcal{I}_{4,n}(p)&\equiv&
(d-1)\Omega_{d-1}\frac{\partial_s Z(s)(e^{2s}-p^2)+2Z(s)e^{2s}}{\Big(Z(s)(e^{2s}\bar m^2+e^{2s})\Big)^n}\frac{(e^{2s}-p^2)^{\frac{d-1}{2}}}{(d-1)}\cr
&-&(d-1)\Omega_{d-1}\frac{\partial_s Z(s)}{\Big(Z(s)(e^{2s}\bar m^2+e^{2s})\Big)^n}
\frac{(e^{2s}-p^2)^{\frac{d+1}{2}}}{(d+1)}.
\eea
In the same manner the sum \eqref{Vinaurelie} can be integrated in the $d-2$ ball as follows:
\bea
\mathcal{I}_{3,n}(p)&=&(d-2)\Omega_{d-2}\frac{\partial_s Z(s)(e^{2s}-p^2)+2Z(s)e^{2s}}{\Big(Z(s)(e^{2s}\bar m^2+e^{2s})\Big)^n}\frac{(e^{2s}-p^2)^{\frac{d-2}{2}}}{(d-2)}\cr
&-&(d-2)\Omega_{d-2}\frac{\partial_s Z(s)}{\Big(Z(s)(e^{2s}\bar m^2+e^{2s})\Big)^n}\frac{(e^{2s}-p^2)^{\frac{d}{2}}}{(d)}.
\eea
Now we come to the simple expressions of $\mathcal{I}_{4,n}(0)$ and $\mathcal{I}_{3,n}(0)$ as:
\bea
\mathcal{I}_{4,n}(0)=\frac{\pi^2 e^{6s-2ns}}{6Z^{n-1}(\bar m^2+1)^n}(\eta+6),\quad \mathcal{I}_{3,n}(0)=\frac{8\pi e^{5s-2ns}}{15Z^{n-1}(\bar m^2+1)^n}(\eta+5)\\
\mathcal{I}'_{4,n}(0)=-\frac{\pi^2 e^{(4-2n)s}}{2Z^{n-1}(\bar m^2+1)^n}(\eta+4),\quad \mathcal{I}'_{3,n}(0)=-\frac{4\pi e^{(3-2n)s}}{3Z^{n-1}(\bar m^2+1)^n}(\eta+3).
\eea

\section{  $\phi^6$-truncation for mixing -$U(1)$ TGFT}\label{AppA}
In this section the flow equation for the $\phi^4$ model is derived using the usual truncation method of the Weterrich equation. For more detail concerning this process  let us see the references 
\cite{Geloun:2016qyb}-\cite{Lahoche:2018vun} and \cite{Dona:2015tnf}-\cite{Eichhorn:2013isa}  in the case of matrix models. 
\subsection{Truncation procedure}
The truncation is a projection of the RG flow into a finite dimensional subspace of the infinite dimensional full theory space. In the case where $d=5$  we write
\bea
\Gamma_s&=&Z(s)\sum_{\vec p}T_{\vec p}(\vec p\,^2+e^{2s}\bar m(s)^2)\bar T_{\vec p}+Z^2(s)\bar\lambda_{4,1}(s)V_{4,1}\cr
&&+Z^2(s)e^s\bar\lambda_{4,2}(s)V_{4,2}+Z^3(s)\bar\lambda_{6,1}(s)V_{6,1}
\eea
where the renormalized couplings $Z^2(s)\bar\lambda_{4,1}(s)$, $Z^2(s)\bar\lambda_{4,2}(s) e^s$ and $Z^3(s)\bar\lambda_{6,1}(s)$ are used and the functions $V_{4,1}$, $V_{4,2}$ and $V_{6,1}$ are the four and six-point interactions taking into account the two different sectors: melon and pseudo-melon.  The Wetterich flow equation can then be expand as:
\bea
\partial_s \Gamma_s&=&{\rm Tr}\,\dot{r}_s G^{(0)}_s\Big(1-2\lambda_{4,1}(s)V''_{4,1}G^{(0)}_s-2\lambda_{4,2}(s)V''_{4,2}G^{(0)}_s-3\lambda_{6,1}(s)V''_{6,1}(G^{(0)}_s)^2\cr
&+&4(\lambda_{4,1}(s)V''_{4,1} G_s^{(0)})^2+4(\lambda_{4,2}(s)V''_{4,2} G_s^{(0)})^2+4\lambda_{4,1}(s)V''_{4,1} G_s^{(0)}\lambda_{4,2}(s)V''_{4,2} G_s^{(0)}\cr
&+&6\lambda_{4,1}(s)\lambda_{6,1}(s)V''_{4,1}V''_{6,1} (G_s^{(0)})^3+6\lambda_{4,2}(s)\lambda_{6,1}(s)V''_{4,1}V''_{6,1} (G_s^{(0)})^3\cr
&+&9\lambda^2_{6,1}(s)V''_{6,1}(G_s^{(0)})^4+\cdots\Big).
\eea
Now separate the contribution to the mass and couplings, the flow equations involve many contractions of lines which can be represented graphically by the following diagrams: (note that we only considered the leading order contribution taking into account the melons, the pseudo-melons and the intermediate contributions reported in subsection \eqref{sousec4}).
\bea\label{tomate1}
\partial_s\Gamma_s^{(2)}&=&\sum_{i}K_1^{2,i}\vcenter{\hbox{\includegraphics[scale=0.9]{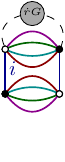} }}+\sum_{i,j}K_2^{2,ij}\vcenter{\hbox{\includegraphics[scale=0.9]{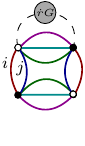} }},
\eea
\bea\label{tomate2}
\partial_s\Gamma_s^{(4)}&=&\sum_{i}K_1^{4,i} \vcenter{\hbox{\includegraphics[scale=0.9]{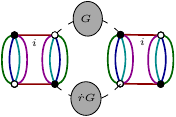} }}+\sum_{i,j}K_2^{4,ij}\vcenter{\hbox{\includegraphics[scale=0.9]{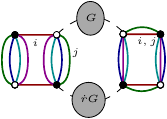} }}\cr
&+&\sum_{ij}K_3^{4,ij}\vcenter{\hbox{\includegraphics[scale=0.9]{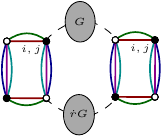} }}
+K_4^{4,ij}\vcenter{\hbox{\includegraphics[scale=0.9]{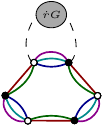} }},
\eea
\bea\label{tomate3}
\partial_s\Gamma_s^{(6)}&=&\sum_{i}K_1^{6,i}\vcenter{\hbox{\includegraphics[scale=0.9]{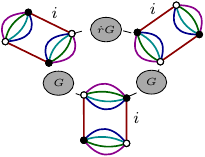} }}+\sum_{i,j}K_4^{6,ij}\vcenter{\hbox{\includegraphics[scale=0.9]{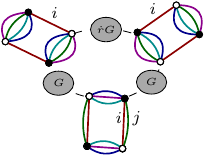} }}\cr
&+&\sum_{i,j}\Bigg[K_2^{6,ij}\vcenter{\hbox{\includegraphics[scale=0.9]{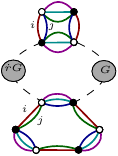} }}+K_3^{6,ij}\vcenter{\hbox{\includegraphics[scale=0.9]{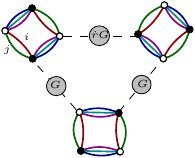} }}+K_5^{6,ij}\vcenter{\hbox{\includegraphics[scale=0.9]{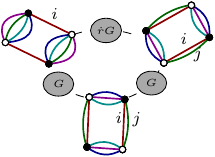} }}\Bigg]\cr
&&
\eea
In \eqref{tomate1} the first term corresponds to the melonic contribution and the second  one to the pseudo-melonic sector. Further,
the first line of equation \eqref{tomate2}  contributes only  to the melonic sector and the second line to the pseudo-melonic one.  On the other hand the first two graphs in the first line  of  equation \eqref{tomate3} contribute to the melonic sector, the first two graphs of the second line contribute to the pseudo-melonic sector and the last one to the intermediate leading order contribution. Note that only  the pseudo-melonic sector  will be  taken into account in the flow equation of the coupling $\lambda_{6,1}(s)$. The coefficients  $K_n^{m,i}$ and $K_n^{m,ij}$ are the combinatorial factors and will be given with detail. 

Now using the results of the section \eqref{sec4}, which  provided the full expressions of the four and six-point functions, and  because of the choice of the truncation, we have the following relations:
\bea
&&m^2(s)=\Gamma^{(2)}_s(\vec 0), \\
&& Z(s)=\frac{d\Gamma^{(2)}_s(\vec p)}{dp_1^2}\Big|_{\vec p=\vec 0},\\
&&\partial_s\Gamma_{\text{melo}}^{(4)}(\vec 0,\vec 0,\vec 0,\vec 0)=4\partial_s\lambda_{4,1}(s),\\
&&\partial_s\Gamma_{\text{pseudo--melo}}^{(4)}(\vec 0,\vec 0,\vec 0,\vec 0)=4\partial_s\lambda_{4,2}(s),\\
&& \partial_s\Gamma^{(6)}_{s\,,\text{pseudo-melo}}(\vec{0},\vec{0},\vec{0},\vec{0},\vec{0},\vec{0})=36\partial_s\lambda_{6,1}(s).
\eea
Then using the above expressions and  expanding in detail the relations  \eqref{tomate1}, \eqref{tomate2} and  \eqref{tomate3}, we get the following flow equations in which the dimensionless parameters $\bar m$, $\bar \lambda_{4,1}$, $\bar \lambda_{4,2}$ and $\bar \lambda_{6,1}$ are used :
\begin{align}\label{systnew2}
\left\{\begin{array}{llllll}
&\beta_m=-(2+\eta)\bar{m}^2-10\,\bar{\lambda}_{4,1} \bar{\mathcal{I}}_{4,2}-20\bar{\lambda}_{4,2}\bar{\mathcal{I}}_{3,2}\\
&\beta_{4,1}=-2\eta\bar{\lambda}_{4,1}+4\bar{\lambda}_{4,1}^2\bar{\mathcal{I}}_{4,3}+16\bar{\lambda}_{4,1}\bar{\lambda}_{4,2}\bar{\mathcal{I}}_{3,3} \\
&\beta_{4,2}=-(1+2\eta)\bar{\lambda}_{4,2}-3\bar{\lambda}_{6,1}\bar{\mathcal{I}}_{3,2}+4\bar{\lambda}_{4,2}^2\bar{\mathcal{I}}_{3,3}\\
&\beta_{6,1}=-3\eta \bar{\lambda}_{6,1}+12\bar{\lambda}_{4,2}\bar{\lambda}_{6,1}\bar{\mathcal{I}}_{3,3}-8\bar{\lambda}_{4,2}^3\bar{\mathcal{I}}_{3,4}
\end{array}\right.,
\end{align}  
In the computation of the anomalous dimension $\eta:=\partial_s Z/Z$ we use the fact that
\bea
\partial_s Z=-2\lambda_{4,1}\frac{d}{dp_1^2} \sum_{\vec p_\bot}\partial_s r_s(\vec p_\bot)G_s^2(\vec p_\bot)\Big|_{\vec p=\vec 0}-8\lambda_{4,2}\frac{d}{dp_1^2} \sum_{\vec p_{\bot'}}\partial_s r_s(\vec p_{\bot'})G_s^2(\vec p_{\bot'})\Big|_{\vec p=\vec 0},
\eea 
and now solve the linear equation of the form $\partial_s Z=L(\partial_s Z)$, which leads to 
\bea
\eta&=&\frac{12\pi(\pi\bar\lambda_{4,1}+8\bar\lambda_{4,2})}{3(\bar m^2+1)^2-\pi(3\pi\bar\lambda_{4,1}+32\bar\lambda_{4,2})}.
\eea

\subsection{Fixed points in the UV regime}\label{acide1}
\noindent
Solving numerically the system \eqref{systnew2}, we find some discrete non-Gaussian fixed points, whose relevant characteristics are summarized in  Table \eqref{table1} below. Also we give the critical exponents $\theta^{(i)},\,i=1,2,3,4$ of these different fixed points, corresponding to the opposite signs of the eigenvalues of the \textit{stability matrix} :$\beta_{ij}:=\partial_i\beta_j\, i\in\{m^2,\lambda_1,\lambda_2,\lambda_3\}$.\\

\noindent
Let us remember that the fixed points are chosen in the  domain $\{D>0\}$ where $D$ is the denominator of the anomalous dimension $\eta$ i.e.
\bea
D=3(\bar m^2+1)^2-\pi(3\pi\bar\lambda_{4,1}+32\bar\lambda_{4,2}).
\eea
 This denominator introduces a singularity in the phase space of the flow.   But further away from the Gaussian fixed point (GFP) , i.e. the non-Gaussian fixed points  $D$ may vanish, creating in the plan $(\lambda_i,\bar m^2),\, i\in\{(4,1), (4,2),(6,1)\}$, a singularity hyper-surface with equation $D=0$. The area below this line where $D < 0$ is thus disconnected from the region $D > 0$ connected to these fixed points. Then, we ignore for our purpose the fixed points in the disconnected region, for which $D < 0$. Note that we get such a singularity surface with the EVE method, get close to the one discussed here for small enough couplings. 
Numerically we get the fixed point given in Table \eqref{table1}.
\begin{table}
\begin{center}
\begin{tabular}{|l|l|l|l|l|l|l|l|l|l|l|l|l|l|}
\hline FP & $\bar{m}^2$&$10^4\bar{\lambda}_{4,1}$&$10^4\bar{\lambda}_{4,2}$ &  $10^{6}\bar{\lambda}_{6,1}$ &  $\eta$&$\theta^{(1)}$&$\theta^{(2)}$&$\theta^{(3)}$&$\theta^{(4)}$\\
\hline  $FP_1$ & -0.64 &0&-5.6 & -4.1 &0.513&12+13i&12-13i&-14+13i&-14-13i\\
\hline  $FP_2$ &-0.52&28&0&0&0.553&2.5&-3&1+0.6i&1-0.6i\\
\hline  $FP_3$ &1.36&0& -370&-70& -0.545&3.08&-0.41&-0.75&-1.4\\
\hline
\end{tabular}
\caption{Summary of  the non-Gaussian fixed points.}\label{table1}.
\end{center}
\end{table}
In this table, let
 us remark that $FP2$ corresponds to what we obtained by taking into account only the melonic contributions which is investigated  in our previous work 
\cite{Lahoche:2018vun}.
Now let us discuss the behavior of our model along these non-Gaussian fixed points. For more explanation see \cite{Nagy:2012ef}. Note that  we have four dimensionless couplings, thus the theory space is 4-dimensional.
The signs of the eigenvalues determine whether we approach
or go away from the fixed point where the linearization is performed.
 Consider the UV limits, i.e. $k\to\infty$. According to the Table \eqref{table1}, we have:
 \begin{enumerate}
 \item For the fixed point FP1, two pairs of complex conjugate critical exponents, with reals parts having opposite signs. The fixed point is then focused attractive in the plan defined from two eigenvectors, and focused repulsive in the complementary plan, spanned from the two remaining eigenvectors. 
 
 \item For the fixed point FP2, we have one relevant direction, one irrelevant direction, and a focusing attractive behavior in the complementary plan. 
 
 \item For the fixed point FP3, we get one relevant and four irrelevant directions. 
 \end{enumerate}

\end{appendices}

\end{document}